\documentclass{aastex631}

\shorttitle{NUACF and NUCCF for Nonuniform Sampling}
\shortauthors{Hu et al.}
\graphicspath{{./}{figures/}}
\usepackage{CJK}
\usepackage{amsmath}

\usepackage{threeparttable}
\usepackage{booktabs}
\usepackage{hyperref}
\usepackage[all]{hypcap}
\usepackage{CJK}
\makeatletter

\newcommand{\Rmnum}[1]{\expandafter\@slowromancap\romannumeral #1@}
\makeatother

\begin{document}

\begin{CJK*}{UTF8}{gbsn}

\title{A Missing Tool for Calculating Auto/Cross-correlation
Function under Nonuniform Sampling Observations}

\author[0000-0002-5238-8997]{Chen-Ran Hu (胡宸然)}
\affiliation{School of Astronomy and Space Science, Nanjing
University, Nanjing 210023, China}

\author[0000-0001-7199-2906]{Yong-Feng Huang (黄永锋)}
\thanks{Email: hyf@nju.edu.cn}
\affiliation{School of Astronomy and Space Science, Nanjing
University, Nanjing 210023, China}
 \affiliation{Key Laboratory of
Modern Astronomy and Astrophysics (Nanjing University), Ministry
of Education, China}

\author[0000-0001-9648-7295]{Jin-Jun Geng (耿金军)}
\affiliation{Purple Mountain Observatory, Chinese Academy of
Sciences,  Nanjing 210023, China}

\author[0000-0003-3230-7587]{Orkash Amat (吾热卡西·艾麦提)}
\affiliation{School of Astronomy and Space Science, Nanjing
University,  Nanjing 210023,  China}

\author[0000-0002-6189-8307]{Ze-Cheng Zou (邹泽城)}
\affiliation{School of Astronomy and Space Science, Nanjing
University, Nanjing 210023, China}

\author[0000-0002-2191-7286]{Chen Deng (邓晨)}
\affiliation{School of Astronomy and Space Science, Nanjing
University, Nanjing 210023, China}

\author[0000-0001-7943-4685]{Fan Xu (许帆)}
\affiliation{Institute of Space Weather, School of Atmospheric Physics,
Nanjing University of Information Science and Technology, Nanjing 210044,
China}

\author[0009-0000-0467-0050]{Xiao-Fei Dong (董小飞)}
\affiliation{School of Astronomy and Space Science, Nanjing
University, Nanjing 210023,  China}

\author[0009-0002-8460-1649]{Chen Du (杜琛)}
\affiliation{School of Astronomy and Space Science, Nanjing
University, Nanjing 210023,  China}

\author[0000-0001-9227-3716]{Nurimangul Nurmamat (努尔曼古丽·努尔麦麦提)}
\affiliation{Guangxi Key Laboratory for Relativistic Astrophysics, 
School of Physical Science and Technology, Guangxi University, Nanning 
530004, China}

\author[0000-0002-3386-7159]{Pei Wang (王培)}
\affiliation{State Key Laboratory of Radio Astronomy and Technology, 
NAOC, Chinese Academy of Sciences, Beijing 100101, China}
 \affiliation{Institute for Frontiers in Astronomy and Astrophysics, Beijing
Normal University, Beijing 102206, China}

\author[0000-0003-0721-5509]{Lang Cui (崔朗)}
\affiliation{State Key Laboratory of Radio Astronomy and Technology, 
Xinjiang Astronomical Observatory, CAS, 150 Science 1-Street, Urumqi, 
Xinjiang, 830011, China}
 \affiliation{Xinjiang Key Laboratory of Radio Astrophysics, 150 Science
1-Street, Urumqi 830011, China}

\author[0000-0002-9159-8129]{Cheng-Ming Li (李程明)}
\affiliation{Institute for Astrophysics, School of Physics, Zhengzhou
University, Zhengzhou 450001, China}

\begin{abstract}

Nonuniform sampling presents a long-standing challenge
in astrophysical time-domain analysis, invalidating the standard
autocorrelation and cross-correlation functions and forcing
researchers to adopt ad-hoc methods like interpolation or binning,
which introduce unquantified biases and lack rigorous error
estimation. Here we introduce a new method for calculating the
nonuniform autocorrelation function (NUACF) and nonuniform
cross-correlation function (NUCCF) for irregularly sampled time
series. Instead of relying on interpolation, it naturally
evaluates the correlation function by incorporating time-interval
weights and misalignment penalties. Monte Carlo simulations provide 
confidence bands for significance assessment and a complete error budget 
for the time delays that accounts for both flux uncertainties and 
sampling irregularity (essential but generally absent from 
existing methods). Through extensive simulations, we
demonstrate that our method outperforms traditional methods
across various conditions, from strictly periodic to complex
repeating variability patterns (e.g., intermittent but aperiodic). 
Its effectiveness is demonstrated
via various real astrophysical data sets, revealing repetitive
variability in stellar light curves, measuring time delays for
multi-band disc reverberation in the AGN Fairall 9, and providing
model-independent validation of time delays for the
gravitationally lensed quasar HE 0435-1223. The method provides a
rigorous and general solution to the ubiquitous problem of
nonuniform sampling, positioning it as a useful tool for
large-scale time-domain survey data analysis. The framework is also
directly applicable to emerging time-domain phenomena such as fast radio
bursts (FRBs), enabling, for example, the study of correlations between
persistent radio source luminosity and repeating FRB activity, or among
the multi-parameter variability curves of FRB emission itself.
\end{abstract}



\section{Introduction}
\label{sec1:Introduction}

Autocorrelation function (ACF) and cross-correlation function
(CCF) are useful mathematical tools for quantifying repetitive
structures of a timing series or the connection between two timing
series. Their core concept of measuring similarity permeates
various branches of astrophysics, finding applications across the
temporal, spatial, and frequency domains.

For continuous functions or uniformly sampled timing series,
ACF/CCF can be conveniently calculated. Classic applications
include characterizing the variability timescales of active
galactic nuclei (AGN) \citep{2009ApJ...698..895K}, describing
charged-particle motion in turbulent magnetic fields relevant to
cosmic-ray propagation \citep{1966ApJ...146..480J}, formulating
analytical models for the gravitational clustering of dark matter
halos \citep{1996MNRAS.282..347M}, developing a framework for
analyzing full-sky cosmic microwave background temperature and
polarization maps \citep{1997PhRvD..55.7368K}, and employing the
two-point correlation function to describe the distribution of
cosmic matter \citep{1980lssu.book.....P, 1983ApJ...267..465D,
1993ApJ...412...64L, 1999MNRAS.303..188K, 2018MNRAS.475..676S}.
Furthermore, since the power spectral density and the ACF form a
Fourier transform pair (the Wiener-Khinchin theorem), theoretical
studies of the power spectrum can be viewed as equivalently
specifying a particular form of ACF. Examples encompass modeling
AGN X-ray variability driven by thermal fluctuations in accretion
disks \citep{1997MNRAS.292..679L} and outlining a method to
generate time series with power-law power spectrum
\citep{1995A&A...300..707T}.


When applied to observational data, ACF/CCF are usually calculated
by introducing an additional normalization factor to get the
normalized correlation function. Representative applications cover
diverse tasks such as measuring stellar variability periods
\citep{2013MNRAS.432.1203M}, disentangling various stellar
oscillation modes in asteroseismology \citep{2009A&A...508..877M},
analyzing long-term periodicity in fast radio bursts (FRBs)
\citep{2025ApJ...983L..15P}, combining with classical reverberation
mapping to constrain AGN broad-line region sizes and central black
hole masses \citep{2004ApJ...613..682P, 2014MNRAS.445.3055P} and
with intensive disc-reverberation mapping to probe AGN accretion
disk structures \citep{2017MNRAS.466.1777P, 2020MNRAS.498.5399H},
determining spatial scales of the solar wind
\citep{2005JGRA..110.2104K}, measuring baryon acoustic oscillations
in survey data via the two-point correlation function
\citep{2005ApJ...633..560E, 2011MNRAS.416.3017B,
2017A&A...608A.130D}, and constraining the sizes of FRB emission regions
based on spectral lags \citep{2025Natur.637...48N}.

In practice, astronomical observations are usually performed
through nonuniform sampling, which poses a major challenge to the
correlation analysis, with the exception of frequency-domain
measurements such as spectral lags [e.g.,
\citet{2025Natur.637...48N}]. This renders the normal methods
involving uniformly sampled ACF/CCF inapplicable, forcing
researchers to seek adaptive methods. In spatial-domain
applications involving the two-point correlation function,
nonuniformity often arises from finite sky coverage due to
Galactic avoidance, field tiling, or cosmic extinction. The
established remedy is the Landy-Szalay estimator
\citep{1993ApJ...412...64L}, which effectively eliminates
systematic biases introduced by finite survey geometry.

\setcounter{figure}{0}
\makeatletter
\renewcommand*{\fnum@figure}{{\normalfont\bfseries \figurename~\thefigure}}
\makeatother

\begin{figure}[!htbp]
\centering
\includegraphics[width=0.75\textwidth]{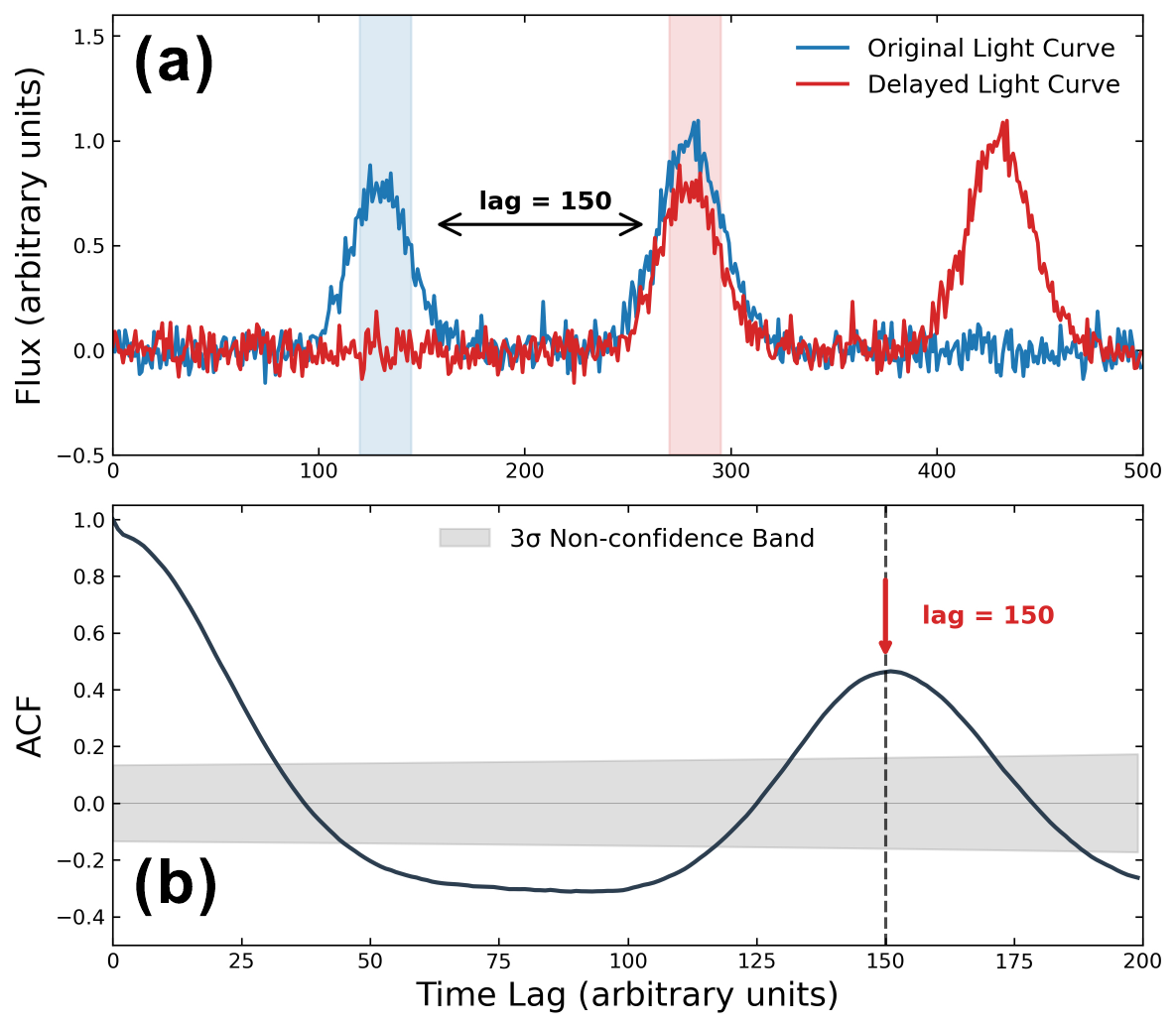}
\caption{Schematic illustration of the standard ACF
procedure under uniform sampling. (a) A light curve (blue), which
exhibits repeating variability patterns, is shifted by a trial
time delay to become the shifted copy (red). (b) Repeating this
procedure over a range of trial time delays yields the ACF
profile. The horizontal gray region indicates the 3$\sigma$
non-confidence band, meaning that an ACF value falling within this
band has a 99.73\% probability of arising from random statistical
fluctuations of the data points. By convention, however, this
region is still commonly referred to as the ``confidence band''. A
significant peak above this band indicates the recurrence
timescale of the variability.} \label{Fig0}
\end{figure}

Throughout this work, we focus on the normalized ACF and
CCF, the forms yielding correlation coefficients in the range
$\left[-1,1\right]$, as these are the quantities typically used
for statistical analysis such as temporal correlation and feature
extraction. This is quite distinct from the unnormalized versions
commonly employed in data processing and theoretical astrophysics,
which differ by a normalization term. We emphasize that such
time-domain correlation analysis is significantly different from
frequency-domain methods such as the Fourier transform or the
Lomb-Scargle periodogram
\citep{1976Ap&SS..39..447L,1982ApJ...263..835S,2018ApJS..236...16V},
which aim to identify periodicities rather than directly
characterize repeating temporal patterns (including intermittent,
aperiodic ones).

The core idea of the normalized ACF is sketched in Figure
\ref{Fig0}(a): one shifts a light curve by a trial time delay,
computes how well it matches itself, and obtains a single
correlation coefficient. Repeating this over a range of trial time
delays yields the ACF profile as shown in Figure \ref{Fig0}(b). A
significant peak (one that rises above the confidence band) then
indicates the timescale of the repeating patterns in the light
curve. The normalized CCF shares the same logic, but compares two
different light curves. We note that Figure \ref{Fig0} illustrates
the standard ACF procedure under uniform sampling, where a
well-defined confidence band is readily available. However, in the
case of nonuniform sampling, such a band is generally unavailable,
which leads to difficulty in the ACF/CCF analysis. 

Temporal sampling irregularity,
stemming from factors such as source visibility, Earth rotation
and orbit motion, instrumental maintenance, and observing schedule
competition, constitutes an impediment distinct from the geometric
effects in spatial sampling. Traditional approaches often rely on
interpolating or resampling the unevenly sampled time series
\citep{2018ascl.soft05032S} or binning time delays, as adopted in
the discrete correlation function (DCF) method
\citep{1988ApJ...333..646E}. While these methods mitigate issues
from irregular sampling, they introduce additional, usually
unquantified artificial biases, leading to three key
shortcomings: \\
(1) Lack of confidence estimation. Standard sample ACF/CCF for
uniform data provides confidence intervals, allowing one to
distinguish genuine peaks and troughs from noise.
Interpolated/resampled ACFs typically lack this, reducing their
utility to mere period detection (via equally spaced
peaks/troughs) and impairing the identification of aperiodic
recurring patterns and their delays. Similarly, while DCF can
indicate a delay at a
peak, it offers no direct statistical assessment of the peak's validity. \\
(2) Inability to assess the significance of identified recurring patterns.
Closely related to the first point, the absence of confidence intervals
precludes a robust significance evaluation for potential recurring
patterns associated with ACF/CCF features. \\
(3) Incomplete error estimation for time-delay measurements. A
complete error budget for a time delay should incorporate both
flux measurement uncertainties and the effects of sampling
irregularity. In traditional approaches, Monte Carlo (MC)
simulations are typically introduced to handle the former but do
not capture the latter. Specifically, interpolation/resampling
effectively removes the sampling-induced error but replaces it
with an artificial, nonanalytic error that cannot be propagated
rigorously. In the case of the DCF, the binning procedure also
introduces artificial biases. The empirically chosen bin width trades
temporal resolution for robustness: narrower bins increase resolution
but reduce the number of matched pairs per bin. The DCF results thus
depend on the bin width, because binning compresses the scatter of time
delays within each bin into a single value (an inherently approximative
step). This prevents the DCF from being fully analytic and hinders a
complete uncertainty estimate, despite partially absorbing the effects
of sampling irregularity.

Some existing approaches seek to mitigate artifacts in correlation
analyses under irregular sampling, for instance, through
conservative interpolation schemes such as zero padding or via
data selection methods like the S-ACF \citep{2023MNRAS.522.5049K}.
While these ideas offer valuable insights, they remain
non-analytic or semi-analytic in nature and therefore cannot fully
resolve the problems outlined above \citep{1992ApJ...385..404P,
1997ASSL..218..163A}.

It is noteworthy that the seminal work of \citet{Franks1981}
derived the theoretical ACF for an ideal Poisson sampling process.
This formalism was applied by \citet{2023MNRAS.522.4907Y} to
model the propagation of FRBs in the magnetosphere,
assuming a coherent curvature radiation mechanism. While this is
viable for theoretical modeling, it is generally unsuitable for
inverse feature extraction from observations because real
observational sampling patterns often deviate significantly from
the Poisson assumption due to limited observing schedules and
other artificial factors.

Consequently, a unified, analytical, tractable and model-agnostic
framework for correlation analysis of nonuniform sampling remains
lacking. This work aims to fill this gap by introducing a
generalized, analytical ACF/CCF method that is applicable to
nonuniformly sampled observations, providing confidence intervals,
significance assessments, and a complete error budget for timing
series. It degenerates to conventional ACF/CCF under uniform
sampling. This approach thus offers a more direct and
statistically rigorous tool for astrophysicists.

In Section \ref{sec2:Nonuniform autocorrelation function},
we present the derivation of the Nonuniform ACF (NUACF), detailing
the method for determining its confidence intervals and the error
budget for time-delay measurements. Section \ref{sec3:Nonuniform
cross-correlation function} extends this framework to the
Nonuniform CCF (NUCCF). Section \ref{sec4:Validation and
application in astronomical observations} demonstrates
the practical application of both the NUACF and NUCCF to real
astrophysical scenarios.


\section{Nonuniform Autocorrelation Function}
\label{sec2:Nonuniform autocorrelation function}


Our derivation of the NUACF follows a generalizable principle: to
extend estimators from the uniform to the nonuniform domain by
replacing simple averages with time-weighted averages. This is
most intuitively illustrated with the simpler case of variance.
For a continuous time series $x\left(t\right)$, the variance over
a duration $T$ is
$\sigma_{\rm{C}}^2=\frac{1}{T}\int_{0}^{T}{\left[x\left(t\right)-
\overline{x\left(t\right)}\right]^2{\rm{d}}t}$. Discretizing this
for observations
$\left\{\left.\left(x_i,t_i\right)\right|i=1,2,\cdots,N\right\}$
naturally leads to a time-weighted estimator:
$\sigma_{\rm{D}}^2=\frac{1}{t_N-t_1}\sum_{i=1}^{N}{\left(x_i-
\bar{x}\right)^2\Delta t_i}$. For uniformly sampled data, where
all time intervals are equal ($\Delta t_i\equiv\Delta t$, implying
$t_N-t_1=\left(N-1\right)\Delta t$), this reduces to the standard
sample variance. For nonuniform sampling, applying a numerical
quadrature rule (e.g., the trapezoidal rule) provides a robust
nonuniform variance estimator, $\sigma_{\rm{D,NU}}^2$. This
transition from an integral to a time-weighted discrete sum forms
the foundational template for generalizing the ACF. When applied
to a uniform data set, we get the uniform sample ACF,
${\rm{acf_{D,U}}}\left(k\right)$.

\begin{figure}[!htbp]
\centering
\includegraphics[width=0.75\textwidth]{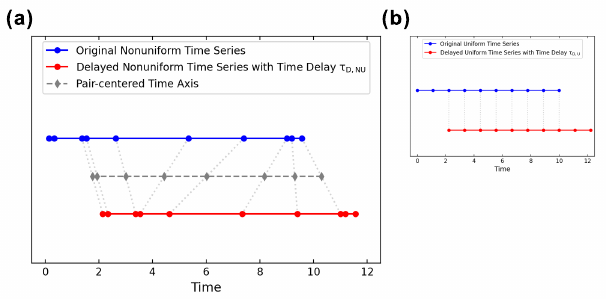}
\caption{Temporal misalignment in nonuniform
autocorrelation. (a) A nonuniformly sampled time series is
illustrated. The blue dots mark the original observation times.
Shifting the sequence by a lag of $k$ data points produces the red
series, whose time stamps do not align with the original ones. To
compare these misaligned pairs (linked by gray dotted lines), a
pair-centered time axis (gray dashed line) is constructed, defined
as the midpoint between the two times in each matched pair.
This axis is a virtual computational device solely for locating
each pair in time and is algebraically eliminated in the final estimator
(see Appendix \ref{appendix A:Derivation of the nonuniform
autocorrelation function}). (b) For comparison, under uniform sampling,
the $k$-lag delayed series (red) can always align perfectly with the
original one (blue) in the time domain, yielding a fixed trial time
delay $k\Delta t$ and requiring no pair-centered time axis or
misalignment penalty.}
\label{Fig1}
\end{figure}

For nonuniform sampling, a shift by $k$ indices does not
temporally align data pairs (Figure \ref{Fig1}). The NUACF framework
addresses this by introducing a pair-centered time axis as a
derivation device and
solving two linked problems: (i) defining an optimal trial time delay
$\tau_{\rm{D,NU}}\left(k\right)$ that minimizes the
overall temporal misalignment for a given trial index lag $k$, and (ii) formulating a correlation measure
that incorporates both time-interval weights and a penalty for
temporal misalignment (Appendix \ref{appendix A:Derivation of the
nonuniform autocorrelation function}).

The optimal trial time delay is defined as the average temporal
separation for the corresponding $k$-th lag:
\begin{eqnarray}
\label{eq1}
\tau_{\rm{D,NU}}\left(k\right)=\frac{1}{N-k}\sum_{i=1}^{N-k}h_{i+k,i},
\end{eqnarray}
where $h_{m,n}=t_m-t_n$. This ensures $\tau_{\rm{D,NU}}\left(k\right)$
reduces to $k\Delta t$ under uniform sampling and minimizes the expected
residual misalignment to zero (Appendix \ref{appendix A:Derivation of
the nonuniform autocorrelation function}).

The correlation measure is then calculated by applying the
time-weighted averaging principle to both the numerator and
denominator of the dimensionless sample ACF. The local sampling
density is incorporated via discrete weight factors
$H_i^{\left(1\right)}$ and $H_i^{\left(2\right)}$ derived from the
trapezoidal rule, while a Gaussian kernel weight $w_i$ penalizes
residual misalignment within each pair. This leads to the compact,
final form of the NUACF:
\begin{eqnarray}
\label{eq2}
{\rm{acf_{D,NU}}}\left(k\right)=\frac{h_{N,1}}{h_
{N-k,1}+h_{N,k+1}}\frac{\sum_{i=1}^{N-k}{\left(x_i-
\bar{x}\right)\left(x_{i+k}-\bar{x}\right)H_i^{\left(2\right)}w_i}}
{\sum_{i=1}^{N}{\left(x_i-\bar{x}\right)^2H_i^{\left(1\right)}}},\quad
k\in\mathbb{N},k\le N-10.
\end{eqnarray}
The detailed definitions of $H_i^{\left(1\right)}$,
$H_i^{\left(2\right)}$ and $w_i$ are provided in
Appendix \ref{appendix A:Derivation of the nonuniform autocorrelation
function}. This formulation is self-consistent and degenerates exactly to
the standard sample ACF under uniform sampling.

To assess whether a peak in the NUACF signifies a real
correlation, we require confidence intervals under the null
hypothesis of white noise. For uniformly sampled data, this leads
to the standard confidence estimation of $\pm
z_{\alpha/2}/\sqrt{N-k}$. Extending this analytically to
nonuniform sampling is complex. Under the assumption that the data
points are sampled following a Poisson process, a theoretical
NUACF confidence interval can be derived (see
Appendix \ref{appendix B:Confidence intervals for the nonuniform
autocorrelation function} for the complete derivation) as
\begin{eqnarray}
\label{eq3}
{\rm{acf_{D,NU}^{noise}}}\left(k\right)\in\left[
-z_{\alpha/2}V\left(k\right),z_{\alpha/2}V\left(k\right)\right],
\end{eqnarray}
where $V\left(k\right)$ is a function of the lag $k$, the observed
timestamps $\left\{t_i\right\}$, and the data variances.

\begin{figure}[!htbp]
\centering
\includegraphics[width=0.7\textwidth]{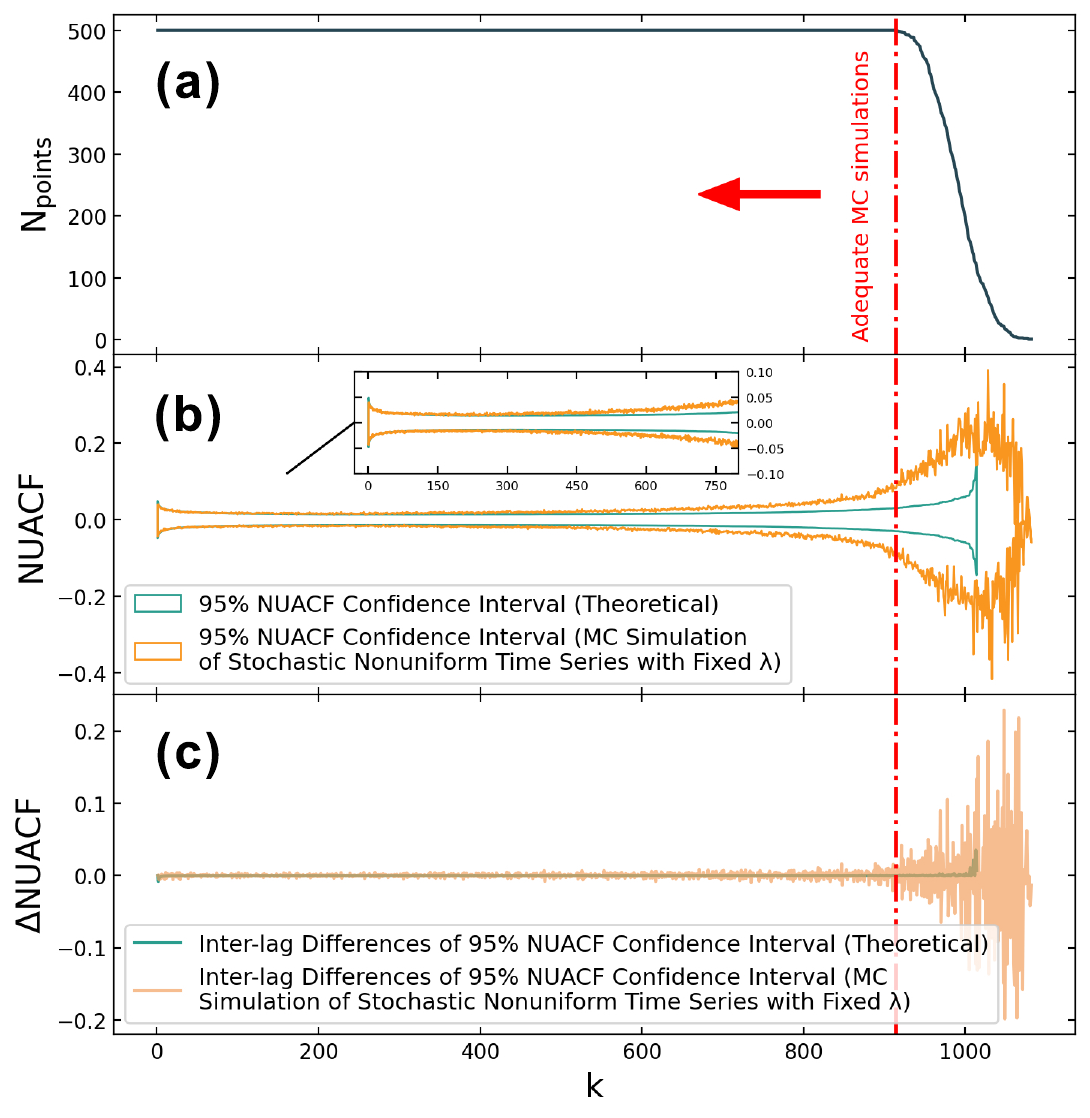}
\caption{Validation of the theoretical nonuniform
autocorrelation function (NUACF) confidence interval under ideal
Poisson sampling. (a) Number of available data
points ($N_{\rm{points}}$) used to calculate the Monte Carlo (MC)
confidence interval as a function of lag $k$. The vertical red
dash-dotted line marks the lag beyond which $N_{\rm{points}}$
falls below the total number of simulations (500); only the region
to its left is considered meaningful for estimation.
(b) Comparison of the theoretical $95\%$ confidence interval
[solid green line, from Equation (\ref{eq3})] with that derived
from 500 MC simulations (solid orange line) where both observation
times and fluxes were randomly generated following a Poisson
process. The two intervals agree well within a reliable lag range.
(c) Differences of the upper confidence bounds, shown
for both the theoretical (green) and MC (orange) cases. The
discrepancy of the simulated confidence interval increases with
$k$, which is due to fewer matched pairs. The NUACF loses its
effectiveness beyond the reliable range.}
\label{Fig2}
\end{figure}

Figure \ref{Fig2} compares the theoretical NUACF confidence interval
with that obtained via MC simulations using white-noise sequences.
In the MC case, for a given event rate $\lambda$, Poisson sampling
is applied to randomly generate observation times and flux values
in each simulation, yielding a white-noise series. The NUACF
values at all lags $k$ are recorded in the simulation. After
completing all runs, the NUACF values at each $k$ follow a normal
distribution, from which the $100\left(1-\alpha\right)\%$
confidence interval is directly extracted to get the MC-based
confidence estimation. A total of 500 simulations were conducted,
providing 500 NUACF values at each $k$. However, due to Poisson
sampling, the sample size differs across different simulations
even under the same $\lambda$, leading to variation in the maximum
computable $k$ in each run. As shown in Figure \ref{Fig2}(a), the
number of available data points ($N_{\rm{points}}$) for
constructing the MC confidence interval decreases as $k$
increases. To ensure reliability, we consider a range of $k$ as
adequate where $N_{\rm{points}}$ remains equal to the total number
of simulations (500), indicated by the vertical dash-dotted line
at the left side.

Figure \ref{Fig2}(b) demonstrates that within this effective range,
the theoretical NUACF confidence interval given by Equation
(\ref{eq3}) agrees well with the MC-derived interval. Although
minor deviations emerge at larger $k$, they still remain small
enough (about 0.05 at most). These deviations originate from the
finite number of simulations because, as $k$ grows, the number of
matched pairs available for NUACF computation decreases, reducing
the robustness of the estimated NUACF [Equation (\ref{eq35}) in
Appendix \ref{appendix B:Confidence intervals for the nonuniform
autocorrelation function}] and thus causing the MC confidence interval to gradually
depart from the theoretical one. In principle, such a deviation would be
nonexistent for an infinite sample size case.

Figure \ref{Fig2}(c) displays the inter-lag backward differences of the upper confidence
bounds (the lower bounds behave symmetrically), comparing the theoretical
and simulated cases. The discrepancy increases with $k$,
confirming the aforementioned decrease in matched-pair counts. To
the right of the vertical red dash-dotted line, the instability
rises sharply.

\begin{figure}[!htbp]
\centering
\includegraphics[width=0.75\textwidth]{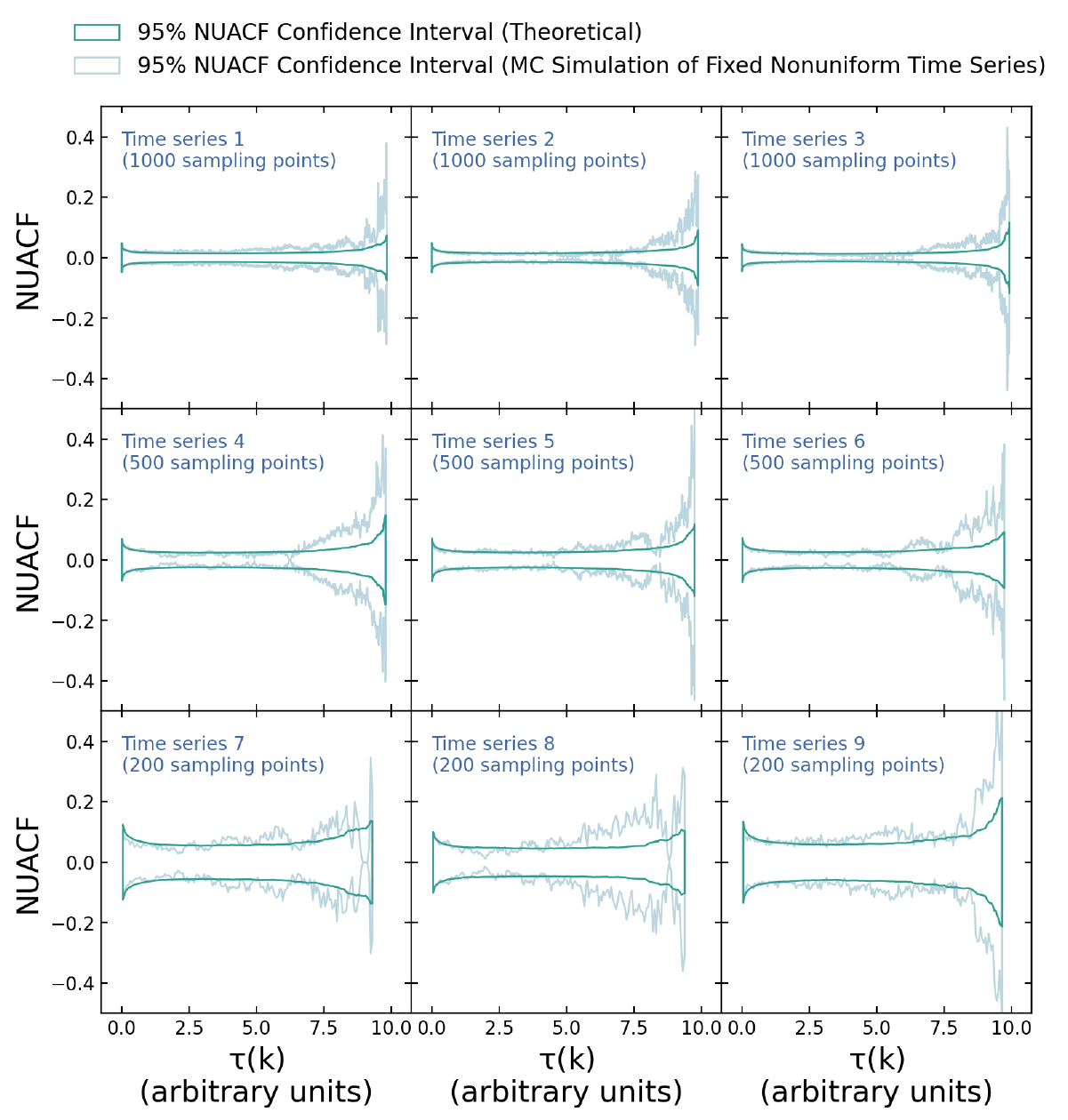}
\caption{Limitation of the theoretical confidence interval
for fixed observational timestamps. Nine independent sets of
simulations are shown, where the observation times are fixed and
only the flux values are randomized as white noise in each MC
realization. In each panel, the solid green line shows the
theoretical interval [Equation (\ref{eq3})], and the solid blue
line shows the MC case constructed from 500 realizations. While
the theoretical interval follows the general trend, it does not
capture the detailed, structured deviations of the MC interval,
which is a persistent discrepancy regardless of the number of
sampling points. It demonstrates that a confidence interval
derived from fixed-time MC simulations is more reliable than the
purely analytic result derived under a Poisson-sampling
assumption.}
\label{Fig3}
\end{figure}

At first glance, the theoretical NUACF confidence interval
provided by Equation (\ref{eq3}) appears to provide a good fit to
the simulations in Figure \ref{Fig2}. However, it is crucial to
note that the simulations in Figure \ref{Fig2} randomly vary with
both observation times and fluxes. In real observations, the
sampling timestamps are actually fixed. Therefore, for an actual
time series, the observation times cannot be randomized, and the
theoretical interval will not perfectly match an MC interval
constructed with fixed times. To clarify this point, a second type
of simulation was performed, where observation times were fixed
across runs while the flux values were randomized following a
white-noise process. Figure \ref{Fig3} presents nine sets of such
simulations, each set corresponding to a distinct fixed time
series. While the theoretical interval generally captures the
overall trend, it does not align closely with the MC interval, a
behavior that persists regardless of the number of sampling
points. Thus, although the theoretical confidence interval derived
under ideal Poisson sampling is theoretically sound (Figure
\ref{Fig2}), it is less practical than an MC interval constructed
with fixed observation times, because real observational times are
fixed and thus not fully amenable to analytic modeling.
Consequently, we suggest that confidence intervals for NUACF
should be derived via MC simulation, which is adopted in all our
calculations below.

When a significant NUACF peak is identified, determining the
corresponding time delay $\tau_{\rm{D,NU}}$ with a definite
uncertainty is useful. Two factors may contribute to the
uncertainty: one is the inherent irregularity of the sampling
times, and the other is the flux measurement error.

The uncertainty due to temporal irregularity can be derived from
the dispersion of the pairwise time differences used in Equation
(\ref{eq1}):
\begin{eqnarray}
\label{eq4}
\varepsilon_t\left[\tau_{\rm{D,NU}}\left(k\right)\right]=\frac{1}
{N-k}\sqrt{\sum_{i=1}^{N-k}\left(h_{i+k,i}-\overline{h_{i+k,i}}\right)^2}.
\end{eqnarray}
The flux error contribution can be assessed via MC simulations,
where fluxes are randomly perturbed within their measurement
uncertainties. For a significant peak ${\rm{P}}_\varphi$, once all
MC runs are completed, the time delay of
$\left\{\left.\tau_{\rm{D,NU}}^{(\xi)}\left(k_{{\rm{P}}
_\varphi}\right)\right|\xi=1,2,\cdots,S\right\}$ and the
corresponding temporal irregularity errors
$\left\{\left.\varepsilon_t\left[\tau_{\rm{D,NU}}^
{(\xi)}\left(k_{{\rm{P}}_\varphi}\right)\right]\right|\xi=1,2,\cdots,S\right\}$
can be obtained. The final total uncertainty synthesizes the
contributions from the two factors, which reads
(Appendix \ref{appendix C:Uncertainty of time delays})
\begin{eqnarray}
\label{eq5}
\varepsilon_{\rm{total}}\left[\tau_{\rm{D,NU}}\left(k_{{\rm{P}}_\varphi}
\right)\right]=\sqrt{\frac{\sum_{\xi}\left[\tau_{\rm{D,NU}}^{(\xi)}\left(k_
{{\rm{P}}_\varphi}\right)-
\overline{\tau_{\rm{D,NU}}^{(\xi)}\left(k_{{\rm{P}}_\varphi}
\right)}\right]^2}{S\left(S-1\right)}+\overline{\left\{\varepsilon_t
\left[\tau_{\rm{D,NU}}^{(\xi)}\left(k_{{\rm{P}}_\varphi}\right)\right]\right\}^2}},
\end{eqnarray}
where
$\overline{\tau_{\rm{D,NU}}^{(\xi)}\left(k_{{\rm{P}}_\varphi}\right)}$
is the final mean time delay (the ensemble mean). The first term
in the above Equation corresponds to the standard error of the
mean from the MC realizations, and the second corresponds to the
root-mean-square temporal irregularity error.

To conclude, our NUACF provides a model-independent framework for
autocorrelation analysis of irregularly sampled time series. It
delivers a well-defined time delay and correlation measure, along
with a robust estimate of the significance level and a useful
uncertainty for the time delay.



\section{Nonuniform Cross-correlation Function}
\label{sec3:Nonuniform cross-correlation function}

We generalize the NUACF framework to calculate the CCF of two
irregularly sampled time series,
$\left\{\left.\left(x_i,t_i^x\right)\right|i=1,2,\cdots,N\right\}$
and
$\left\{\left.\left(y_i,t_i^y\right)\right|i=1,2,\cdots,M\right\}$,
which presents a useful method for nonuniform cross-correlation
function (NUCCF) analysis. It provides a model-independent
estimate of both the correlation measure and the time delay
$\tau_{\rm{D,NU}}^{xy}\left(k\right)$ between the two series,
\begin{equation}
\label{eq6}
\begin{gathered}
{\rm{ccf_{D,NU}}}\left(k\right)=\frac{\sqrt{h_{N,1}^xh_{M,1}^y}}
{h_{i_{\rm{max}},i_{\rm{min}}}^x+h_{i_{\rm{max}}+k,i_{\rm{min}}+k}^y}
\frac{\sum_{i=i_{\rm{min}}}^{i_{\rm{max}}}{\left(x_i-
\bar{x}\right)\left(y_{i+k}-\bar{y}\right)H_i^{xy,\left(2\right)}w_i^{xy}}}
{\sqrt{\sum_{i=1}^{N}{\left(x_i-
\bar{x}\right)^2H_i^{x,\left(1\right)}}\sum_{i=1}^{M}{\left(y_i-
\bar{y}\right)^2H_i^{y,\left(1\right)}}}},\\
k\in\mathbb{Z},-\left(N-10\right)\le k\le M-10,
\end{gathered}
\end{equation}
where $i_{\rm{min}}={\rm{max}}\left(1,1-k\right)$,
$i_{\rm{max}}={\rm{min}}\left(N,M-k\right)$,
$h_{m,n}^x=t_m^x-t_n^x$ and $h_{m,n}^y=t_m^y-t_n^y$. The discrete
weight factors $H_i^{x,\left(1\right)}$, $H_i^{y,\left(1\right)}$,
and $H_i^{xy,\left(2\right)}$, derived via the trapezoidal rule,
and the misalignment weight $w_i^{xy}$ (see Appendix \ref{appendix
D:Extending to the nonuniform cross-correlation function}
for definitions) can effectively correct for nonuniform sampling
and temporal misalignment. The associated time delay at lag $k$ is
the mean temporal offset,
\begin{eqnarray}
\label{eq7}
\tau_{\rm{D,NU}}^{xy}\left(k\right)=\overline{t_{i+k}^y-t_i^x}=\frac{1}
{i_{\rm{max}}-i_{\rm{min}}+1}\sum_{i=i_{\rm{min}}}^
{i_{\rm{max}}}\left(t_{i+k}^y-t_i^x\right).
\end{eqnarray}

Similar to NUACF, significance of NUCCF is also assessed via MC
confidence interval, and the final uncertainty synthesizes
flux-error and nonuniform-sampling contributions. Note that for
the two series, we construct a conservative, envelope-based
confidence interval from two complementary MC procedures (fixing
one series and replacing the other with a white noise series, then
vice versa). The final uncertainty for a significant delay is:
\begin{eqnarray}
\label{eq8}
\varepsilon_{\rm{total}}\left[\tau_{\rm{D,NU}}^{xy}\left(k\right)\right]=
\sqrt{\frac{\sum_{\xi}\left[\tau_{\rm{D,NU}}^{xy,
(\xi)}\left(k_{{\rm{P}}_\varphi}\right)-\overline{\tau_{\rm{D,NU}}^{xy,
(\xi)}\left(k_{{\rm{P}}_\varphi}\right)}\right]^2}{S\left(S-
1\right)}+\overline{\left\{\varepsilon_t\left[\tau_{\rm{D,NU}}^{xy,
(\xi)}\left(k_{{\rm{P}}_\varphi}\right)\right]\right\}^2}},
\end{eqnarray}
where the temporal irregularity error is
\begin{eqnarray}
\label{eq9}
\varepsilon_t\left[\tau_{\rm{D,NU}}^{xy}\left(k\right)\right]=\sqrt{\frac
{\sum_{i=i_{\rm{min}}}^{i_{\rm{max}}}\left(t_{i+k}^y-t_i^x-
\overline{t_{i+k}^y-t_i^x}\right)^2}{\left(i_{\rm{max}}-
i_{\rm{min}}+1\right)\left(i_{\rm{max}}-i_{\rm{min}}\right)}}.
\end{eqnarray}

Our NUCCF framework provides a useful cross-correlation tool for
analyzing multi-instrument, multi-epoch astronomical data.


\section{Validation and Application in Astronomical Observations}
\label{sec4:Validation and application in astronomical observations}

\begin{figure}[!htbp]
\centering
\includegraphics[width=0.8\textwidth]{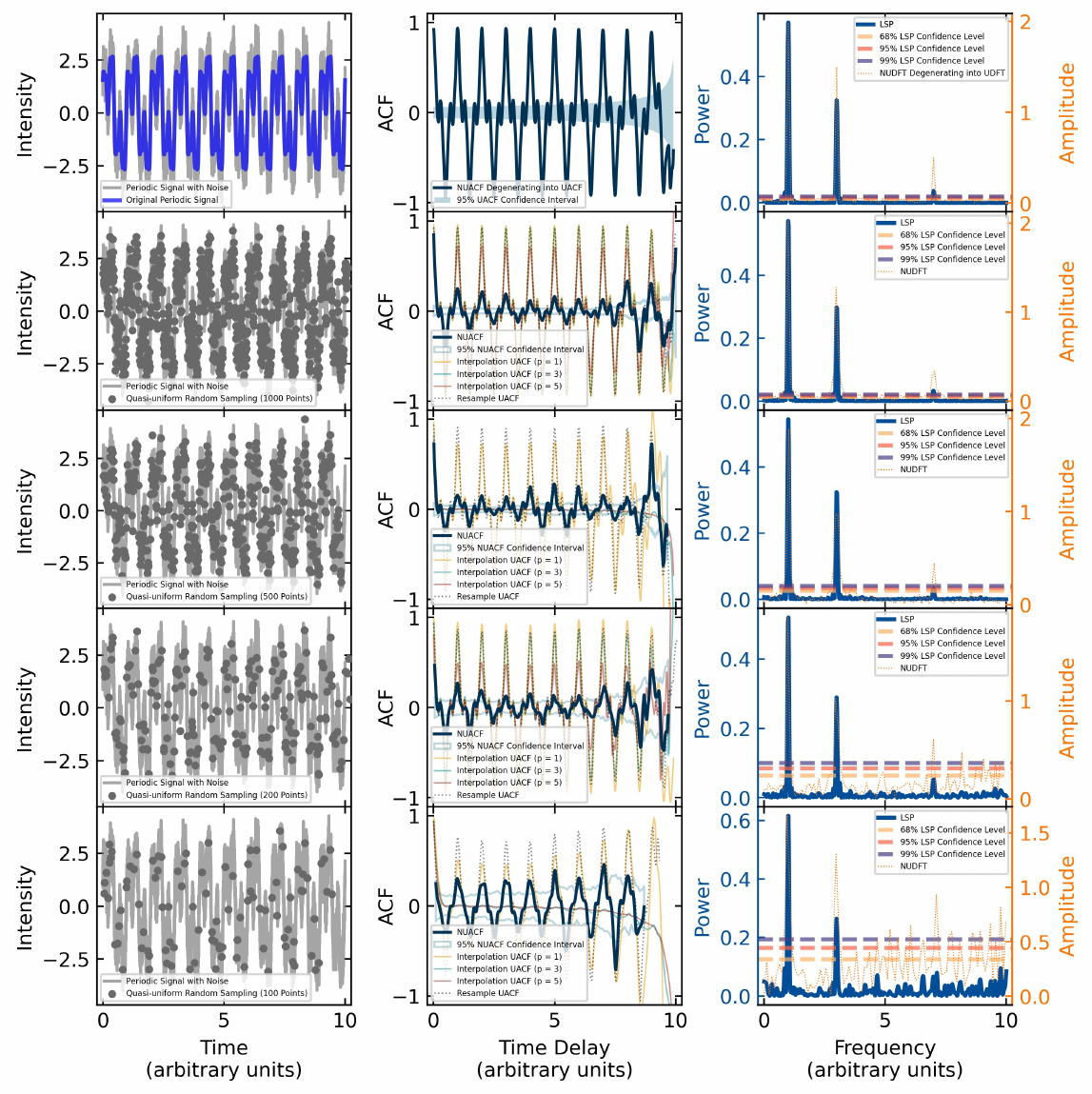}
\caption{Exemplary synthetic signal (Pattern A) under quasi uniform
sampling. The complete figure set (18 images) is available in the
online journal. For the first image, the description is: Shown here is
the ACF analysis of a quasi uniformly sampled periodic signal with noise
added, designated as Pattern A. Left column: The input periodic
signal, shown from densely uniform sampling (top) to progressively
sparser quasi-uniform sampling. Middle column: ACF analysis of the
corresponding signals to the left. The NUACF (dark blue line, with the
$95\%$ confidence interval shown as the blue region), resampled ACF
(dotted gray line), and interpolated ACF of orders $p=1,3,5$ (orange,
green, red) are shown. The top panel demonstrates that the NUACF reduces
exactly to the standard sample ACF under uniform sampling. Right
column: Power spectra of the signals to the left, from LSP (blue solid
line, left axis) and a nonuniform discrete Fourier transform (NUDFT,
orange dotted line, right axis) implemented via the trapezoidal rule.
[The complete figure set (18 images) is available in the online journal.]}
\label{Fig4}
\end{figure}

As the fist step, we have conducted a series of exhaustive
simulation tests to evaluate the performance of our NUACF and
NUCCF methods systematically. We first generated three different
synthetic signals exhibiting repetitive variability patterns
(e.g., simple periodic signals) along with injected noise. They
are sampled under both quasi-uniform (Figures
\ref{Fig4}.1--\ref{Fig4}.3) and nonuniform
(Figures \ref{Fig4}.4--\ref{Fig4}.6) schemes.
We then analyzed these signals using the resampled
ACF (which reconstructs a uniformly sampled series by assigning
the flux value of the temporally nearest observation to each grid
point, thus generating no new data), the interpolated ACF, and our
NUACF method. The results demonstrate that the resampled ACF, the
first-order interpolated ACF, and the NUACF can all successfully
identify the time delays corresponding to the repetitive patterns.
Note that our NUACF assesses the significance not by its curve
morphology but by considering whether the peaks/troughs lie
outside the corresponding confidence interval. We also notice that
for a fixed total time span, a higher sampling density yields
lower absolute NUACF values, but the confidence interval also
narrows accordingly, thus its ability to identify significant
features is not affected.

\newpage

\figsetstart
\figsetnum{5}
\figsettitle{Systematic simulation tests for NUACF performance under
diverse sampling and windowing conditions}

\figsetgrpstart
\figsetgrpnum{5.1}
\figsetgrptitle{Pattern A under quasi-uniform sampling}
\figsetplot{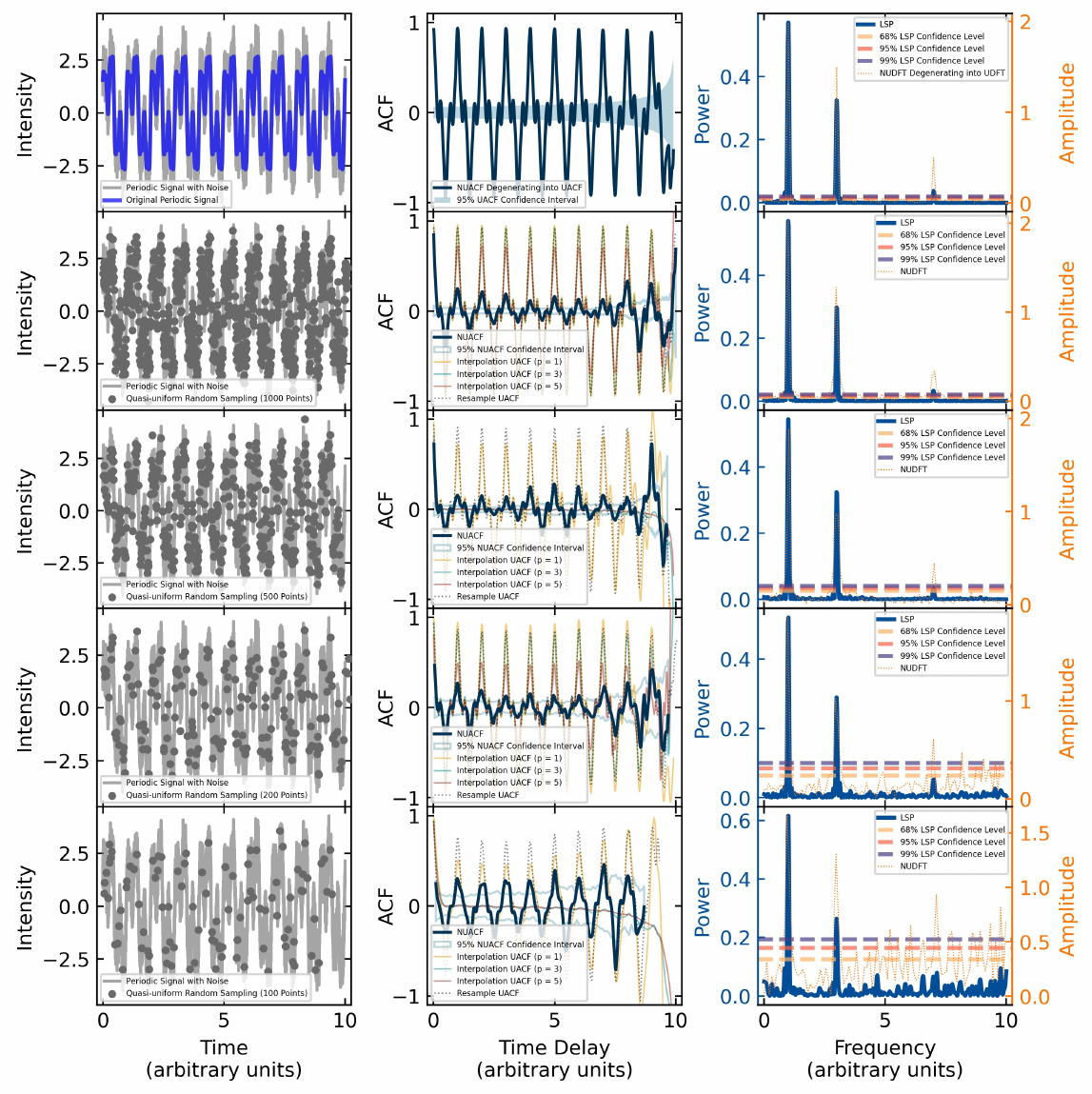}
\figsetgrpnote{Exemplary synthetic signal (Pattern A) under quasi
uniform sampling. Shown here is the ACF analysis of a quasi uniformly
sampled periodic signal with noise added, designated as Pattern A.
Left column: The input periodic signal, shown from densely uniform
sampling (top) to progressively sparser quasi-uniform sampling.
Middle column: ACF analysis of the corresponding signals to the
left. The NUACF (dark blue line, with the $95\%$ confidence interval
shown as the blue region), resampled ACF (dotted gray line), and
interpolated ACF of orders $p=1,3,5$ (orange, green, red) are shown. The
top panel demonstrates that the NUACF reduces exactly to the standard
sample ACF under uniform sampling. Right column: Power spectra of
the signals to the left, from LSP (blue solid line, left axis) and a
nonuniform discrete Fourier transform (NUDFT, orange dotted line, right
axis) implemented via the trapezoidal rule.}
\figsetgrpend

\figsetgrpstart \figsetgrpnum{5.2} \figsetgrptitle{Pattern B under
quasi-uniform sampling} \figsetplot{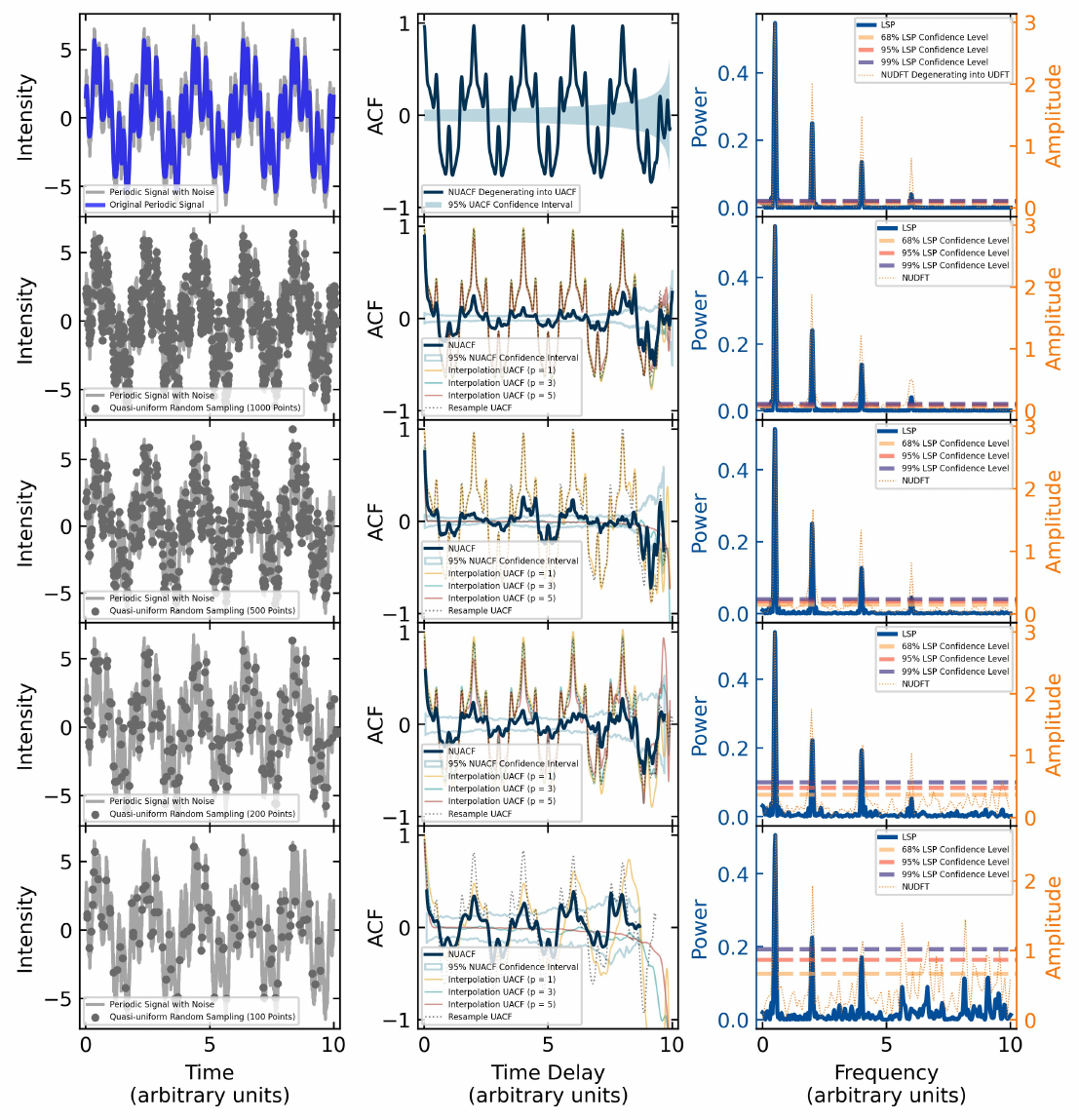}
\figsetgrpnote{Exemplary synthetic signal (Pattern B) under
quasi-uniform sampling. This figure presents the same analysis as
Figure \ref{Fig4}.1, applied to a different repetitive pattern
(Pattern B). The NUACF's significance is assessed via its
confidence interval, while traditional methods rely on profile
morphology.} \figsetgrpend

\figsetgrpstart \figsetgrpnum{5.3} \figsetgrptitle{Pattern C under
quasi-uniform sampling} \figsetplot{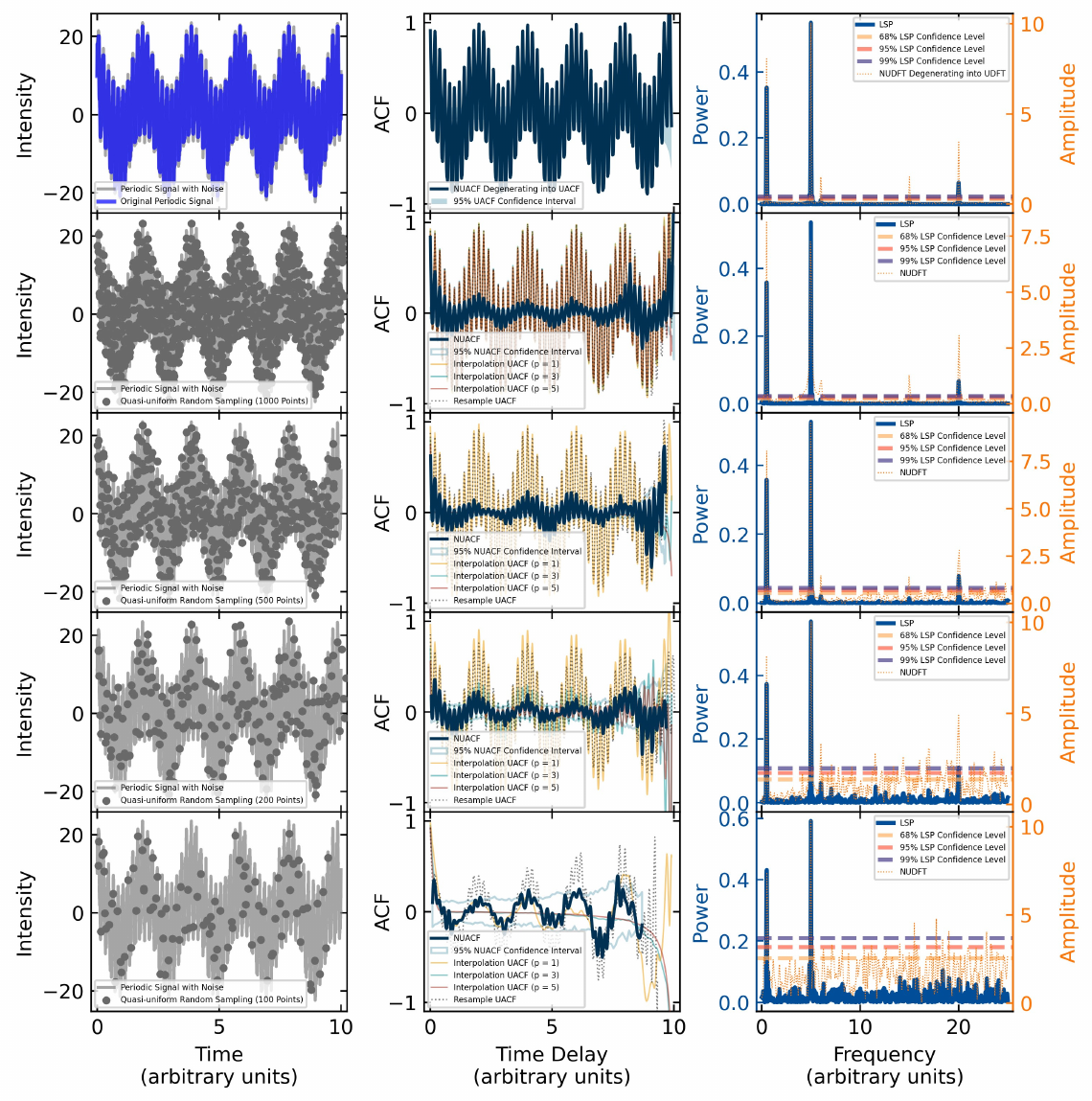}
\figsetgrpnote{Exemplary synthetic signal (Pattern C) under
quasi-uniform sampling. Following Figures \ref{Fig4}.1 and
\ref{Fig4}.2, this figure shows the analysis for Pattern C, a
third repetitive pattern under quasi-uniform sampling. The NUACF,
resampled ACF, and first-order interpolated ACF exhibit more
stable performance in identifying repetitive variability under low
sampling conditions.} \figsetgrpend

\figsetgrpstart
\figsetgrpnum{5.4}
\figsetgrptitle{Pattern A under nonuniform sampling}
\figsetplot{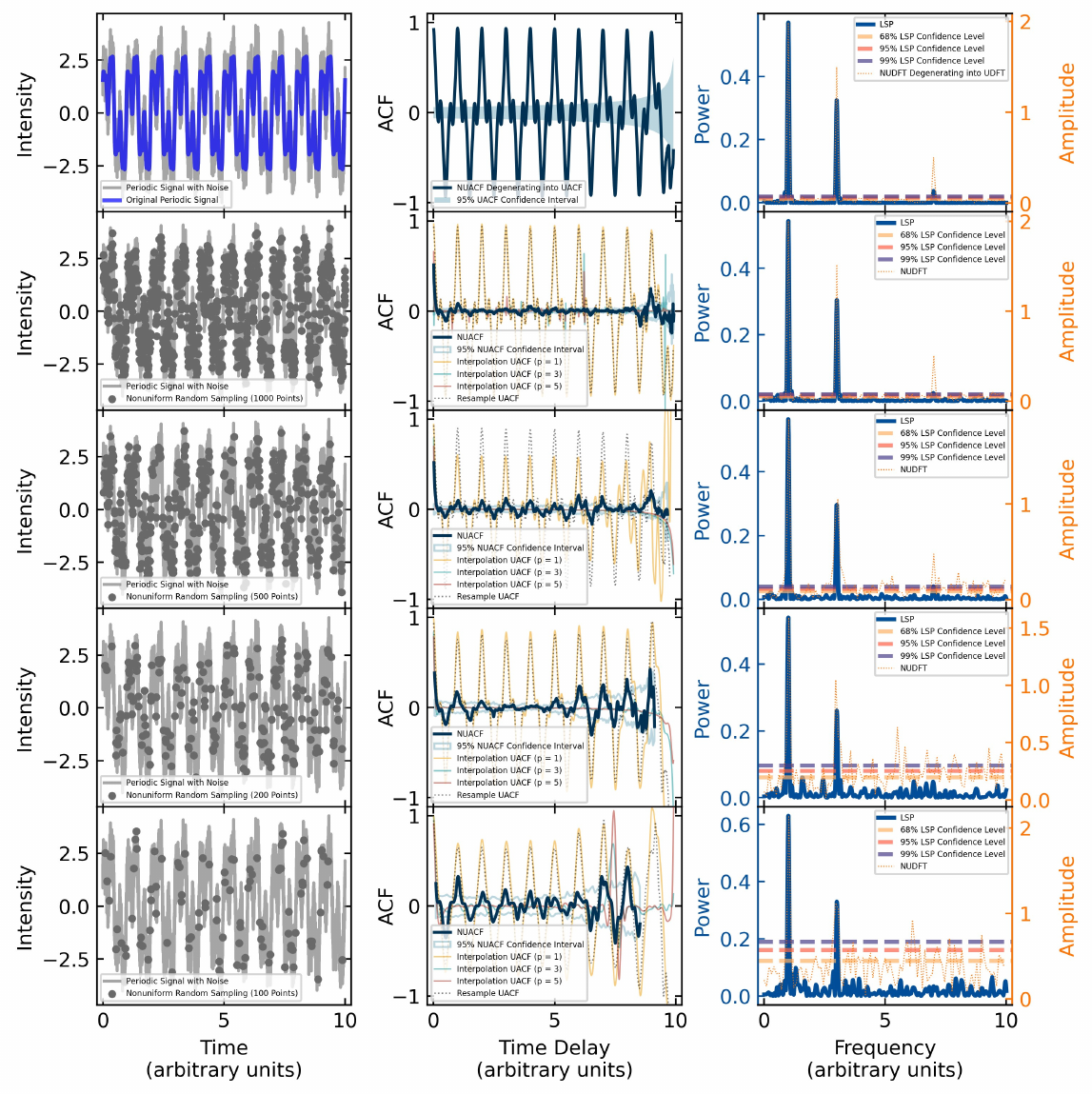}
\figsetgrpnote{Pattern A under nonuniform sampling. This figure
shows the same signal as in Figure \ref{Fig4}.1,
but under nonuniform sampling. The NUACF still performs reliably,
demonstrating its adaptability to irregular time grids.}
\figsetgrpend

\figsetgrpstart
\figsetgrpnum{5.5}
\figsetgrptitle{Pattern B under nonuniform sampling}
\figsetplot{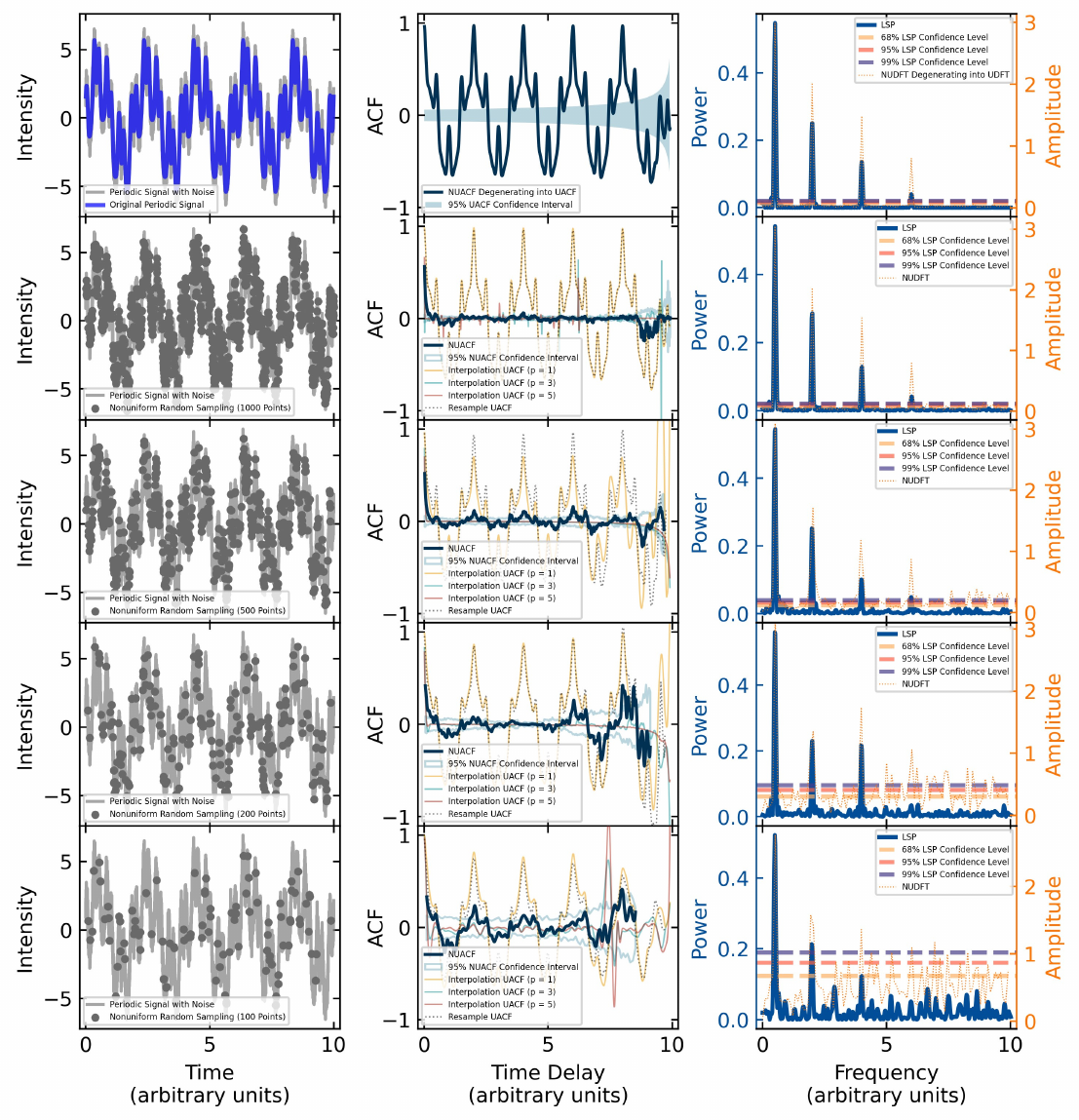}
\figsetgrpnote{Pattern B under nonuniform sampling. This figure
presents the nonuniformly sampled version of Pattern B. The NUACF's
confidence interval adjusts with sampling density, preserving the
advantage of revealing significant features.}
\figsetgrpend

\figsetgrpstart
\figsetgrpnum{5.6}
\figsetgrptitle{Pattern C under nonuniform sampling}
\figsetplot{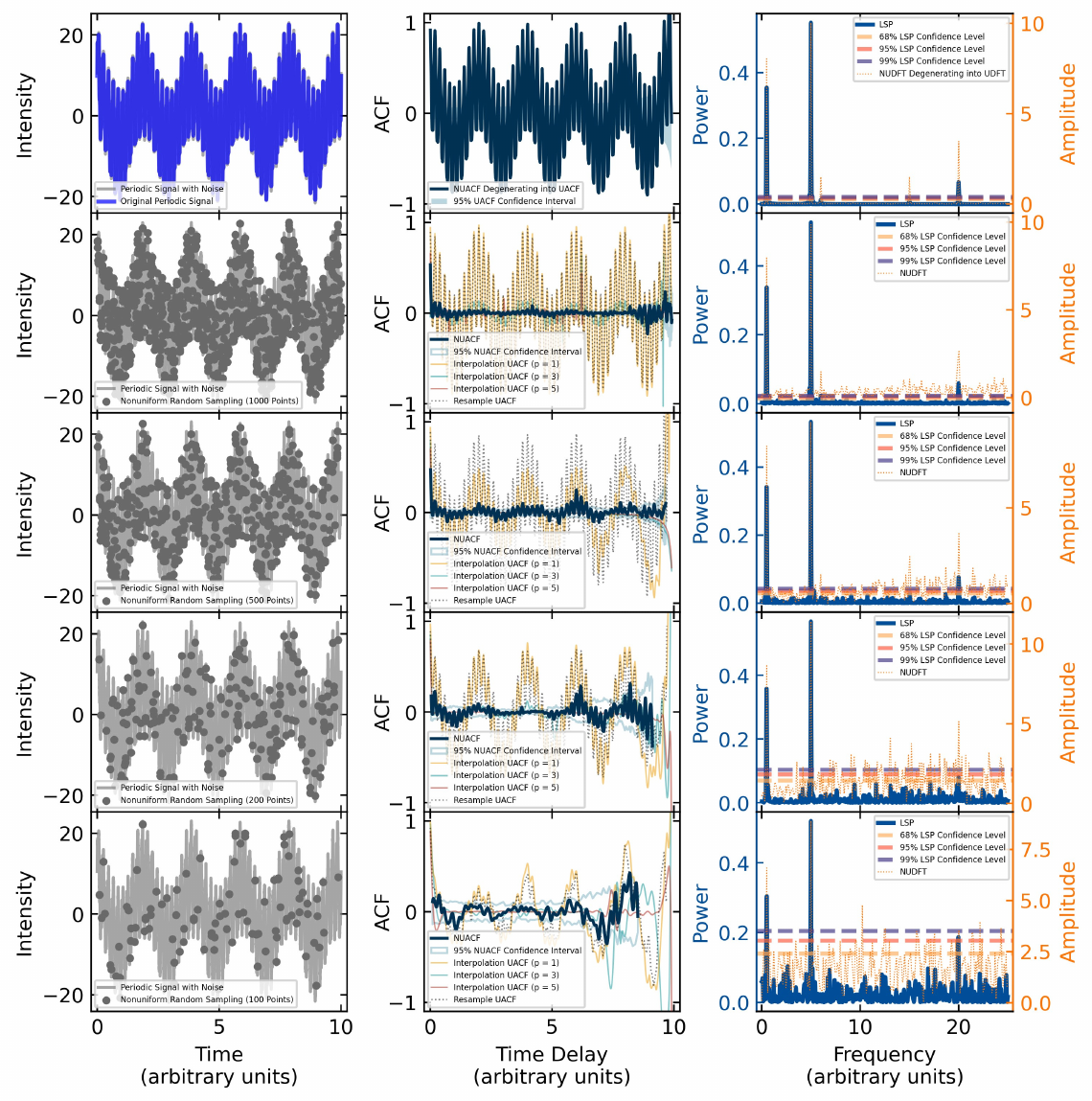}
\figsetgrpnote{Pattern C under nonuniform sampling. This figure
presents the nonuniformly sampled version of Pattern C. The NUACF's
confidence interval adjusts with sampling density, preserving the
advantage of revealing significant features.}
\figsetgrpend

\figsetgrpstart
\figsetgrpnum{5.7}
\figsetgrptitle{Constant number of sampling points per window}
\figsetplot{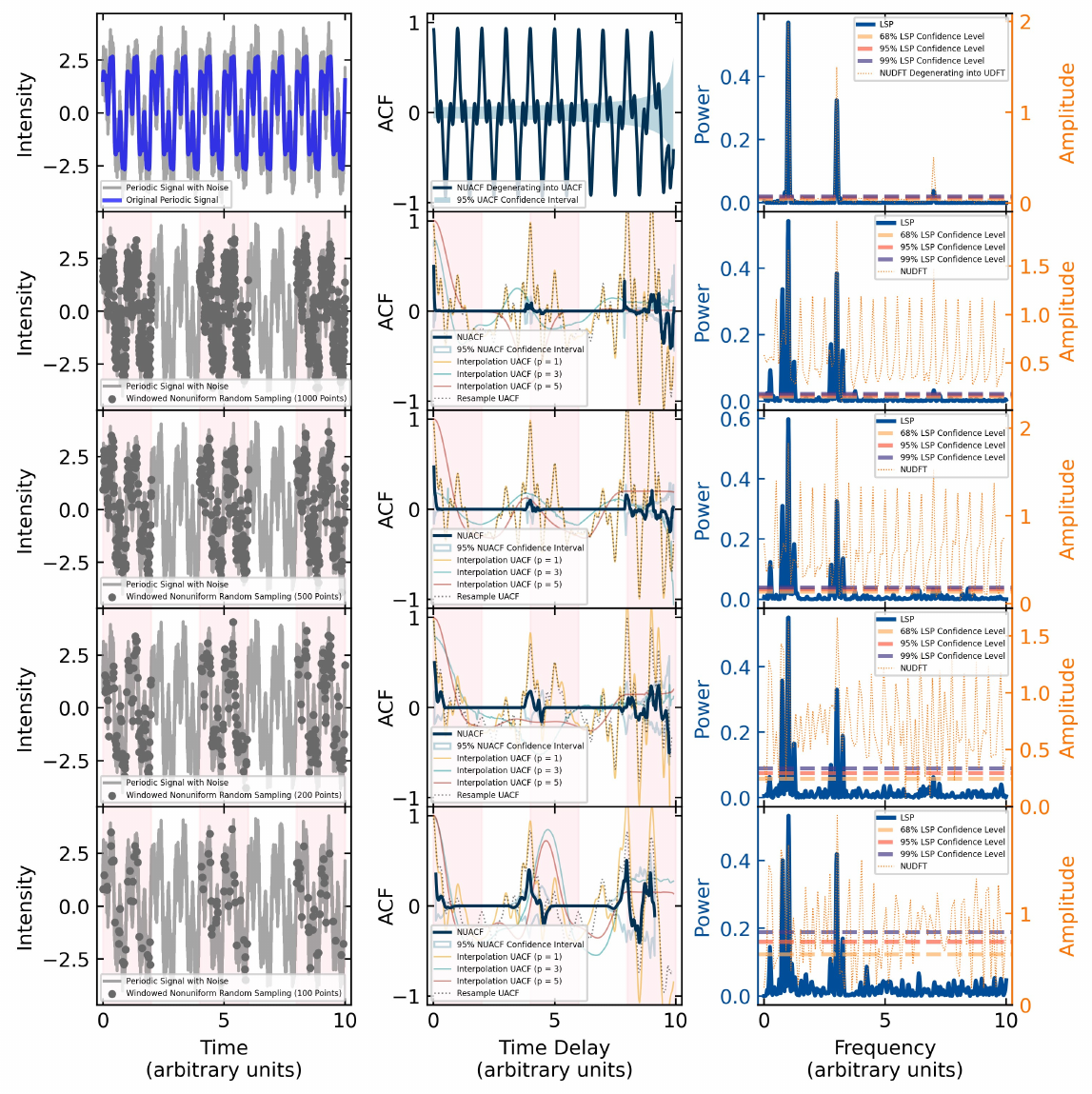}
\figsetgrpnote{Windowing test: varying number of sampling points per
window (Case 1). Shown here is a simulated periodic signal (Patter A)
observed in discrete windows. This figure presents a case with a constant
number of sampling points per window. Windowing introduces distortion in
the resampled and interpolated ACF amplitudes.}
\figsetgrpend

\figsetgrpstart
\figsetgrpnum{5.8}
\figsetgrptitle{Increasing number of sampling points per window}
\figsetplot{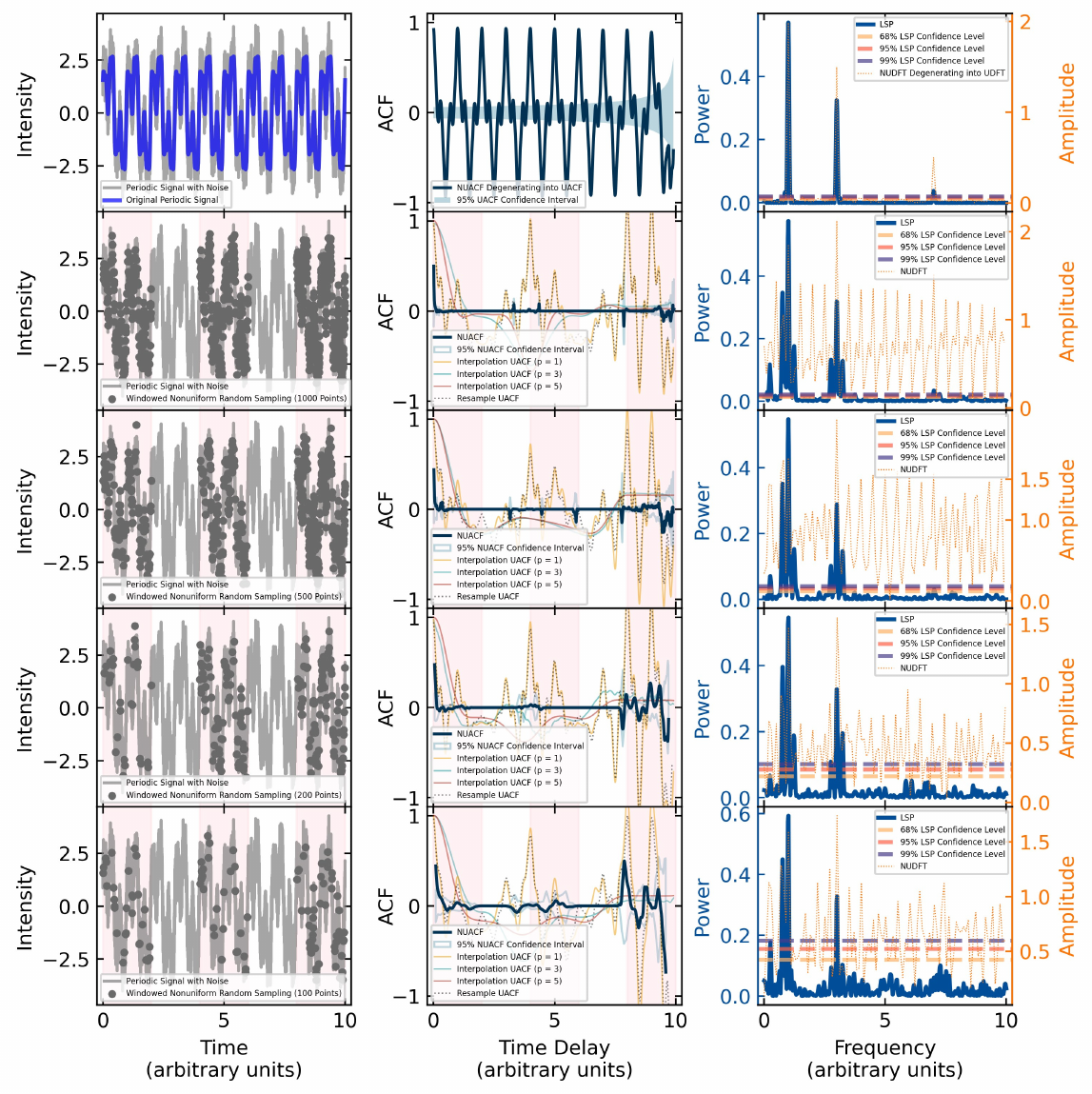}
\figsetgrpnote{Windowing test: varying number of sampling points per
window (Case 2). Shown here is a case with a variable number of
sampling points per window where the count increases from one window to
the next, for Pattern A. The NUACF's confidence interval provides a
stable significance criterion despite the gappy sampling.}
\figsetgrpend

\figsetgrpstart
\figsetgrpnum{5.9}
\figsetgrptitle{Decreasing number of sampling points per window}
\figsetplot{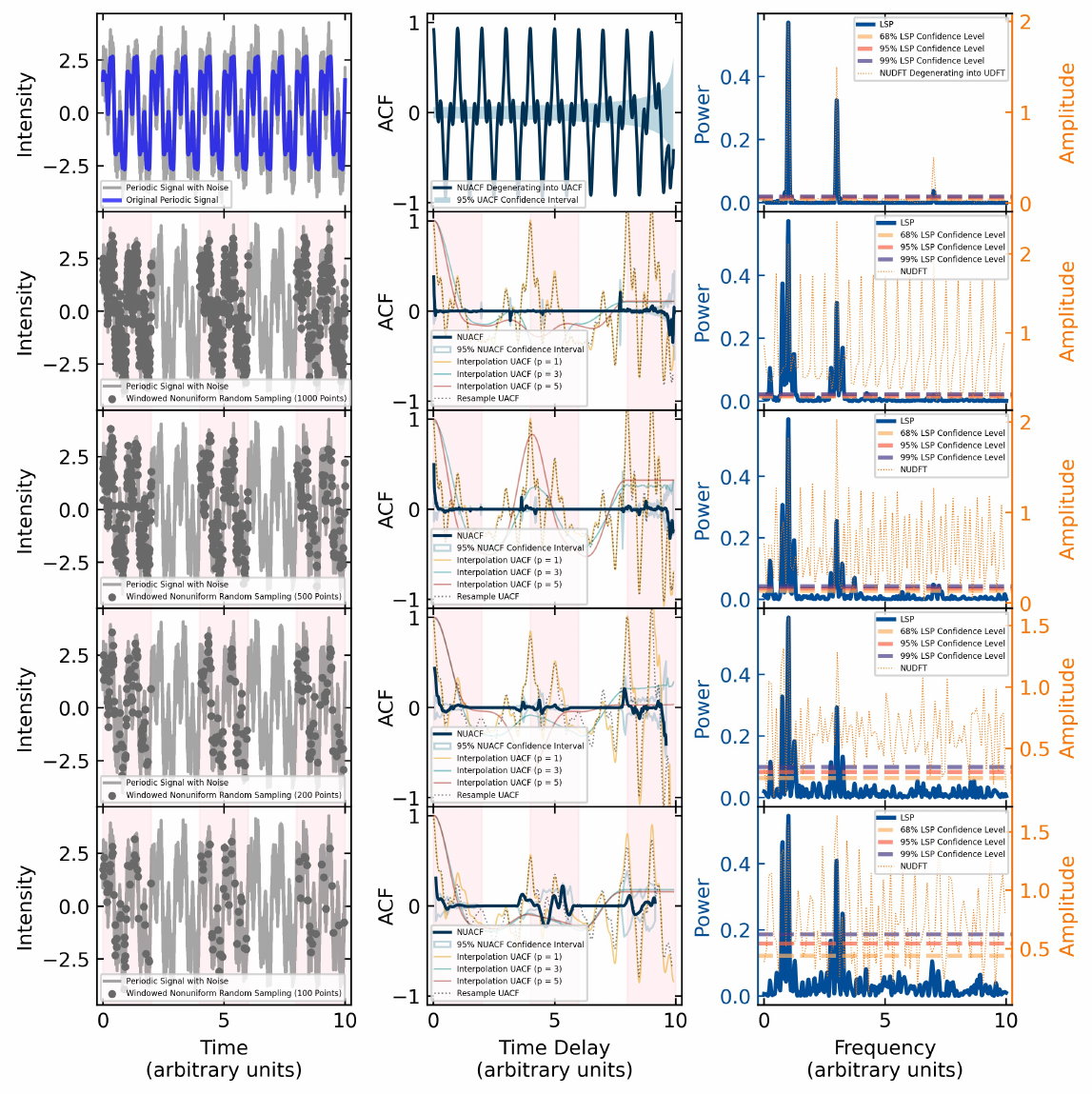}
\figsetgrpnote{Windowing test: varying number of sampling points per
window (Case 3). We show another case with a variable number of
sampling points per window where the count decreases from one window to
the next, for Pattern A. Traditional ACF profiles become distorted, while
the NUACF can still reveal statistically significant features.}
\figsetgrpend

\figsetgrpstart
\figsetgrpnum{5.10}
\figsetgrptitle{Small number of observing windows}
\figsetplot{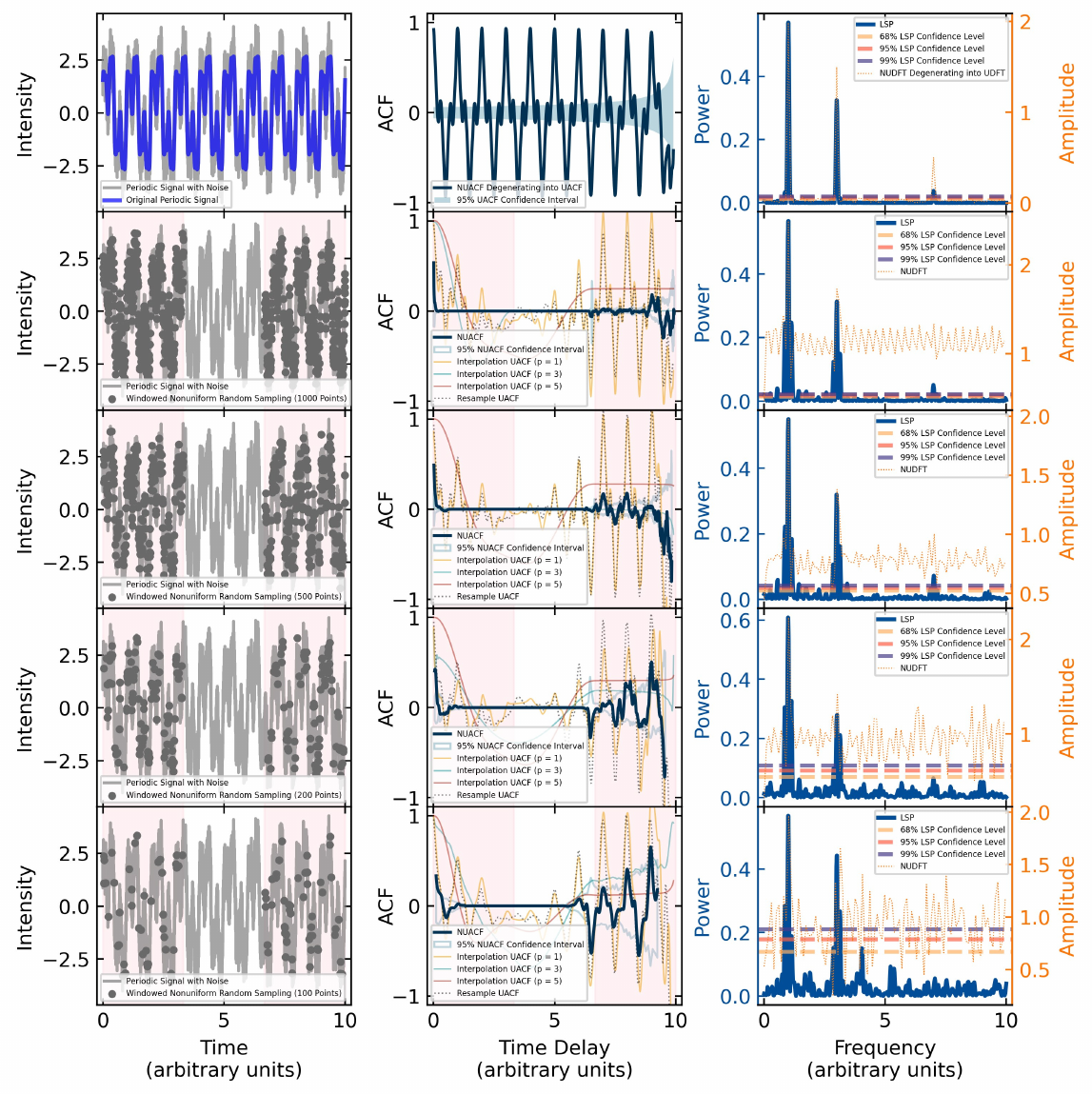}
\figsetgrpnote{Windowing test: fewer observing windows. This
figure shows a periodic signal (Patter A) observed in a small number of
widely separated windows. The resampled and interpolated ACFs show strong
amplitude distortion. Our NUACF identifies fewer but still significant
peaks/troughs.}
\figsetgrpend

\figsetgrpstart
\figsetgrpnum{5.11}
\figsetgrptitle{Moderate number of observing windows}
\figsetplot{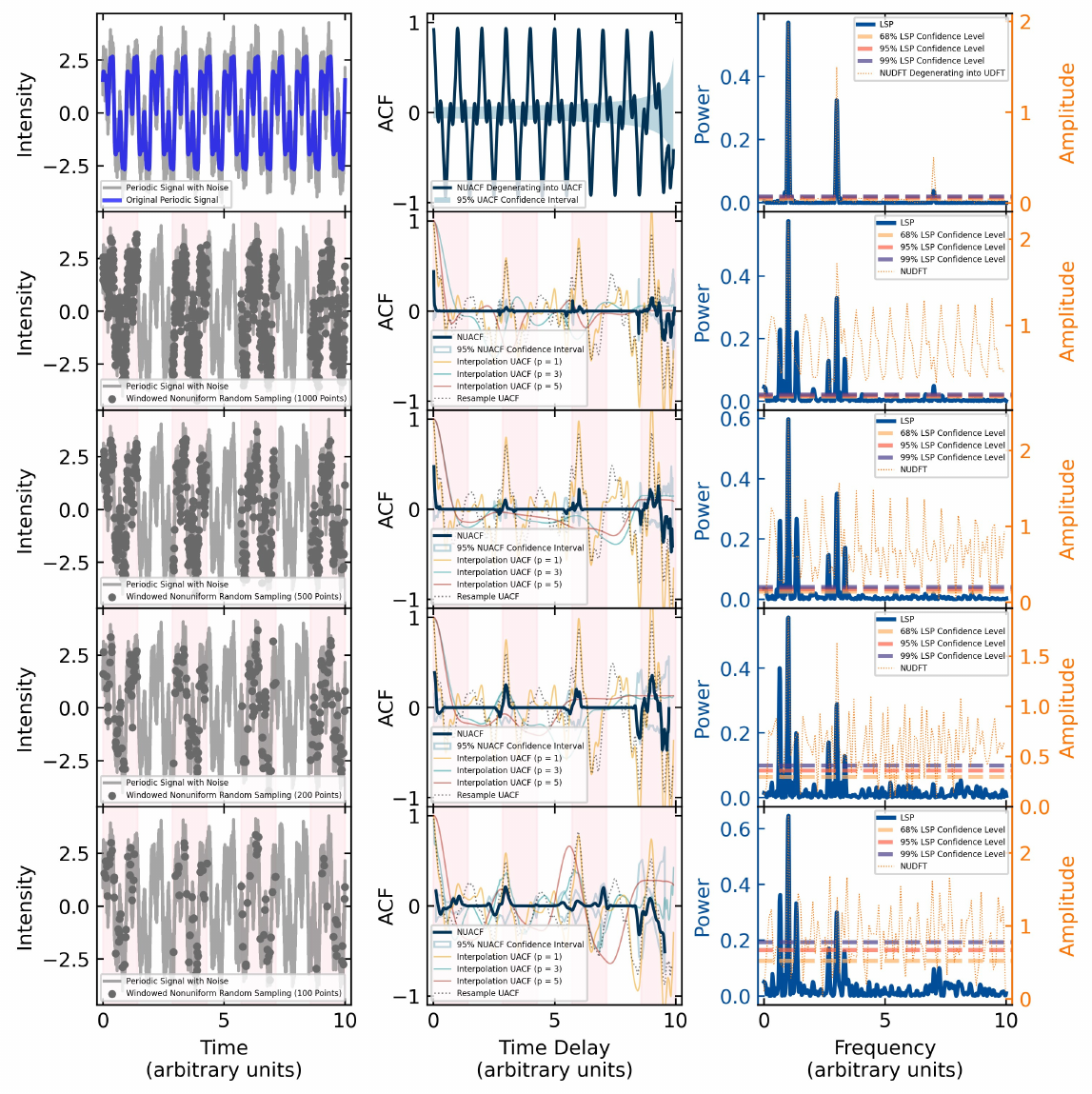}
\figsetgrpnote{Windowing test: moderate number of windows.
Performance with an intermediate number of observing windows is shown,
still for Pattern A. All the methods tested show partial recovery of the
repeated patterns, with our NUACF offering the only quantitative
significance assessment.}
\figsetgrpend

\figsetgrpstart
\figsetgrpnum{5.12}
\figsetgrptitle{Large number of observing windows}
\figsetplot{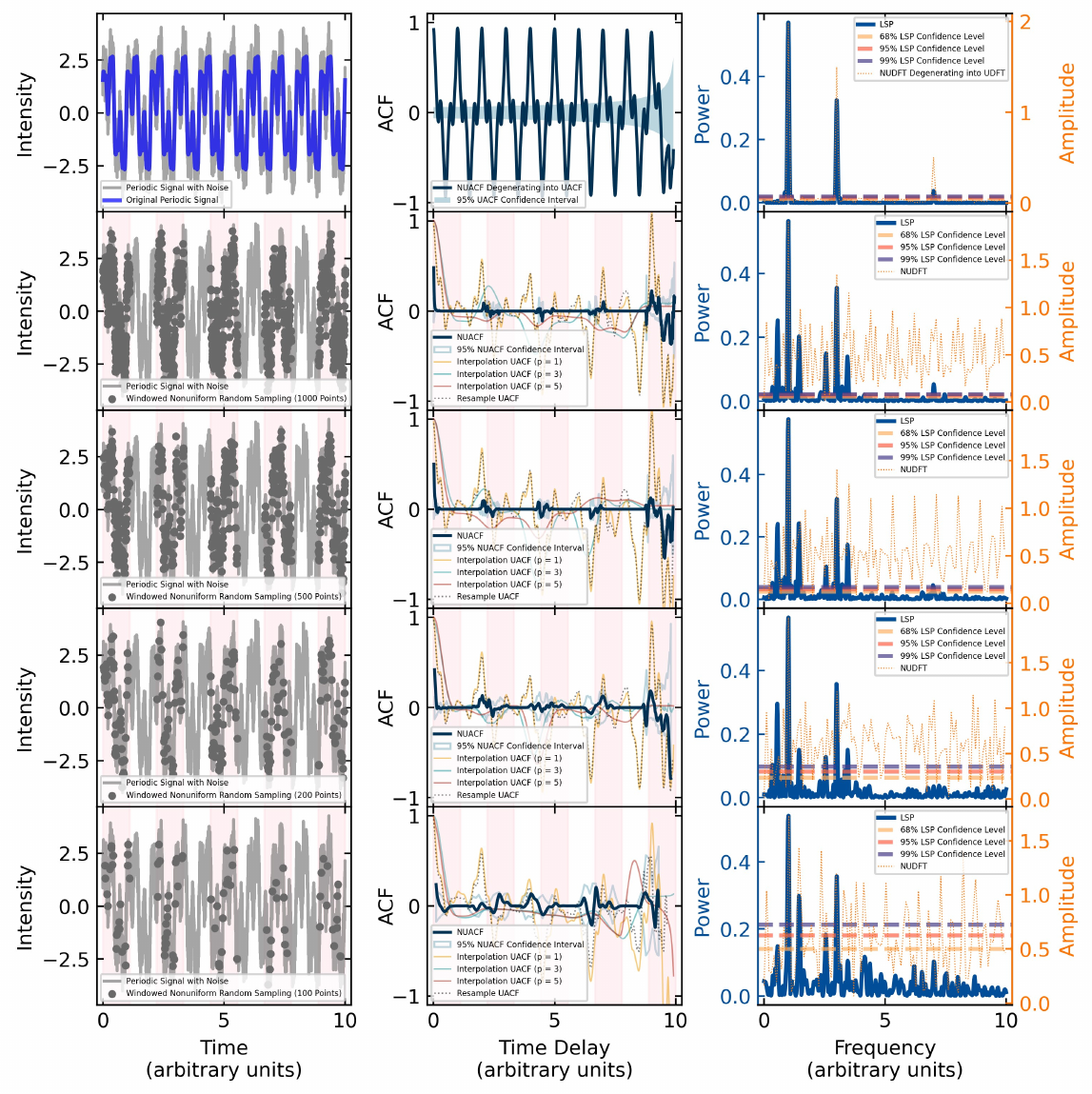}
\figsetgrpnote{Windowing test: many observing windows. Performance
with a larger number of windows (i.e., with more continuous coverage) is
shown, still for Pattern A. Distortion in traditional methods persists,
while our NUACF's confidence-based approach remains uniquely robust.}
\figsetgrpend

\figsetgrpstart
\figsetgrpnum{5.13}
\figsetgrptitle{Irregular window spacing case 1}
\figsetplot{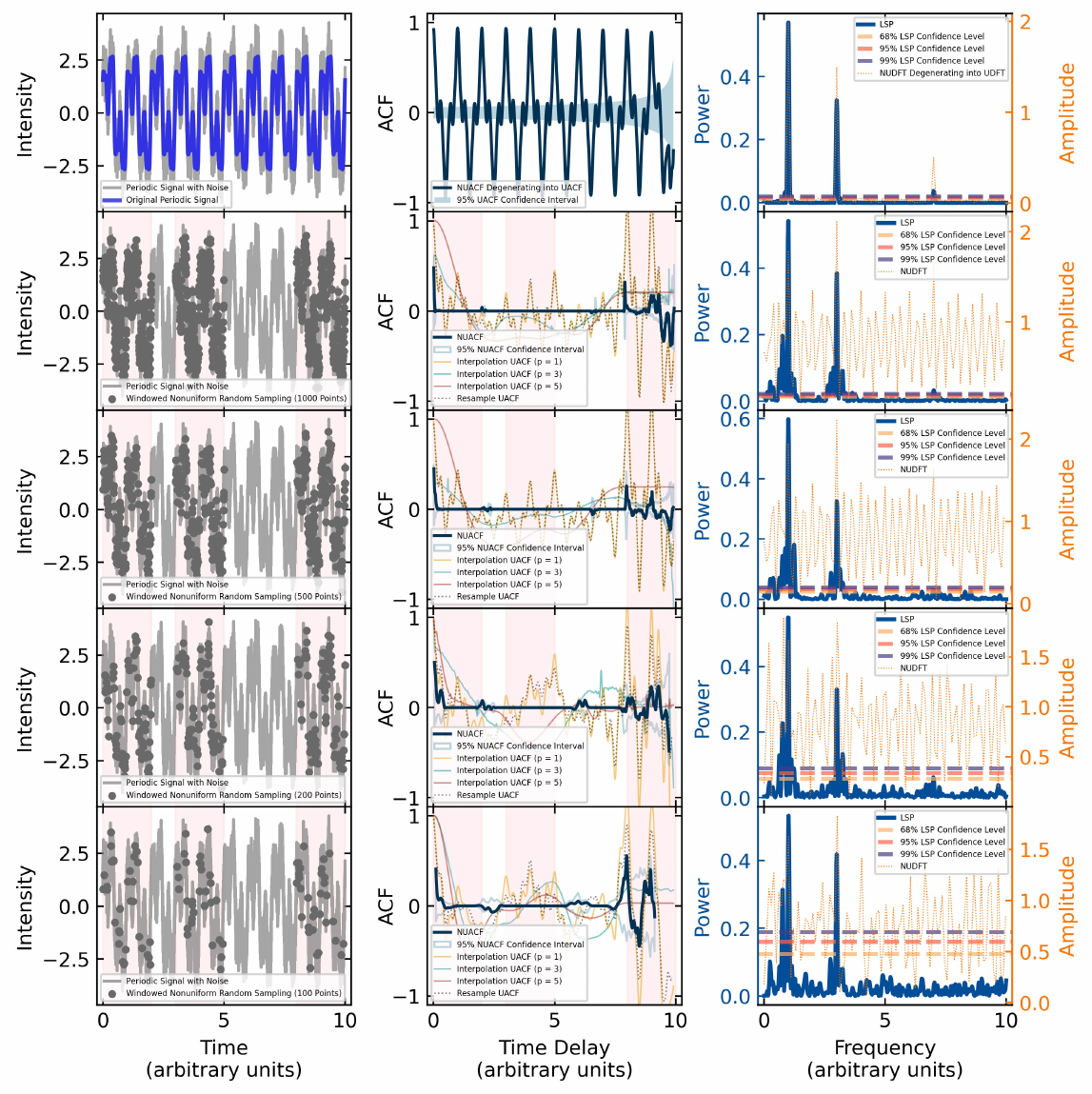}
\figsetgrpnote{Windowing test: irregular window spacing (Case 1).
For Pattern A, now we present a situation that the windows are unevenly
spaced in time. The resampled and interpolated ACF profiles retain
approximate periodicity but with irregular amplitudes.}
\figsetgrpend

\figsetgrpstart
\figsetgrpnum{5.14}
\figsetgrptitle{Irregular window spacing case 2}
\figsetplot{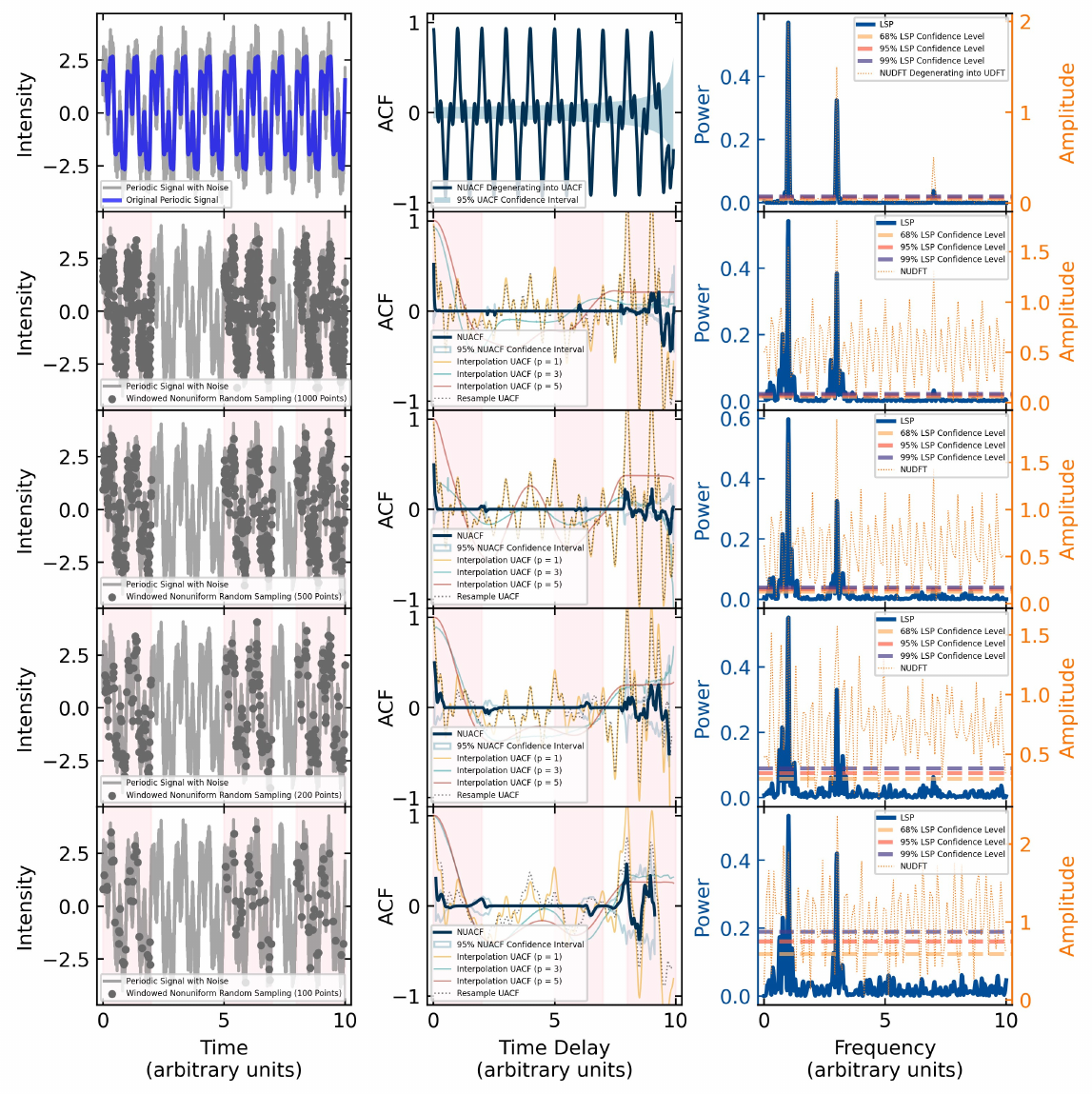}
\figsetgrpnote{Windowing test: irregular window spacing (Case 2).
Another case of windows being separated by irregular time intervals is
shown.}
\figsetgrpend

\figsetgrpstart
\figsetgrpnum{5.15}
\figsetgrptitle{Irregular window spacing case 3}
\figsetplot{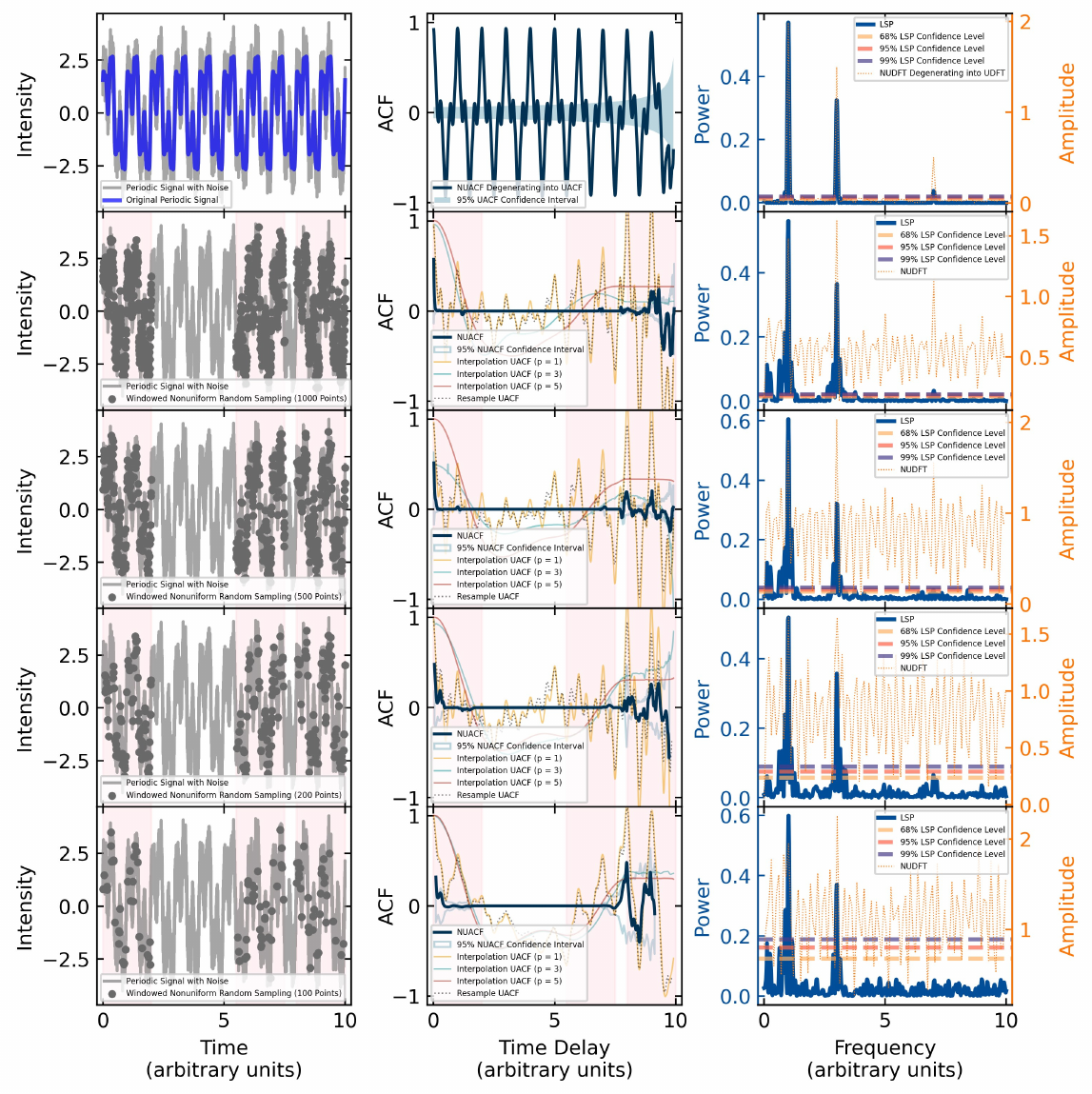}
\figsetgrpnote{Windowing test: irregular window spacing (Case 3).
In this figure, the windows are placed with stronger irregularity. This
highlights the NUACF's advantage in gappy regimes: it can still reveal
the timing structure, whereas traditional ACF profiles become severely
distorted.}
\figsetgrpend

\figsetgrpstart
\figsetgrpnum{5.16}
\figsetgrptitle{Variable window duration case 1}
\figsetplot{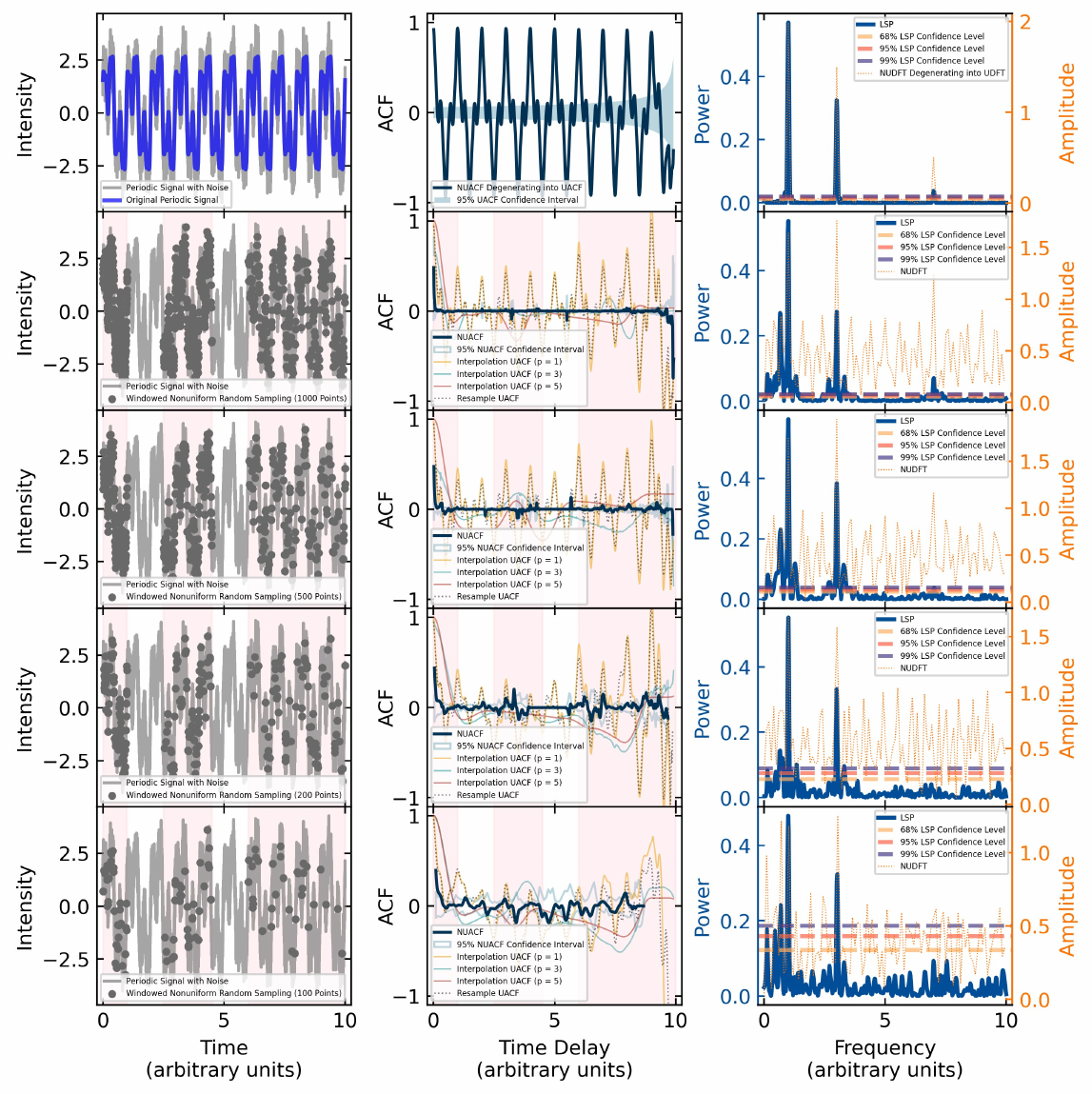}
\figsetgrpnote{Windowing test: variable window duration (Case 1).
For Pattern A, now we present a situation that the observing windows
differ in length. Although the performance of all methods is degraded to
some extent especially when the sampling points are sparse, our NUACF
method, through its explicit time weighting and confidence interval,
handles this heterogeneity more reliably than traditional methods.}
\figsetgrpend

\figsetgrpstart
\figsetgrpnum{5.17}
\figsetgrptitle{Variable window duration case 2}
\figsetplot{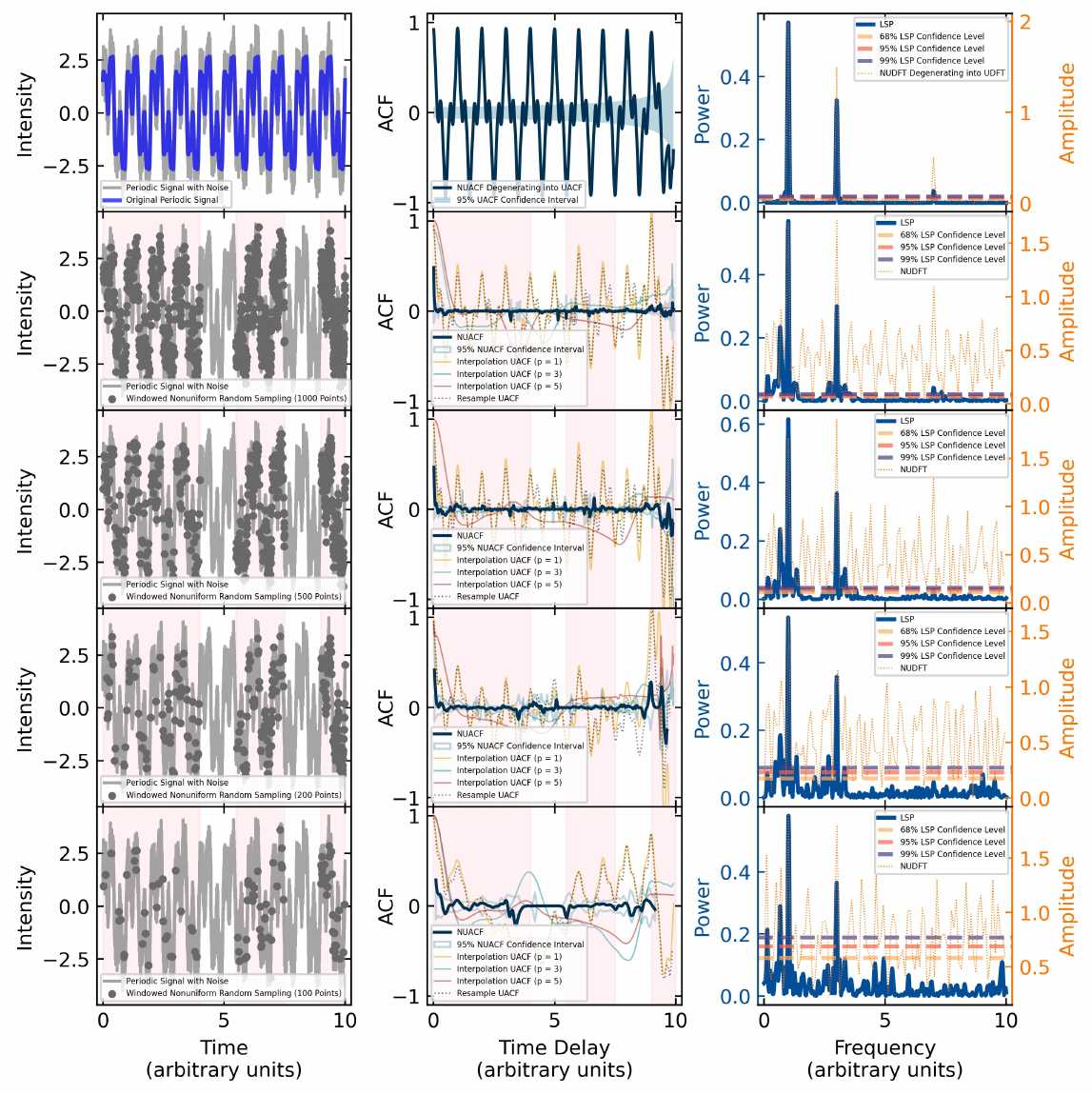}
\figsetgrpnote{Windowing test: variable window duration (Case 2).
Shown here is another case of windows varying in length.}
\figsetgrpend

\figsetgrpstart
\figsetgrpnum{5.18}
\figsetgrptitle{Variable window duration case 3}
\figsetplot{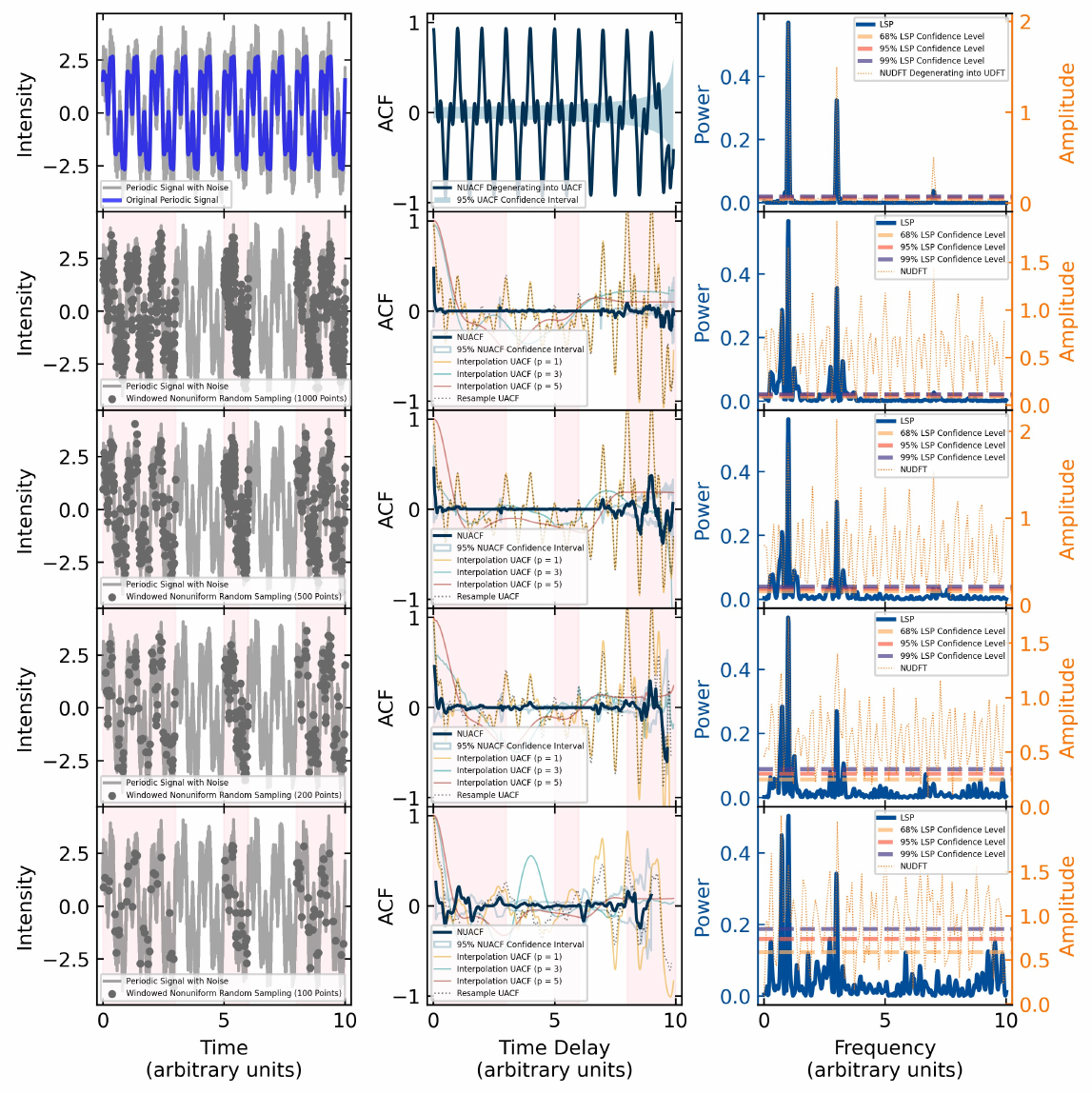}
\figsetgrpnote{Windowing test: variable window duration (Case 2).
This figure shows the cases that the window lengths vary more
significantly.}
\figsetgrpend

\figsetend

\clearpage

\figsetstart
\figsetnum{6}
\figsettitle{Robustness comparison between resampled ACF and NUACF under
complex signal conditions}

\figsetgrpstart
\figsetgrpnum{6.1}
\figsetgrptitle{Linear baseline trend}
\figsetplot{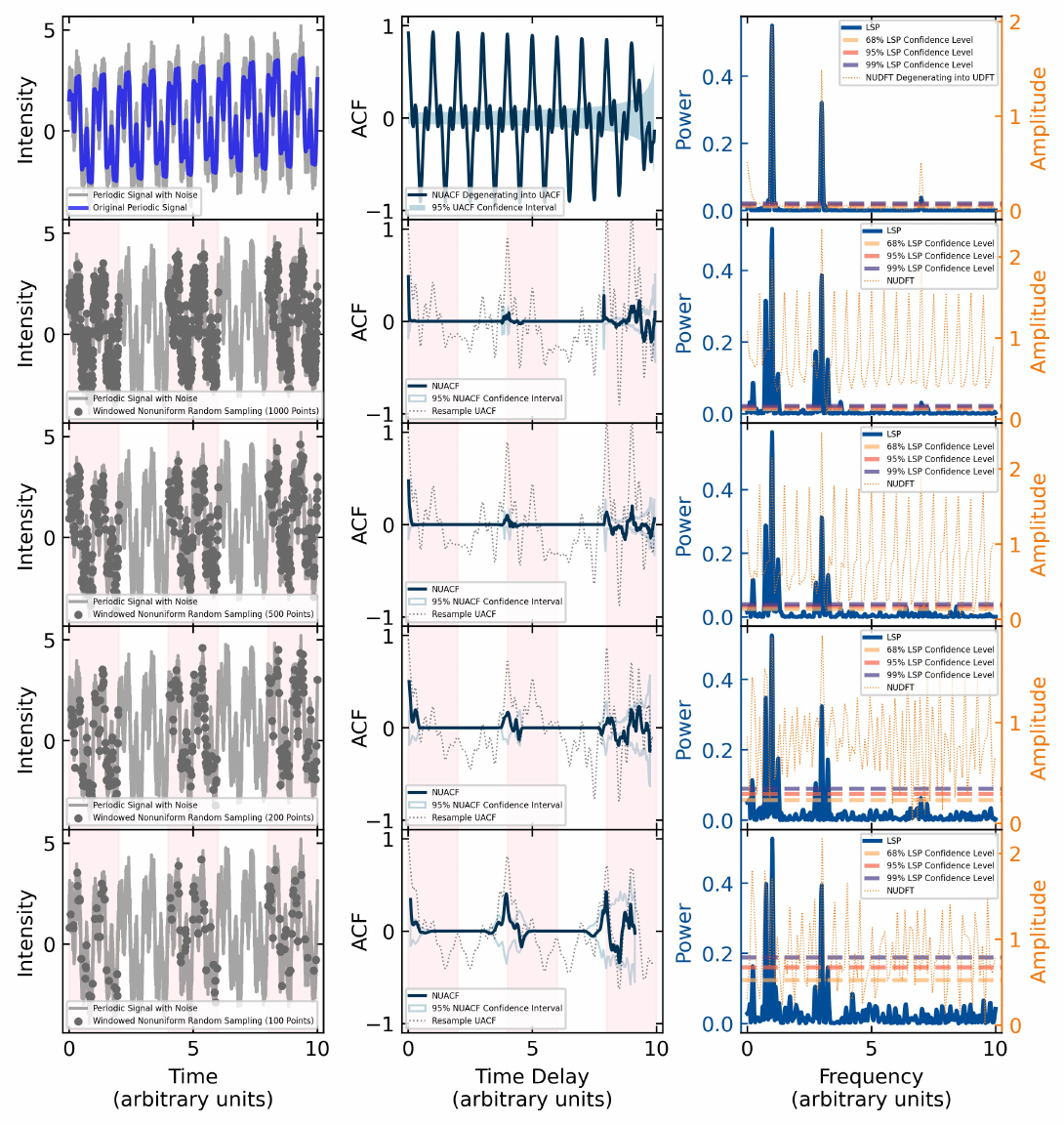}
\figsetgrpnote{Comparison of resampled ACF and our NUACF: signal
with a linear baseline trend. Shown here is a periodic signal (Pattern
A) with an additive linear evolution trend. Our NUACF successfully
identifies a portion of the repeated variability (i.e. those NUACF
peaks/troughs beyond the NUACF's confidence interval), while the
resampled ACF profile is biased.}
\figsetgrpend

\figsetgrpstart
\figsetgrpnum{6.2}
\figsetgrptitle{Exponential baseline trend}
\figsetplot{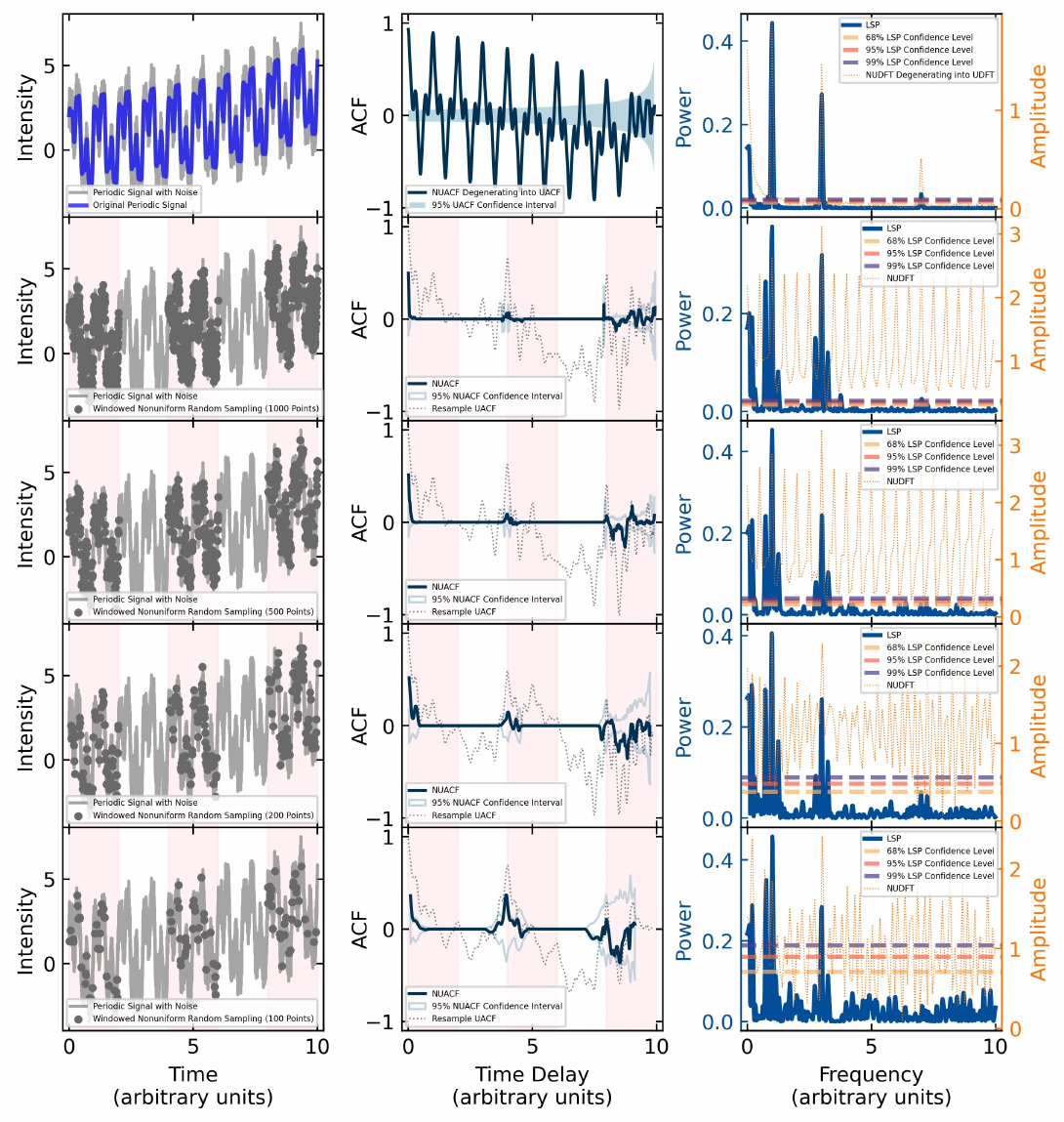}
\figsetgrpnote{Comparison of resampled ACF and our NUACF: signal
with an exponential baseline trend. Pattern A with an additive
exponential evolution trend is shown. Even the standard sample ACF (the
uniform-sampling limit of the NUACF, top panel in the middle column) is
significantly altered by the baseline component. Our NUACF remains
effective at detecting the underlying repetitive patterns.}
\figsetgrpend

\figsetgrpstart
\figsetgrpnum{6.3}
\figsetgrptitle{Compound baseline trend}
\figsetplot{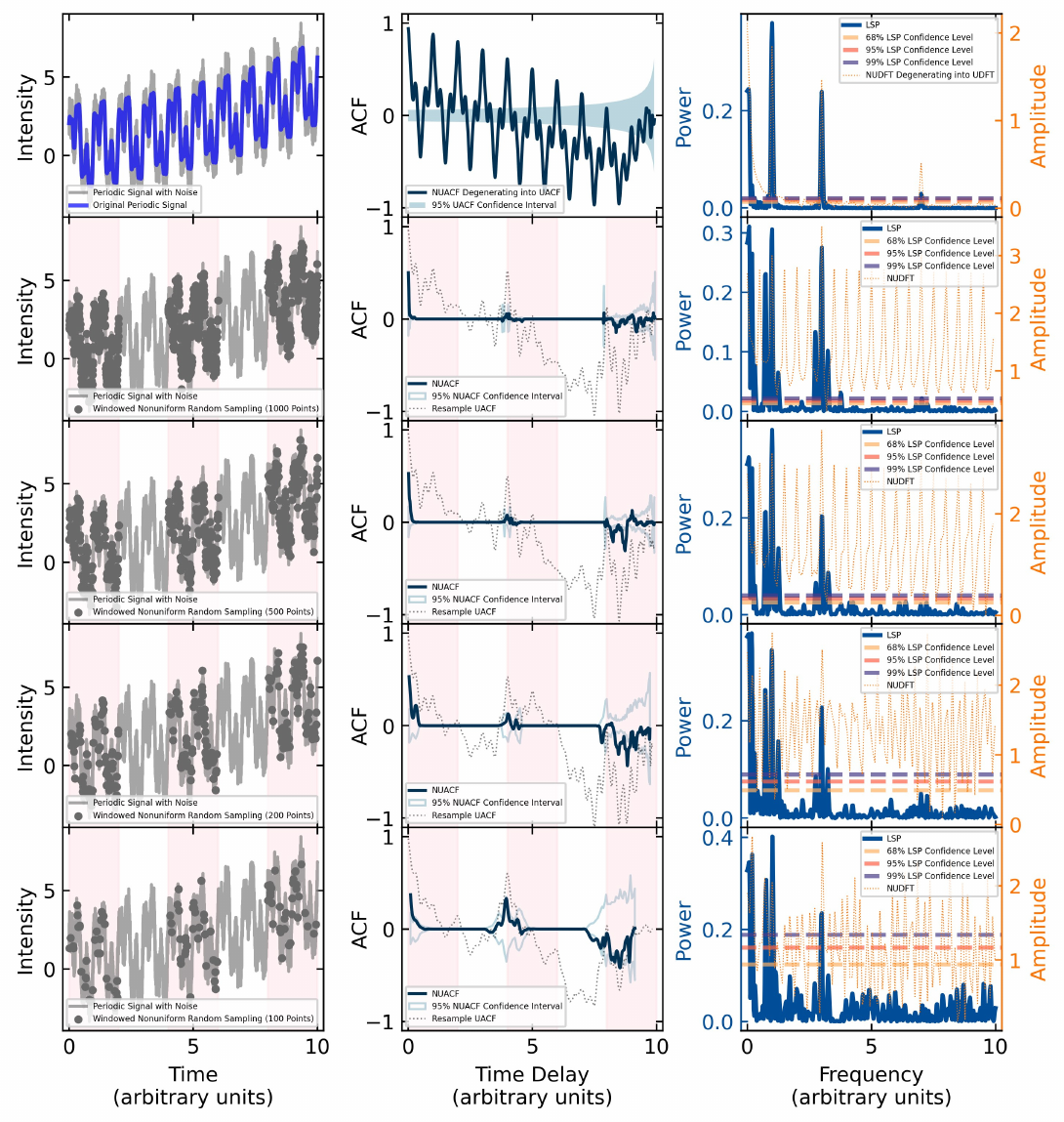}
\figsetgrpnote{Comparison of resampled ACF and our NUACF: signal
with a compound baseline trend. Pattern A with a strong additive
composite (linear $+$ exponential) baseline trend is shown. Our NUACF
still robustly reveals a subset of the underlying repetitive patterns. In
contrast, the resampled ACF profile is heavily distorted.}
\figsetgrpend

\figsetgrpstart
\figsetgrpnum{6.4}
\figsetgrptitle{High noise level (SNR $\approx$ 3)}
\figsetplot{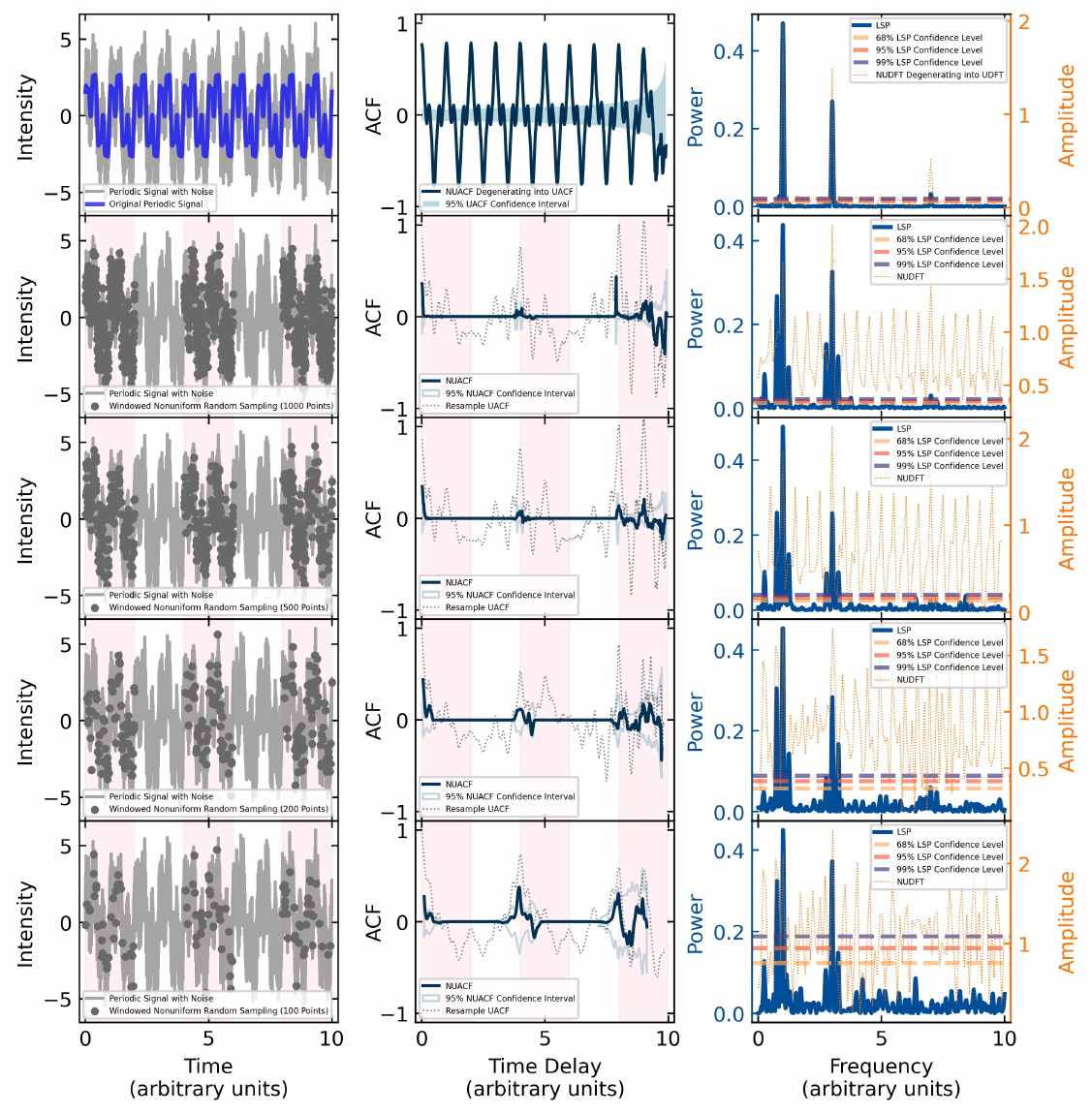}
\figsetgrpnote{Comparison of resampled ACF and our NUACF: high noise
level cases (${\rm{SNR}} \approx 3$). This figure presents a periodic
signal (Pattern A) contaminated by a high noise [i.e. low signal-to-noise
ratio (SNR)]. Our NUACF identifies a subset of significant features via
its confidence interval; the resampled ACF profile appears ambiguous
particularly under low-sample-size conditions.}
\figsetgrpend

\figsetgrpstart
\figsetgrpnum{6.5}
\figsetgrptitle{Very high noise level (SNR $\approx$ 1.5)}
\figsetplot{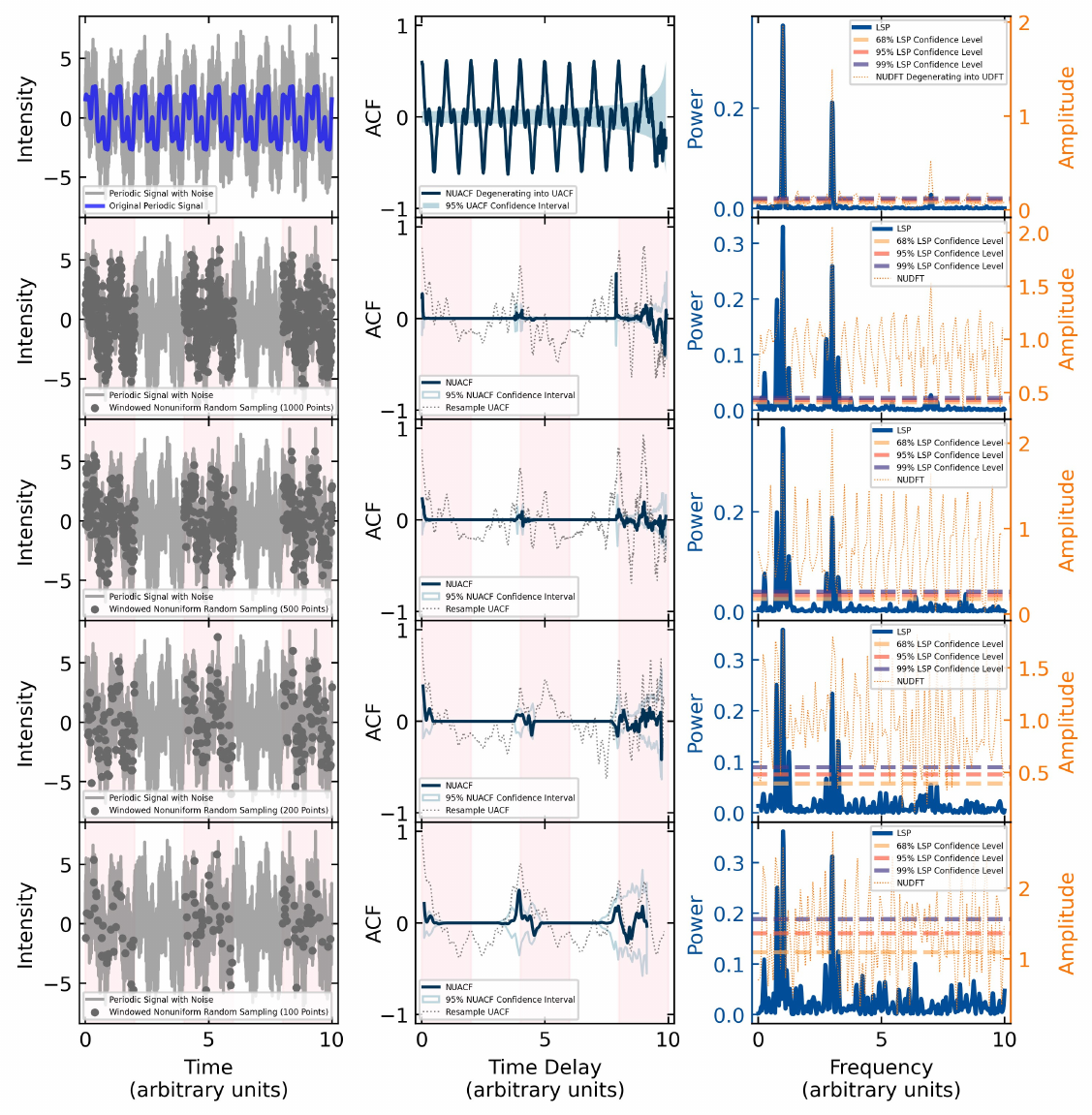}
\figsetgrpnote{Comparison of resampled ACF and our NUACF: very high
noise level cases (${\rm{SNR}} \approx 1.5$). For Pattern A,
performance under even higher noise contamination is shown. Our NUACF
still can reveal a few statistically significant features, while the
resampled ACF profile is largely indistinguishable.}
\figsetgrpend

\figsetgrpstart
\figsetgrpnum{6.6}
\figsetgrptitle{Extreme noise level (SNR $\approx$ 1)}
\figsetplot{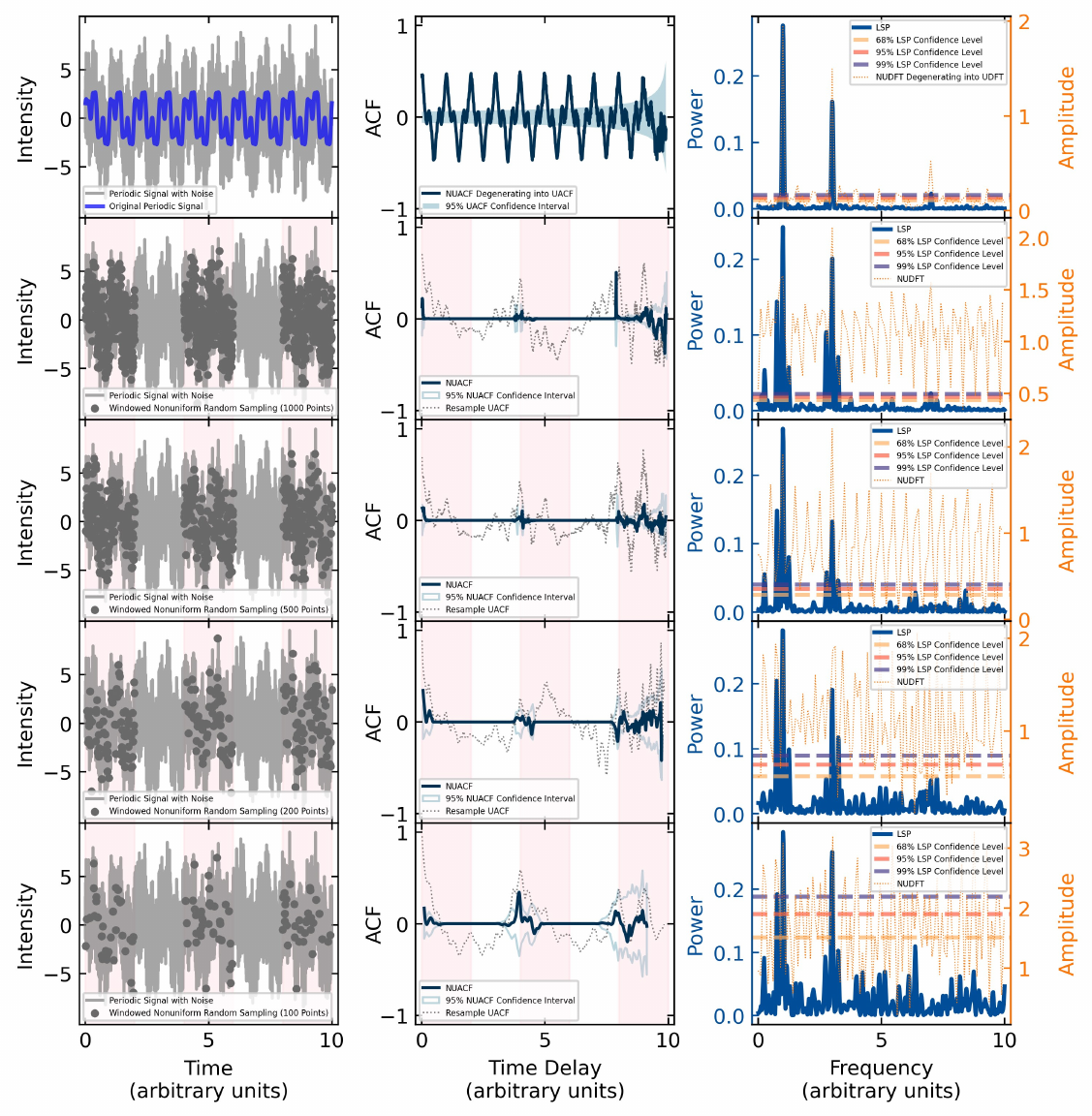}
\figsetgrpnote{Comparison of resampled ACF and our NUACF: extreme
noise level cases (${\rm{SNR}} \approx 1$). Here the signal is barely
distinguishable from the noise for Pattern A. Our NUACF still reveal one
or two marginally significant features. The resampled ACF shows no clear
pattern.}
\figsetgrpend

\figsetgrpstart
\figsetgrpnum{6.7}
\figsetgrptitle{Sinusoidal modulation}
\figsetplot{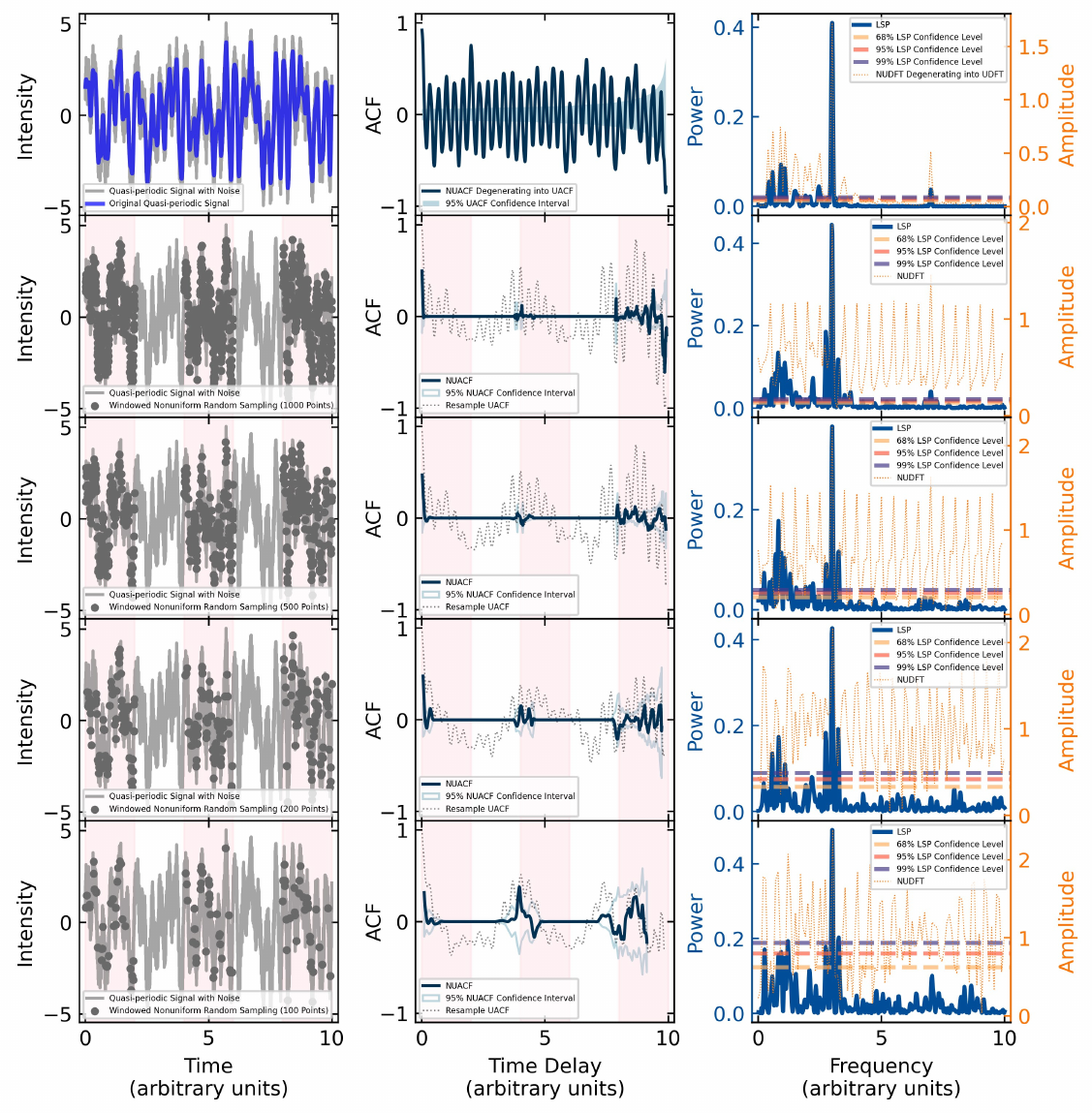}
\figsetgrpnote{Comparison of resampled ACF and our NUACF:
quasi-periodic signal (sinusoidal modulation) cases. Shown here is a
quasi-periodic signal modulated by a sinusoidal function. Our NUACF
reveals significant, but not strictly equally spaced, peaks corresponding
to the underlying modulated repeated pattern.}
\figsetgrpend

\figsetgrpstart
\figsetgrpnum{6.8}
\figsetgrptitle{Uniform-distributed modulation}
\figsetplot{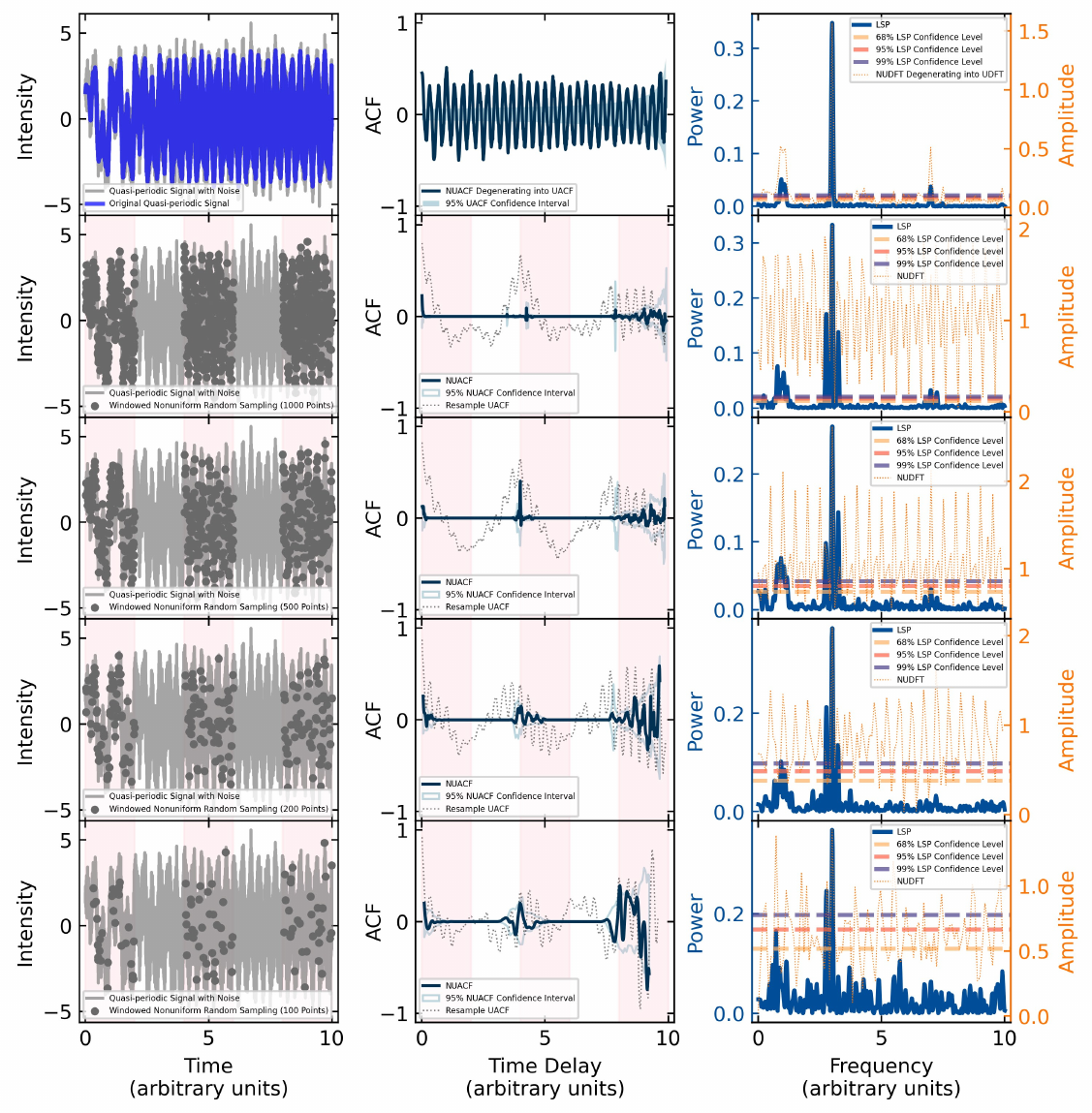}
\figsetgrpnote{Comparison of resampled ACF and our NUACF:
quasi-periodic signal (uniformly-distributed modulation) cases. A
different modulation pattern is shown. The resampled ACF fails to
identify any clear structures, while our NUACF identifies significant
features of repeated variability.}
\figsetgrpend

\figsetgrpstart
\figsetgrpnum{6.9}
\figsetgrptitle{Gaussian-distributed modulation}
\figsetplot{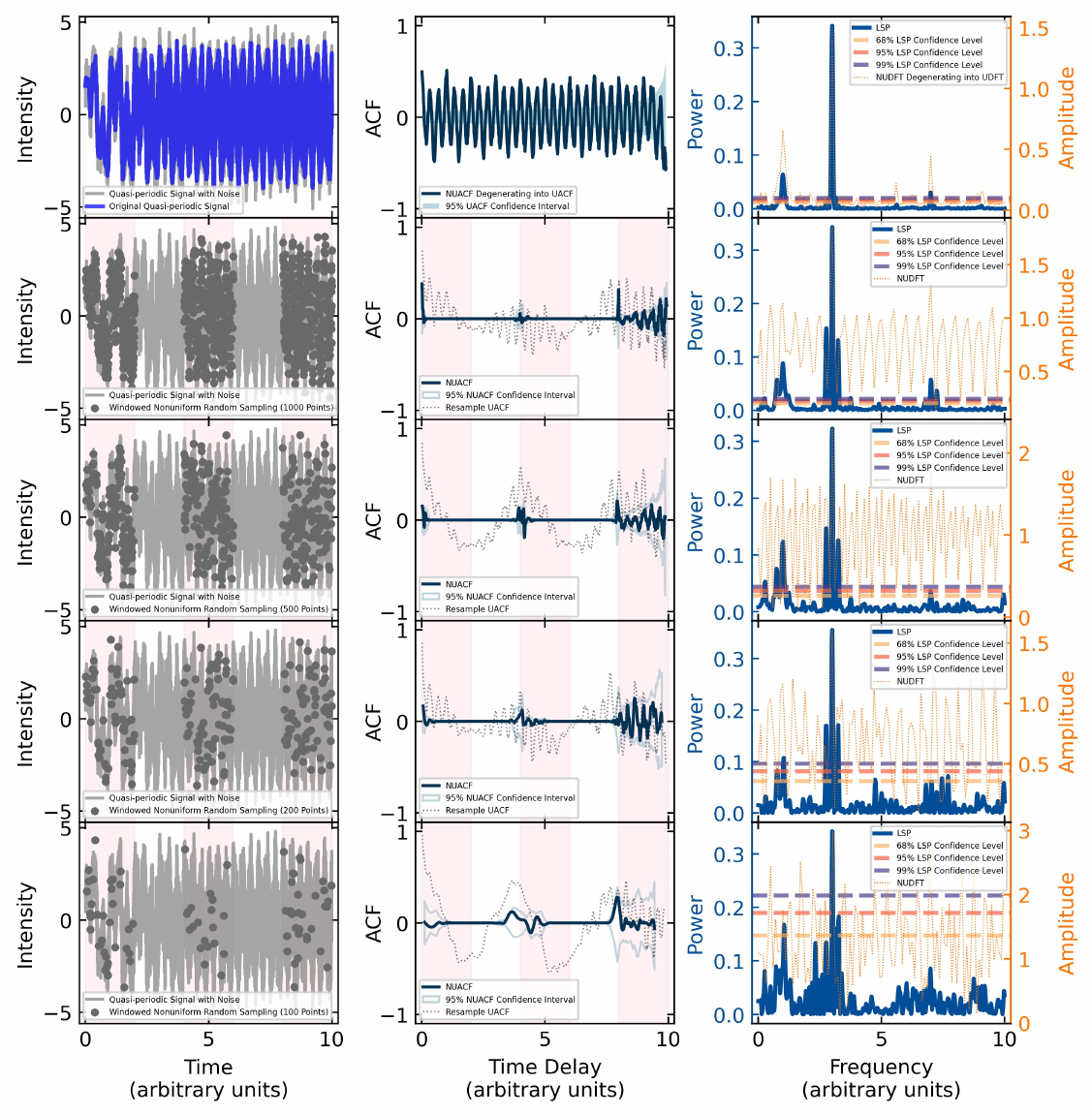}
\figsetgrpnote{Comparison of resampled ACF and our NUACF:
quasi-periodic signal cases (Gaussian-distributed modulation). These
cases of quasi-periodic signals demonstrate our NUACF's performance to
find significant, non-strictly periodic, repetitive patterns, a task
where traditional ACF methods fail.}
\figsetgrpend

\begin{figure}[!htbp]
\centering
\includegraphics[width=0.8\textwidth]{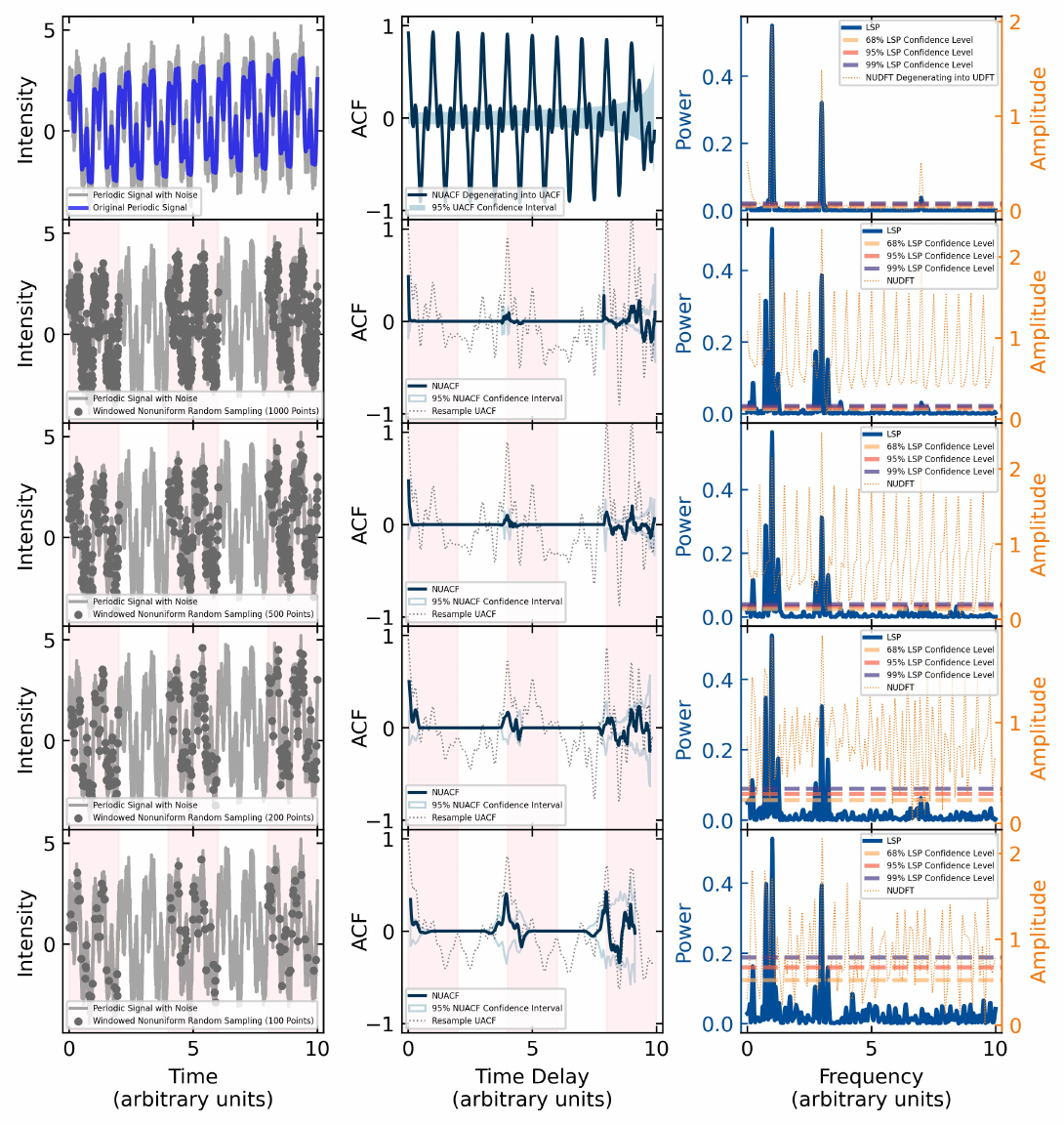}
\caption{Comparison of resampled ACF and our NUACF: signal with a
linear baseline trend. Shown here is a periodic signal (Pattern A) with
an additive linear evolution trend. Our NUACF successfully identifies a
portion of the repeated variability (i.e. those NUACF peaks/troughs
beyond the NUACF's confidence interval), while the resampled ACF profile
is biased. [The complete figure set (12 images) is available in the
online journal.]}
\label{Fig5}
\end{figure}

\figsetgrpstart
\figsetgrpnum{6.10}
\figsetgrptitle{Pure Gaussian noise}
\figsetplot{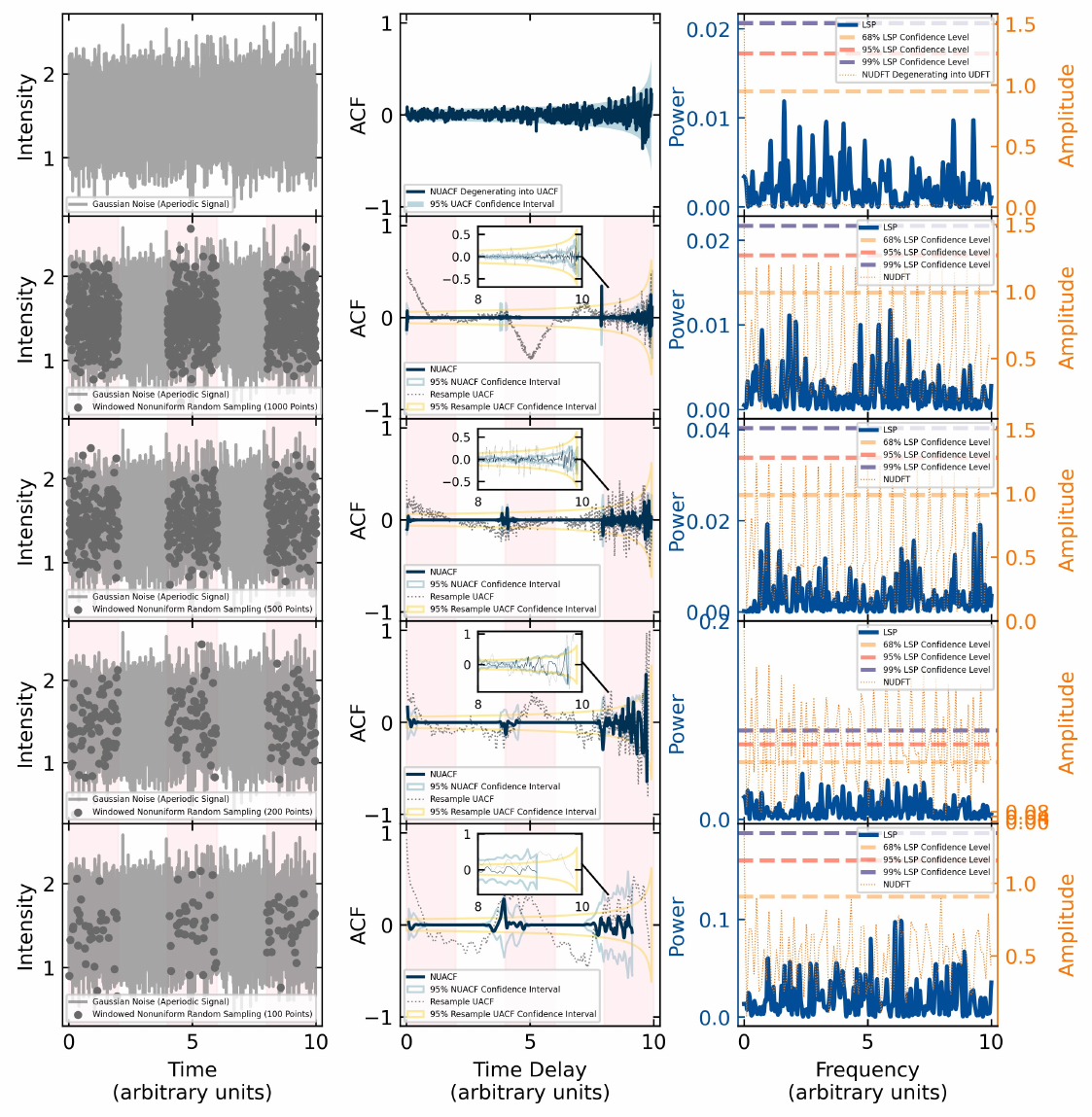}
\figsetgrpnote{Comparison of resampled ACF and our NUACF: pure
Gaussian noise cases (aperiodic signal). Shown here is a time series of
pure Gaussian noise, which can be regarded as an aperiodic signal. Such
critical test reveals a key weakness of the resampled ACF: it produces a
deceptively periodic-looking profile from pure noise, particularly under
low-sampling-rate conditions. Such spurious profiles closely resemble the
resampled ACF results of quasi-periodic signals shown in
Figures \ref{Fig5}.7--\ref{Fig5}.9. On the
country, our NUACF shows no significant features outside its confidence
band, which is consistent with the original signal.}
\figsetgrpend

\figsetgrpstart
\figsetgrpnum{6.11}
\figsetgrptitle{Pure uniform noise}
\figsetplot{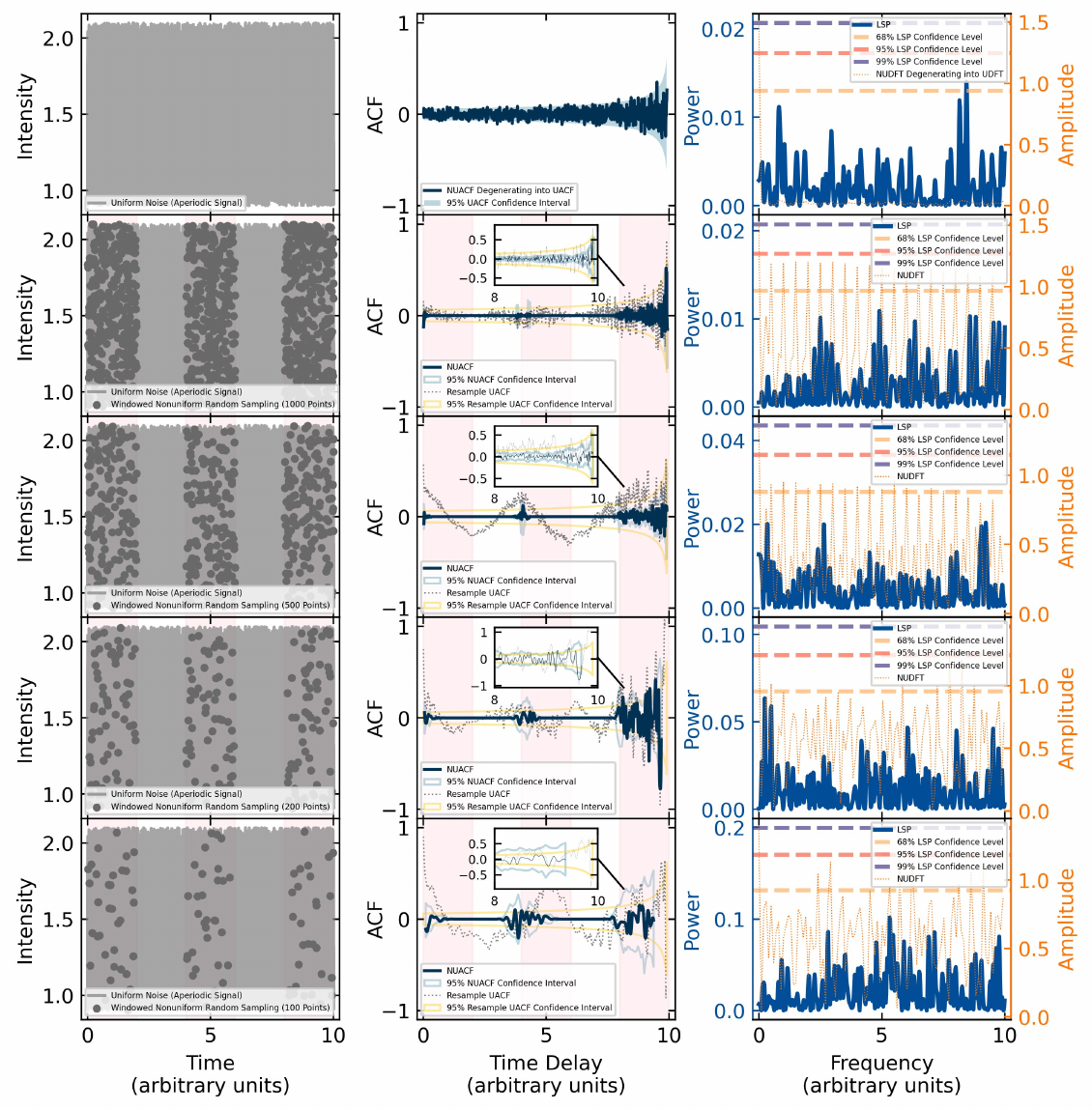}
\figsetgrpnote{Comparison of resampled ACF and our NUACF: pure
uniform noise case(aperiodic signal). The resampled ACF shows spurious
pseudo-periodic wiggles. In the top panel of the middle column, the
standard sample ACF (the degenerate form of NUACF under uniform sampling)
for pure noise appropriately matches its confidence band, as expected
from Equation (\ref{eq29}). For the resampled ACF applied to nonuniformly
sampled data, we tentatively adopt the confidence interval designed for
uniformly sampled data (area enclosed by the yellow line) to test its
validity. The deviation of the resampled ACF profile from this interval
demonstrates that such an extrapolation is unreliable, highlighting the
artifactual distortions introduced by the resampling procedure itself.}
\figsetgrpend

\figsetgrpstart
\figsetgrpnum{6.12}
\figsetgrptitle{Pure compound noise}
\figsetplot{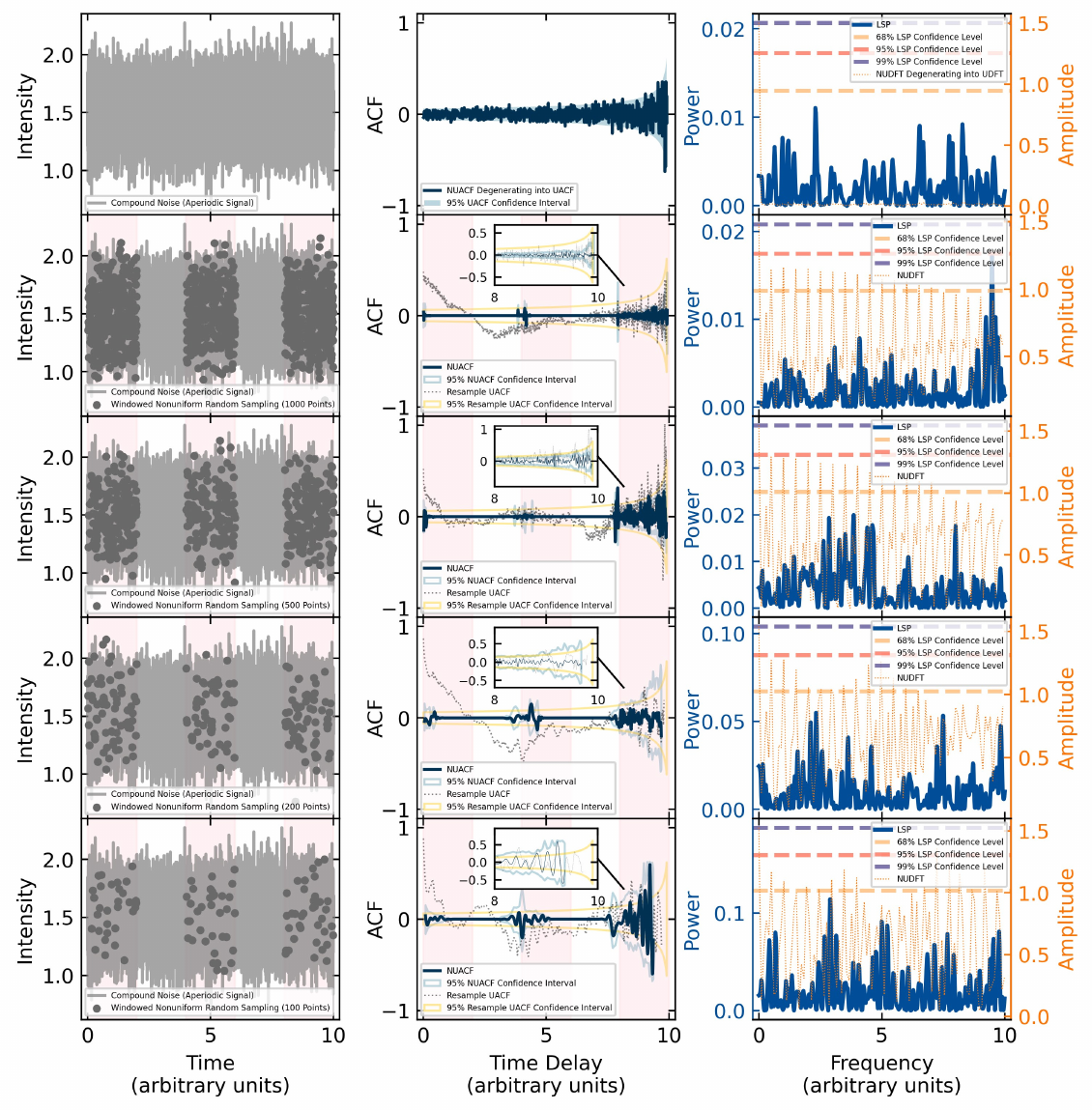}
\figsetgrpnote{Comparison of resampled ACF and our NUACF: pure
compound noise cases (aperiodic signal). A time series of pure
composite (Gaussian $+$ uniform) noise is shown. Our NUACF's confidence
interval correctly indicates no significant correlation.}
\figsetgrpend

\figsetend

\clearpage

We further simulated more realistic ``windowed'' observations,
characteristic of astronomical data sets which are interspersed
with gaps due to observing constraints. For a periodic signal, we
tested the impact of various windowing conditions: constant versus
variable number of data points per window (Figures
\ref{Fig4}.7--\ref{Fig4}.9), different
number of windows (Figures \ref{Fig4}.10--\ref{Fig4}.12),
irregular window spacing
(Figures \ref{Fig4}.13--\ref{Fig4}.15),
and variable window duration (Figures
\ref{Fig4}.16--\ref{Fig4}.18). The
existence of windows degrades the performance of all three methods
to some extent. For the resampled and interpolated ACF, this
manifests as irregular distortion in the amplitude of
peaks/troughs, although their temporal spacing remains roughly
periodic; this distortion becomes severe with sparse sampling.
When data are sparse, they both produce severely distorted
profiles, making it difficult to quantify the presence of a
repetitive pattern and the associated time delay. For our NUACF,
some intrinsic peaks/troughs may lose significance. From a
robustness standpoint, the NUACF framework holds a distinct
advantage thanks to its significance assessment by engaging the
confidence interval.

Based on these initial simulations, the resampled and first-order
interpolated ACF also seem to give a good performance. Given the
smaller distortion of the resampled ACF under low-count
conditions, we performed additional simulations to compare it
directly with our NUACF. We tested scenarios including a baseline
trend (Figures \ref{Fig5}.1--\ref{Fig5}.3), high noise levels (Figures
\ref{Fig5}.4--\ref{Fig5}.6),
quasi-periodic signals (Figures \ref{Fig5}.7--\ref{Fig5}.9),
and pure noise (i.e., aperiodic signal)
(Figures \ref{Fig5}.10--\ref{Fig5}.12). The NUACF's performance, while
degraded, remained robust in detecting a subset of significant
features, confirming its applicability even in these complex
cases. In contrast, the resampled ACF applied to pure noise,
particularly under sparse sampling, can produce spurious
periodic-like profiles. This ambiguity makes it challenging to
distinguish between a signal with a genuine repetitive variability
pattern and aperiodic signal when sampling is limited.

We next evaluated the NUACF's performance on real stellar light
curves from the dataset of An Expandable Light Curve Dataset for
Automatic Classification of Variable Stars (LEAVES)
\citep{2024ApJS..275...10F}. This dataset, hosted by China's
National Astronomical Data Center, provides a homogeneous
collection of stellar light curves by integrating data from
several major surveys: the All-Sky Automated Survey for Supernovae
(ASAS-SN) Catalog of Variable Stars X \citep{2023MNRAS.519.5271C},
Gaia Data Release 3 \citep{2023A&A...674A...1G}, and the Zwicky
Transient Facility (ZTF) Data Release 2
\citep{2019PASP..131a8003M}. Despite the elaborately designed plan
of these surveys, their light curves invariably exhibit
significant nonuniform sampling due to observational constraints,
intrinsic survey modes, and quality control. This makes them an
ideal testbed for our method.

\begin{figure}[!htbp]
\centering
\includegraphics[width=0.64\textwidth]{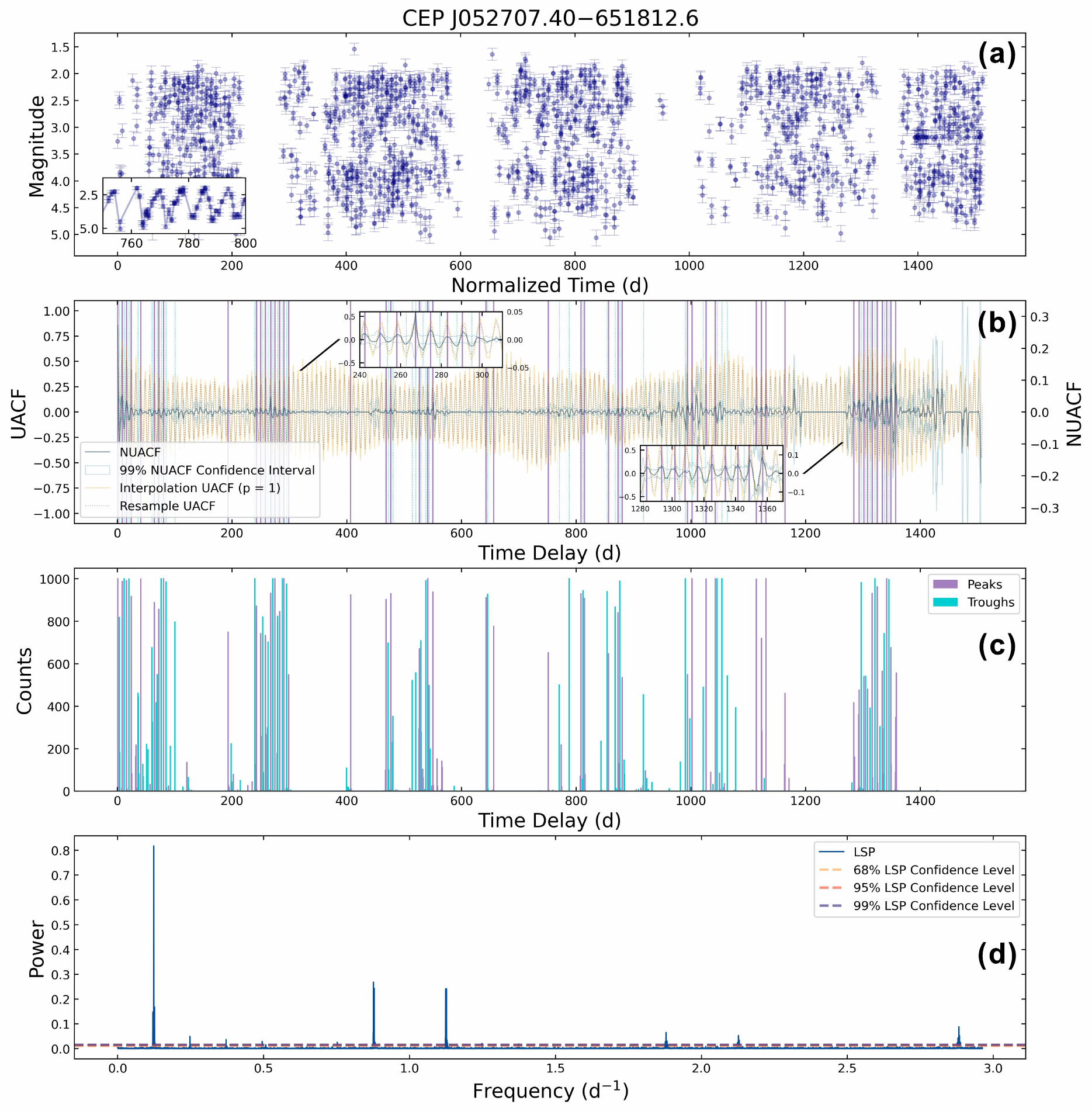}
\caption{Application of our NUACF to a strictly periodic
Cepheid variable, CEP J052707.40$-$651812.6. (a)
Four-year light curve taken from the LEAVES dataset, with the
inset zooming in on the $\sim$10-day period. (b)
Comparison of the different autocorrelation function (ACF)
methods: NUACF (dark blue line, with the $99\%$ confidence
interval shown as a blue region), resampled ACF (gray dotted
line), and first-order interpolated ACF (orange line). The
resampled and interpolated ACFs (collectively labeled UACF) are
shown through the left vertical axis, while the NUACF refers to
the right axis. The NUACF identifies significant peaks (solid
purple vertical lines) and troughs (dotted blue vertical lines)
outside its confidence band. The traditional methods show a
periodic profile but lack a statistical criterion for
significance. (c) Histogram of significant features
identified across MC simulations, validating the features in
panel (b) (high-count bins correspond to significant
peaks/troughs). Based on this histogram and Equation (\ref{eq5}),
our NUACF are able to provide a complete error estimate for the
significant time delays, while other methods cannot.
(d) Corresponding Lomb-Scargle periodogram (LSP), confirming the
dominant period of the light curve shown in panel (a).}
\label{Fig6}
\end{figure}

We first analyzed a Cepheid variable, CEP J052707.40-651812.6, a
strictly periodic star crucial to the cosmic distance ladder
\citep{2001ApJ...553...47F, 2022ApJ...934L...7R}. Figure \ref{Fig6}(a)
shows its light curve over four years, with the inset highlighting
the $\sim$10-day period, also evident in its power spectrum [Figure
\ref{Fig6}(d)]. Figure \ref{Fig6}(b) compares the results from NUACF,
resampled ACF, and first-order interpolated ACF. The NUACF, based
on its $99\%$ confidence interval, robustly identifies the
repetitive pattern, with significant peaks denoted by the solid
purple lines and troughs by the dashed blue lines, respectively.
The other methods also reveal the pattern in their profile
morphology but lack built-in significance assessment and error
estimation for the time delays. Figure \ref{Fig6}(c) shows the
histogram of significant features from the MC simulations used for
error estimation; features with high counts correspond to the
significant peaks/troughs in panel (b).

A key methodological difference is that the resampled and
interpolated ACF rely on periodically spaced features in their
profile to infer a repeating pattern, whereas our NUACF identifies
significant features solely based on the confidence interval
(i.e., NUACF peaks exceeding the upper bound or valleys falling
below the lower bound are deemed significant). This allows the
NUACF to detect non-periodic repetitive patterns with irregular
spacing, for which other methods without a clear confidence
estimation would completely fail. The Lomb-Scargle periodogram is
suitable for simple periodicity analysis
\citep{1976Ap&SS..39..447L,
1982ApJ...263..835S,2018ApJS..236...16V}, while our NUACF is a
time-domain tool that can be efficiently applied to identify
repetitive variability and the corresponding time delays, as
illustrated in the above periodic Cepheid case.

\begin{figure}[!htbp]
\centering
\includegraphics[width=0.7\textwidth]{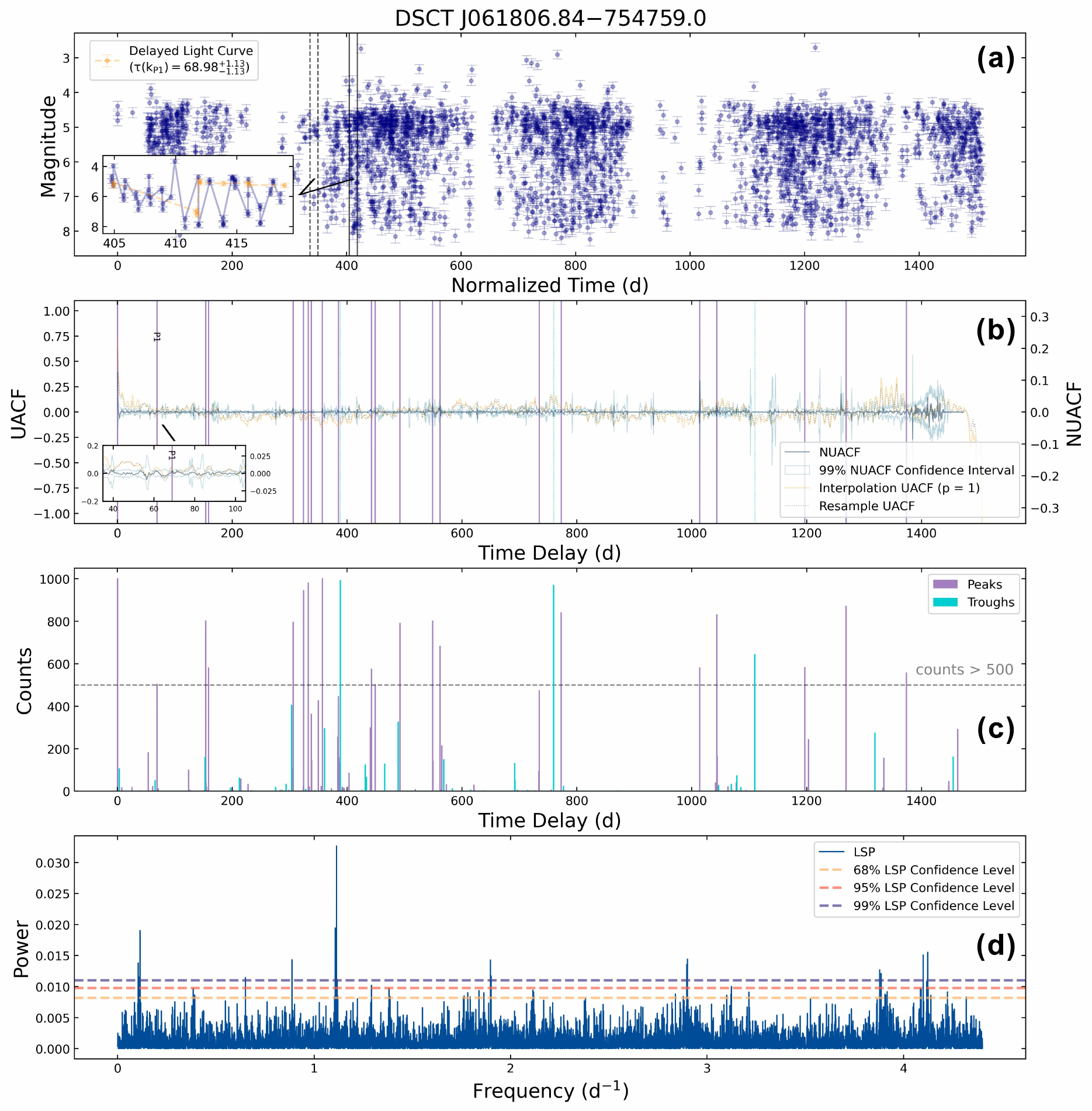}
\caption{NUACF analysis of a multi-periodic Delta Scuti
variable, DSCT J061806.84-754759.0. (a) Light curve
of the star. The inset compares two segments separated by the time
delay of the significant NUACF peak P1, suggesting a repeated
pattern. (b) Results of different ACF methods. The
resampled (gray dotted line) and interpolated (orange line) ACFs
fail to produce a clear interpretable pattern for this complex
signal. In contrast, our NUACF (dark blue line, with the $99\%$
confidence interval shown as the blue region) identifies multiple
significant peaks (solid vertical purple lines) and troughs
(dotted blue vertical lines). (c) MC histogram of
significant NUACF features, corroborating the features in
panel (b) that were derived from the central flux
measurements. (d) LSP analysis revealing eight
frequency components.}
\label{Fig7}
\end{figure}

\begin{figure}[!htbp]
\centering
\includegraphics[width=0.7\textwidth]{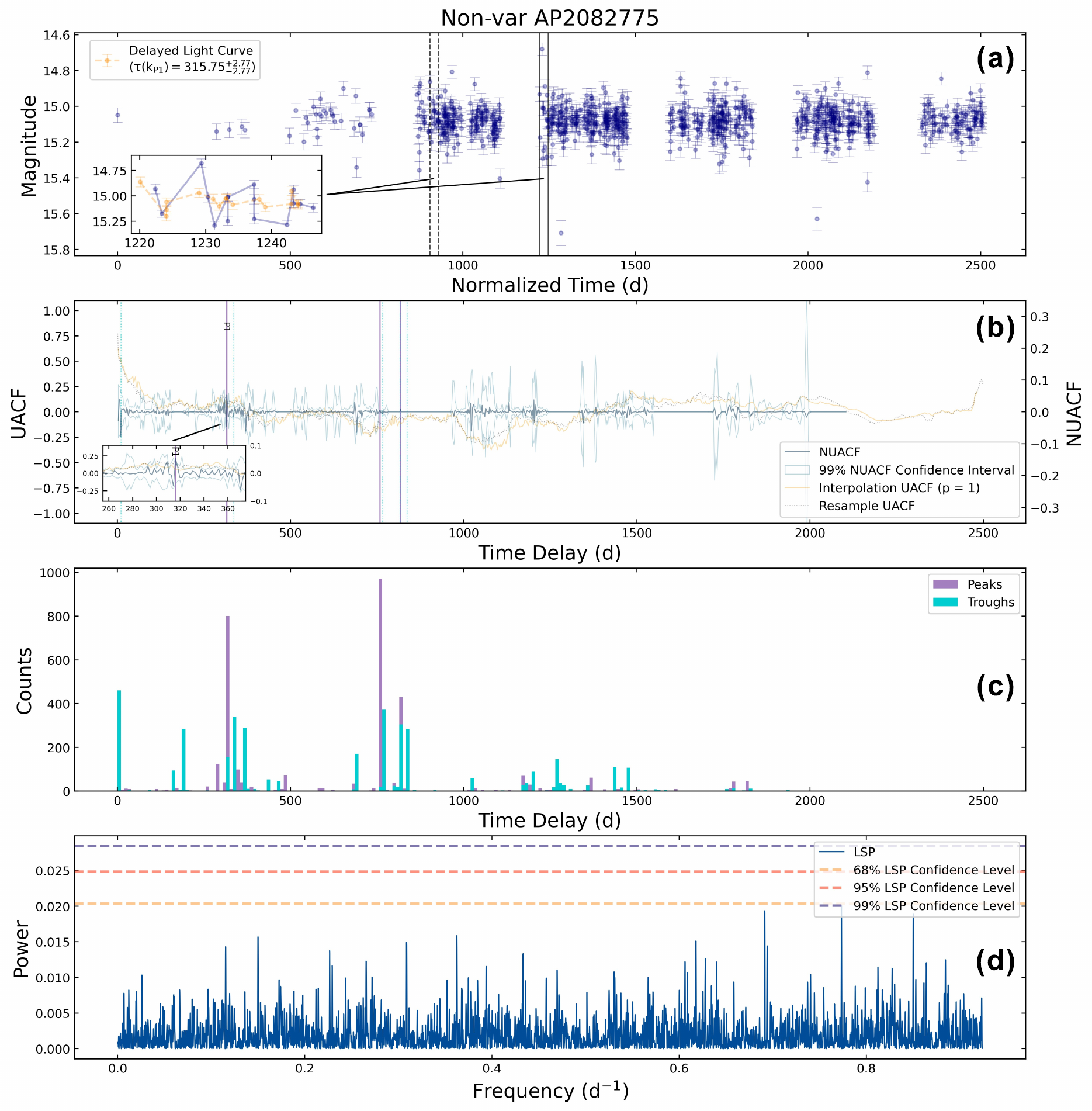}
\caption{NUACF applied to a non-variable star, AP2082775.
(a) Light curve exhibiting small-amplitude,
non-periodic fluctuations. The inset compares two segments
separated by the delay of the most significant NUACF peak, P1,
showing a possible repetition in the activity. (b) The
NUACF (dark blue line, with the $99\%$ confidence interval shown
as the blue region) identifies a number of significant peaks
(e.g., P1) and troughs, whereas the resampled and interpolated
ACFs show no clear pattern. (c) MC histogram
confirming that the detected NUACF features are statistically
robust. (d) The corresponding LSP shows no
periodicity.} \label{Fig8}
\end{figure}

We then analyzed a Delta Scuti variable (DSCT
J061806.84-754759.0), which exhibits multi-periodic pulsations.
Its light curve [Figure \ref{Fig7}(a)] and power spectrum [Figure
\ref{Fig7}(d)] reveal eight frequency components. As shown in Figure
\ref{Fig7}(b), for such complex signals, the resampled and
interpolated ACF become ineffective, while our NUACF successfully
identifies multiple significant peaks and troughs via the
confidence interval method. The inset of Figure \ref{Fig7}(a) compares
two light-curve segments separated by the delay of peak P1,
suggesting a repeated pattern. The corresponding MC histogram is
shown in Figure \ref{Fig7}(c).


Finally, we applied the NUACF to an interesting star, AP2082775,
which exhibits small-amplitude, non-periodic fluctuations [Figure
\ref{Fig8}(a) \& (d)]. Remarkably, our NUACF still identifies a few
significant peaks and troughs [Figure \ref{Fig8}(b)], which are
validated by the MC simulations [Figure \ref{Fig8}(c)]. The inset of
Figure \ref{Fig8}(a) compares two segments separated by the delay of
peak P1, showing a possible match.


\begin{figure}[!htbp]
\centering
\includegraphics[width=0.8\textwidth]{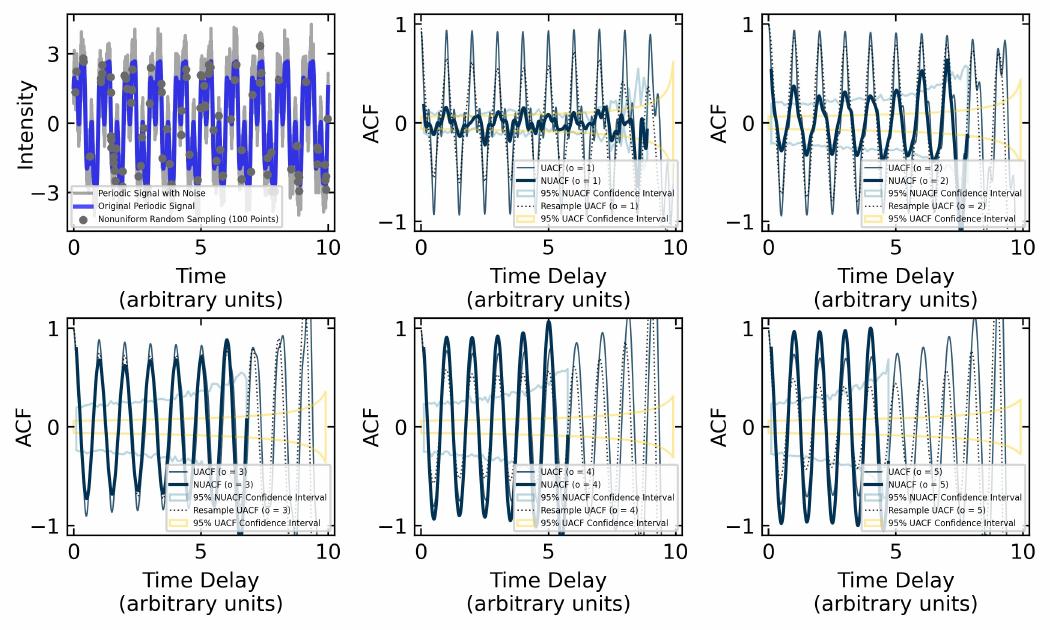}
\caption{Iterated NUACF as a filter on a noisy, nonuniformly
sampled periodic signal. The iterative filtering effects for
different ACF methods are compared. The top-left panel shows the
input: a simulated periodic signal (Pattern A) with added noise,
sampled nonuniformly. The resampled ACF and our NUACF are applied
to this nonuniform data; the standard sample ACF is applied to a
densely and uniformly sampled, quasi-continuous version of the
same underlying signal. Profiles are shown for iteration orders
$o=1$ to $5$ for the three methods, i.e. our NUACF (thick dark
blue line, with a $95\%$ confidence interval shown as light blue
region ), resampled ACF (black dotted line), and standard sample
ACF (thin dark blue line, with a $95\%$ confidence band shown as
yellow region). Here, $o=1$ is the result of applying the ACF
method directly to the data; $o=2$ is obtained by applying the
same method to the $o=1$ result, and so forth. As the iteration
order $o$ increases, the NUACF profile becomes smoother and more
regular, suggesting a purification of the light curve's dominant
frequency.} \label{Fig9}
\end{figure}

\begin{figure}[!htbp]
\centering
\includegraphics[width=0.8\textwidth]{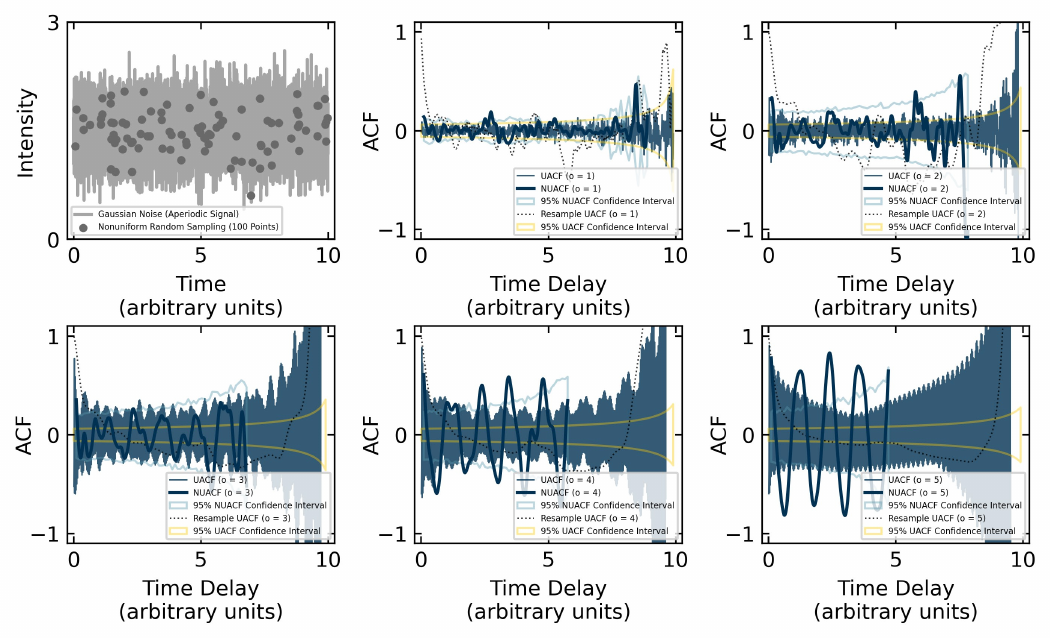}
\caption{Spurious regularity from high-order NUACF applied to
pure noise (aperiodic signal). Shown here is iterative NUACF
applied to a time series of pure Gaussian noise, which can be
treated as a kind of aperiodic signal. While the first-order NUACF
($o=1$) correctly shows no significant structure, higher
iterations ($o\ge3$) produce progressively smoother, regular
profiles. This confirms that beyond its intended role in isolating
the dominant frequency, iterative NUACF acts as a strong low-pass
filter. In practice, we therefore recommend limiting iteration to
order $o=2$. At higher orders, the iterative process
indiscriminately suppresses high-frequency components and
amplifies low-frequency power. Under high iterations, this will
inevitably produce a regular profile even from pure noise.}
\label{Fig10}
\end{figure}

\begin{figure}[!htbp]
\centering
\includegraphics[width=0.8\textwidth]{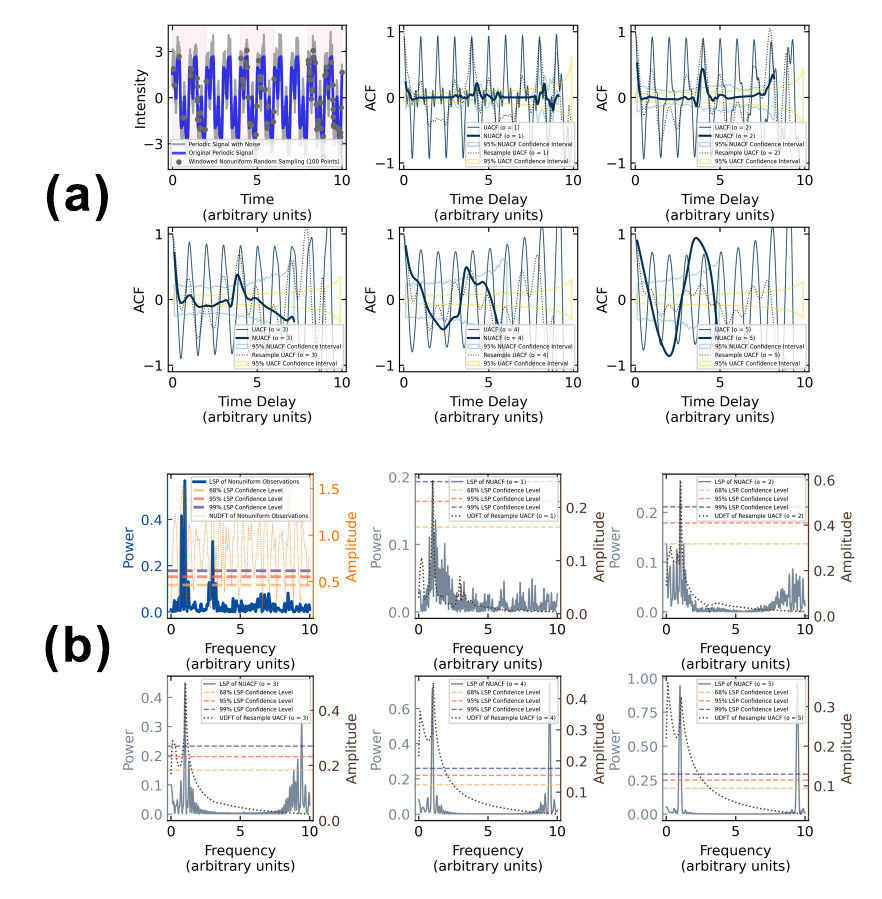}
\caption{Impact of windowed observations on iterated NUACF: periodic
signal. (a) Time-domain profiles of different ACF methods
for iteration orders $o=1$ to $5$ applied to a windowed version of a
periodic signal (Pattern A).
The second-order NUACF ($o=2$) effectively purifies the dominant
frequency; higher orders distort the profile and shift the peak locations.
(b) Corresponding power spectra of panel (a), derived
from LSP (left axis) and NUDFT (right axis). The second-order NUACF
spectrum sharpens the dominant peak, but from the third order onward, the
spectrum distorts with amplified low-frequency power and symmetric
artifacts.}
\label{Fig11}
\end{figure}

\begin{figure}[!htbp]
\centering
\includegraphics[width=0.8\textwidth]{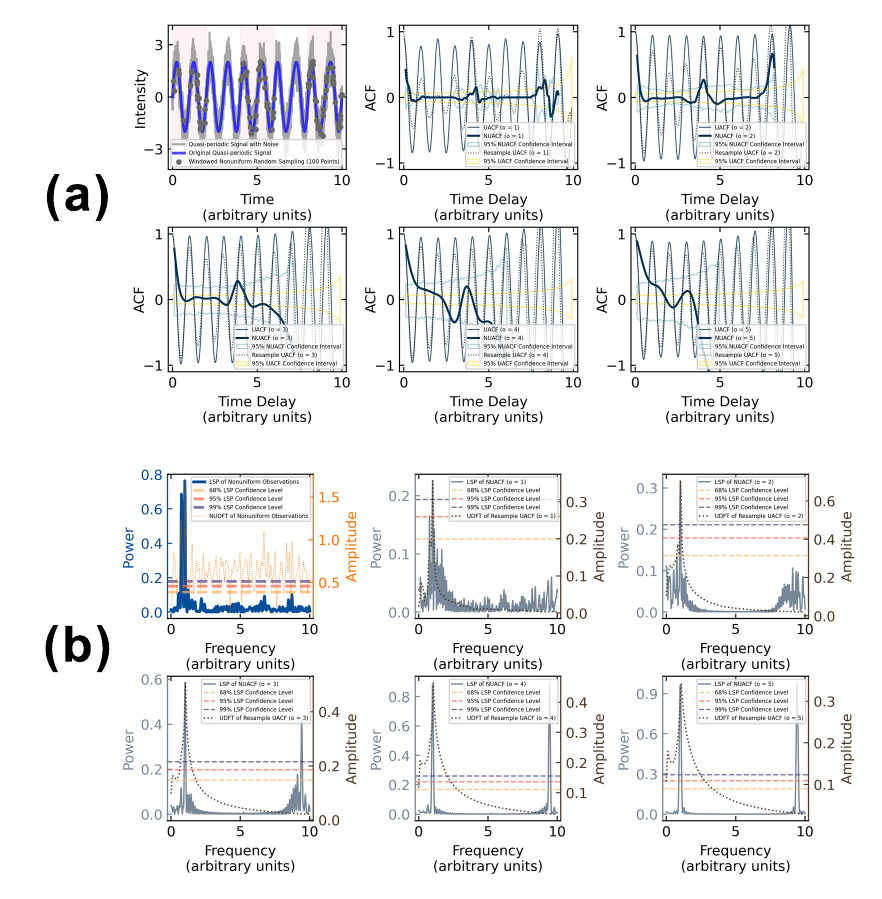}
\caption{Impact of windowed observations on iterated NUACF:
quasi-periodic signal. This figure is similar to
Figure \ref{Fig11}, but for a modulated
(quasi-periodic) signal. The second-order NUACF successfully
enhances the dominant pattern. Higher orders degrade the
time-domain profile and introduce strong spectral distortions.}
\label{Fig12}
\end{figure}

\begin{figure}[!htbp]
\centering
\includegraphics[width=0.8\textwidth]{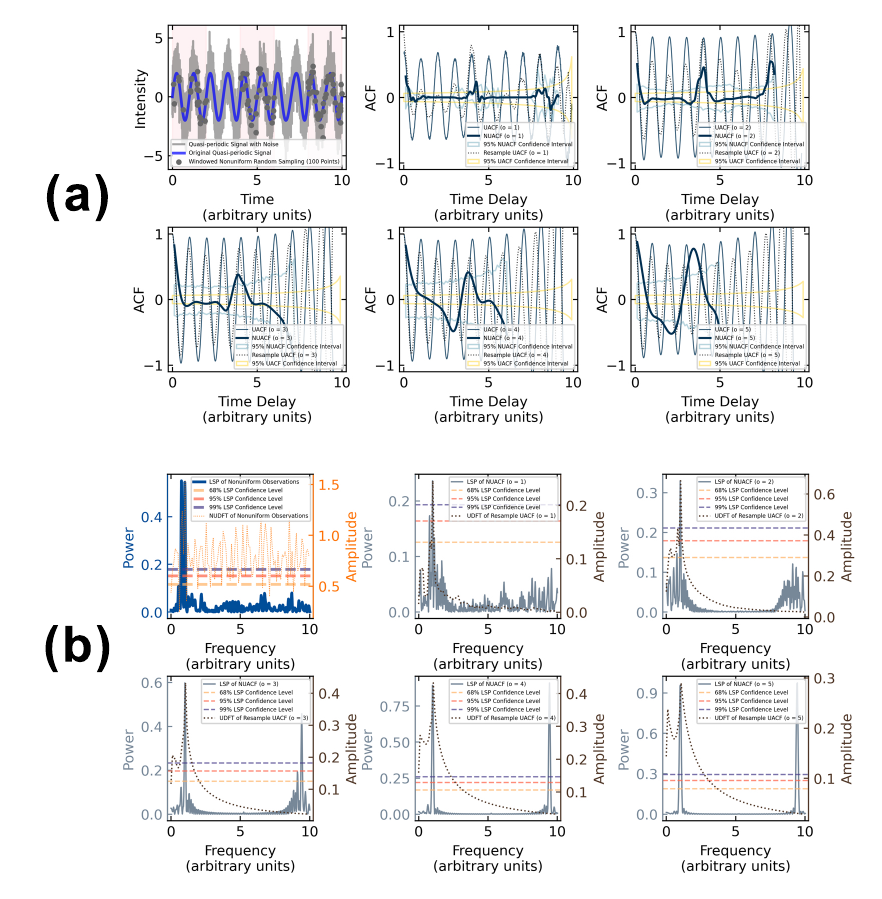}
\caption{Impact of windowed observations on iterated NUACF: severely
noisy quasi-periodic signal. Performance on a quasi-periodic signal (the
same signal as in Figure \ref{Fig12}) with
high noise contamination is shown. The second-order NUACF remains the most
reliable, extracting a cleaner pattern than the first order. Higher
iterations amplify noise artifacts and produce misleadingly smooth
profiles in both time and frequency domains.}
\label{Fig13}
\end{figure}

\begin{figure}[!htbp]
\centering
\includegraphics[width=0.8\textwidth]{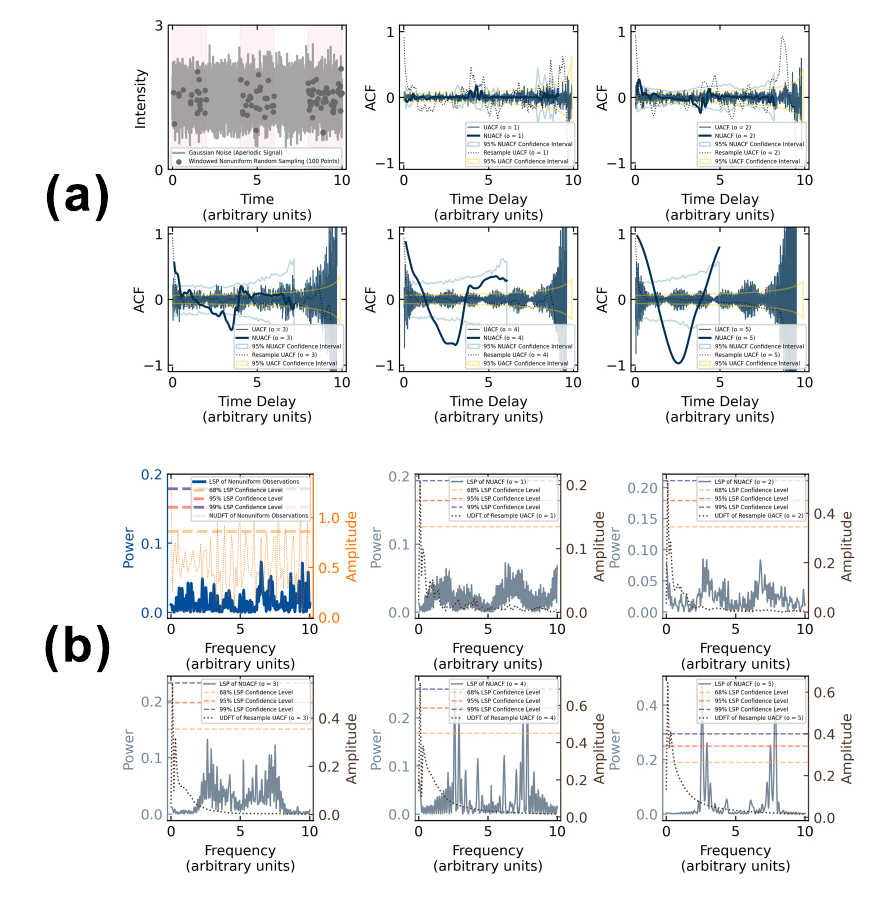}
\caption{Impact of windowed observations on iterated NUACF:
pure noise case (aperiodic signal). Application to pure noise
under windowed sampling is shown. The first- and second-order
NUACF correctly show no significant correlated structure. For
orders $o\ge3$, the iterative filtering creates a smooth,
apparently regular profile and an asymmetric power spectrum,
confirming that high-order iterations generically produce
artifacts.} \label{Fig14}
\end{figure}

Iterative application of the ACF (i.e., applying the ACF to its
own output) can act as a filter, potentially purifying the light
curve's dominant frequency. Figure \ref{Fig9}
compares this effect for different methods on a noisy,
nonuniformly sampled periodic signal. As the iteration order ($o$)
increases, the structure of NUACF profile becomes more prominent,
revealing the potential dominant frequency component. However,
higher orders are not always better. Figure
\ref{Fig10} shows that even for pure noise cases, a
high-order NUACF still produces a structured profile. This occurs
because the iterative process acts as a low-pass filter,
progressively attenuating high-frequency components in the light
curve, regardless of whether they are periodic or not. Given
enough iterations, this process inevitably yields a regular shape.
We therefore recommend limiting NUACF iteration to order 2 in
practice. Figures \ref{Fig11}--\ref{Fig14} test the impact of windowed
observations on iterated NUACF for periodic, quasi-periodic,
severely noisy quasi-periodic, and aperiodic signals. The
second-order NUACF performs well across these cases.  In the
frequency domain (panels (b) of these figures), the second-order
NUACF power spectrum purifies the dominant frequency compared to
the first order. However, from the third order onward, the
spectrum distorts, exhibiting amplified low-frequency components
and axis-symmetric artifacts.


\begin{figure}[!htbp]
\centering
\includegraphics[width=0.8\textwidth]{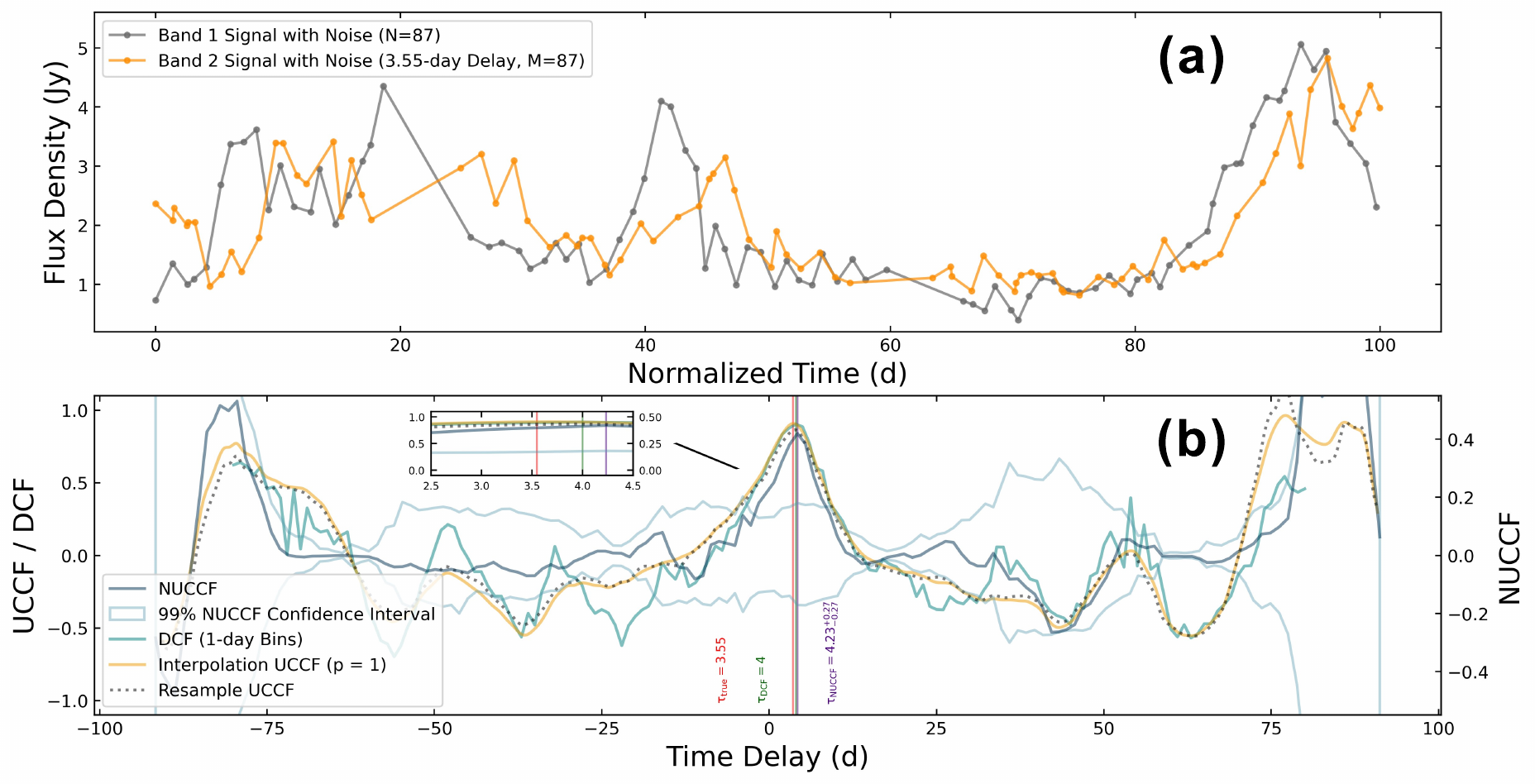}
\caption{Simulation of reverberation mapping with a
pre-assumed time delay. (a) Simulated, nonuniformly
sampled light curves of the seed continuum (Band 1, gray) and the
response emission line (Band 2, orange) with an intrinsic delay of
3.55 days, each with 87 sampling points. (b) CCF
analysis obtained through four different methods: our NUCCF (dark
blue line with $99\%$ confidence interval shown as blue region),
DCF (green line), first-order interpolated CCF (yellow line), and
resampled CCF (gray dotted line). All methods peak near the
pre-assumed delay time, but only the NUCCF provides an error
estimate that directly quantifies the uncertainty arising from
sampling irregularity.}
\label{Fig15}
\end{figure}

\begin{figure}[!htbp]
\centering
\includegraphics[width=0.8\textwidth]{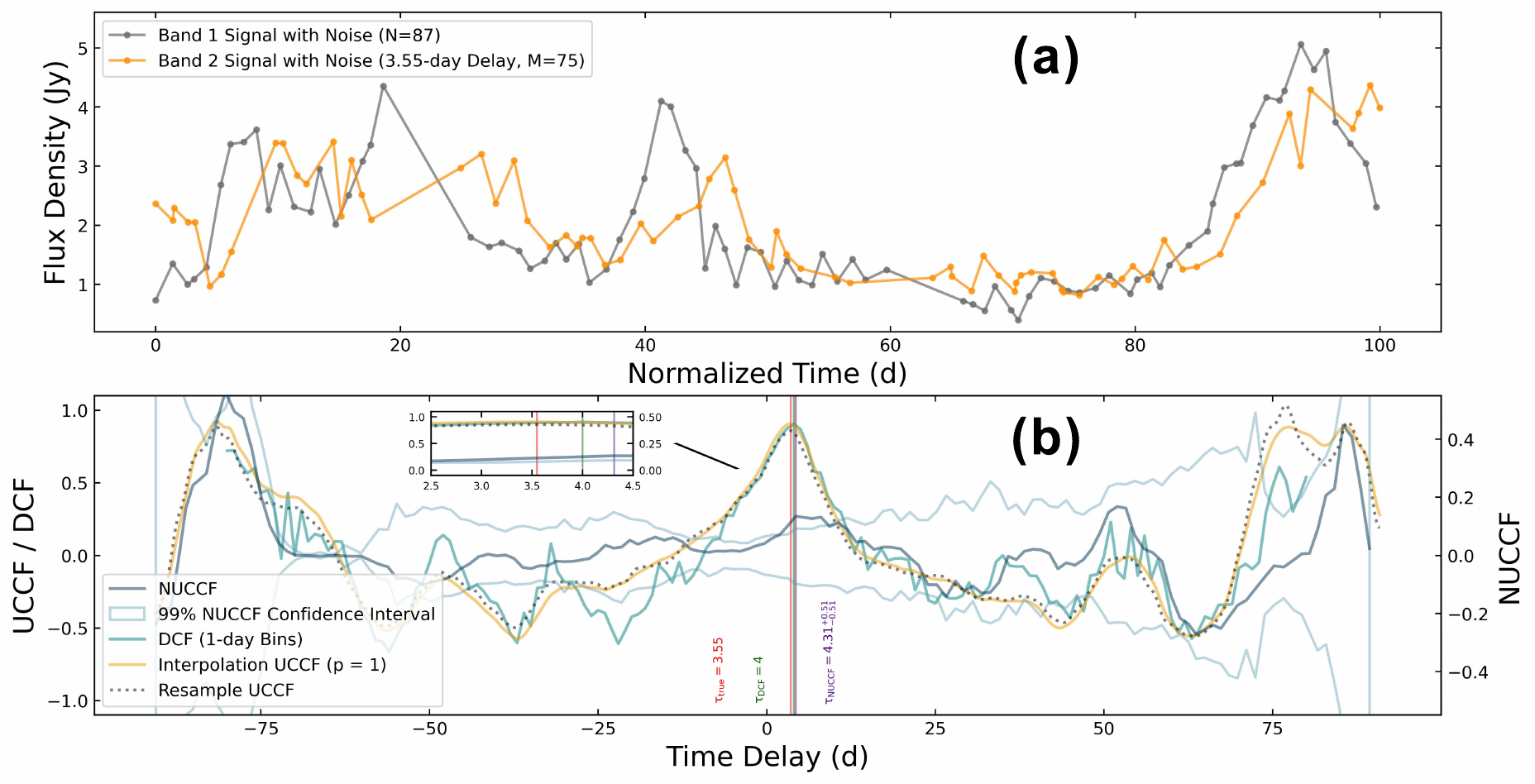}
\caption{Effects of reducing the sampling frequency of the
response band only. This figure is similar to Figure
\ref{Fig15}, but the response band (Band 2) is more
sparsely sampled than the seed band (Band 1). The disparity in
sampling counts between the two bands primarily degrades the
significance of the NUCCF peak (i.e., the height of the peak
relative to its confidence interval), while the peak position
remains close to the intrinsic delay.}
\label{Fig16}
\end{figure}

\begin{figure}[!htbp]
\centering
\includegraphics[width=0.8\textwidth]{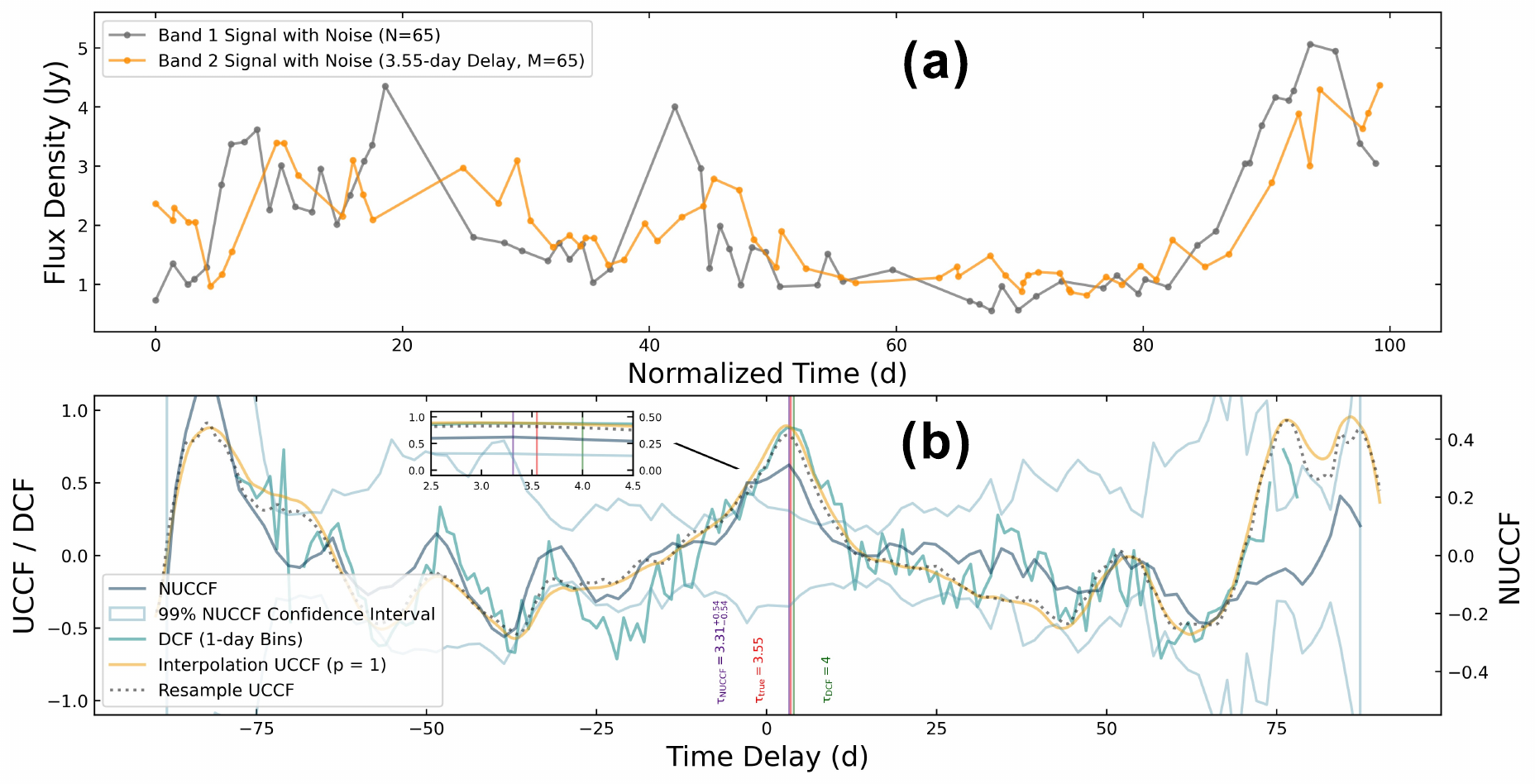}
\caption{Effects of reducing the sampling frequency of both
bands equally. This figure is similar to Figure
\ref{Fig15}, but with equally reduced sampling in
both the seed (Band 1) and response (Band 2) bands. The smaller
total sampling count increases the uncertainty in the time delay
estimate [see Equation (\ref{eq9})]. However, because the sampling
counts of the two bands remain equal here, the significance of the
NUCCF peak is higher than that in Figure \ref{Fig16}
(where the two bands have unequal sampling counts),
though still lower than that in Figure \ref{Fig15} (which has the highest
total sampling counts).}
\label{Fig17}
\end{figure}

Now we turn to the application of NUCCF. Let us consider the
application of CCF in AGN reverberation mapping. In this
framework, the time delay between the UV/optical continuum light
curve (connected to the accretion disk) and the broad
emission-line light curve (relevant to the surrounding gas in the
broad-line region) is measured, providing constraints on the black
hole mass and the size of the broad-line region. We first
performed a test through simulations. Figure
\ref{Fig15}(a) shows the simulated seed (Band 1) and
response (Band 2) light curves with an intrinsic delay of 3.55
days. Figure \ref{Fig15}(b) shows that the
NUCCF, DCF, interpolated CCF, and resampled CCF all peak near the
pre-assumed delay, but only our NUCCF provides an error estimate,
as it quantifies the uncertainty arising from sampling
irregularity itself. Reducing the sampling of the response band
(Figure \ref{Fig16}) or both bands
equally (Figure \ref{Fig17}) reveals two
distinct effects: a large disparity in sampling counts between the
two light curves primarily degrades the NUCCF's significance (the
extent to which the peak exceeds the confidence interval), while a
lower total sampling count is the dominant factor increasing the
uncertainty in the time delay.

\begin{figure}[!htbp]
\centering
\includegraphics[width=0.7\textwidth]{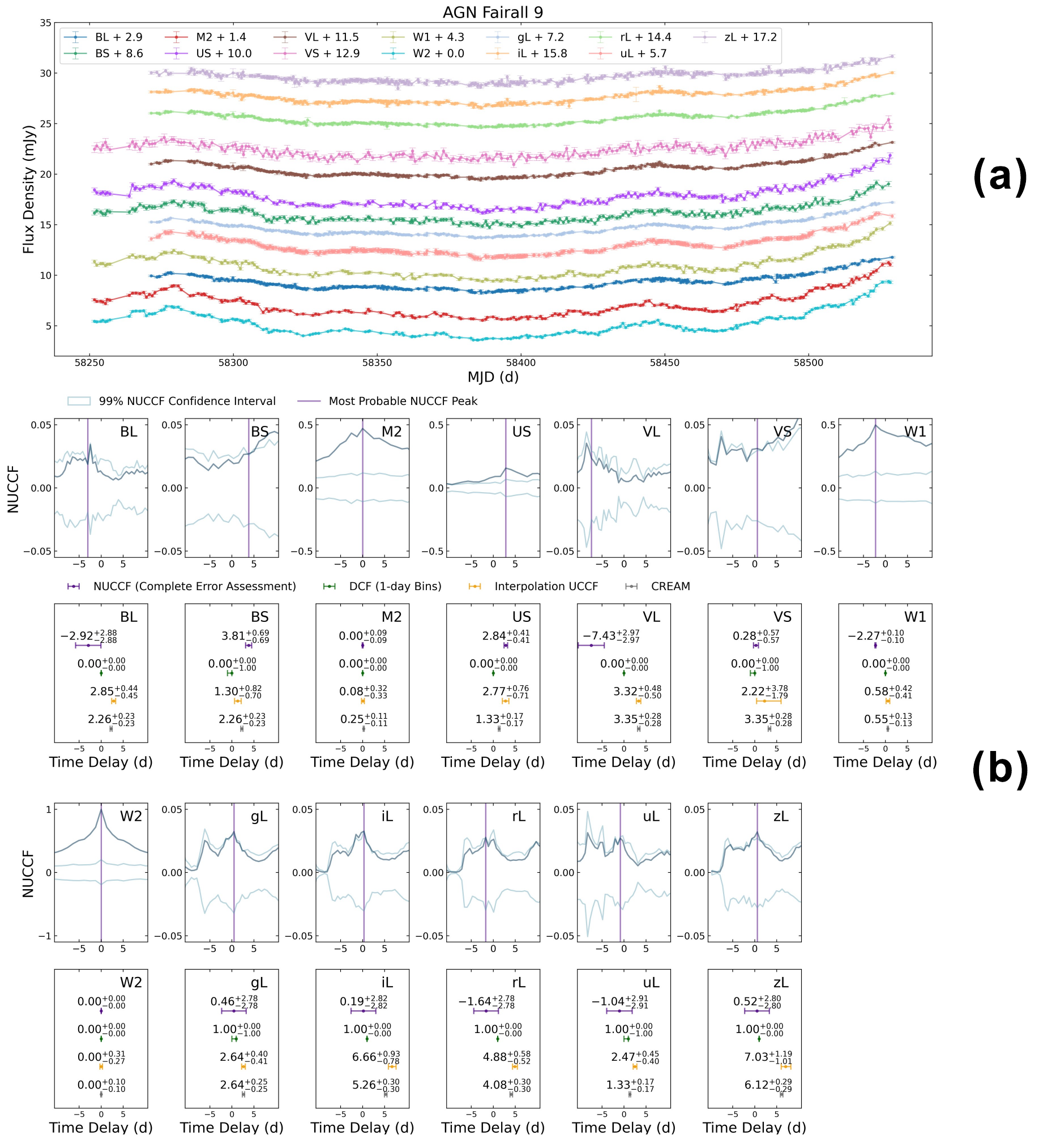}
\caption{Application of our NUCCF to a special AGN, Fairall
9. (a) Intensive, multi-wavelength continuum light
curves of Fairall 9 from a disc-reverberation mapping campaign
\citep{2020MNRAS.498.5399H}. (b) Comparison of time
delays of 13 bands relative to the W2 band, derived by using
four different methods. First/third rows: NUCCF (dark blue line)
with $99\%$ confidence interval (blue region) and the highest
MC-derived peak (purple line). Second/fourth rows: NUCCF delay
(purple) vs. DCF (green), interpolated CCF (yellow, original-study
method), and CREAM model-fitting (gray, original-study method).
The DCF yields nearly zero delays for all bands, failing to
provide meaningful constraints. The NUCCF agrees with the
interpolated CCF and CREAM results in 7 of the 13 bands. In
several bands (most significantly in W1), our NUCCF indicates a
delay of the opposite sign.}
\label{Fig18}
\end{figure}

To demonstrate NUCCF's utility for modern reverberation mapping,
which increasingly employs dense, multi-band monitoring to probe
smaller length scales of the central engine
\citep{2017MNRAS.466.1777P, 2020MNRAS.498.5399H}, we apply the
NUCCF to the intensive, multi-wavelength light curves of the AGN
Fairall 9 [Figure \ref{Fig18}(a); \citealt{2020MNRAS.498.5399H}]. Figure
\ref{Fig18}(b) compares the measured time delays of other bands
relative to the W2 band obtained with different methods, including
the DCF and our NUCCF, as well as the interpolated CCF and the
CREAM model-fitting method  as employed in the original authors'
study \citep{2016MNRAS.456.1960S, 2017ApJ...835...65S}. The DCF
fails to provide meaningful constraints, as it gives nearly zero
delay for all the bands. Our NUCCF agrees with the other two
methods in 7 out of 13 bands. In some bands, notably W1, it
suggests a delay of opposite sign with high significance. While
our NUCCF is intrinsically data-faithful, this case further
highlights that conclusions drawn from finite data can be
contentious.


\begin{figure}[!htbp]
\centering
\includegraphics[width=0.8\textwidth]{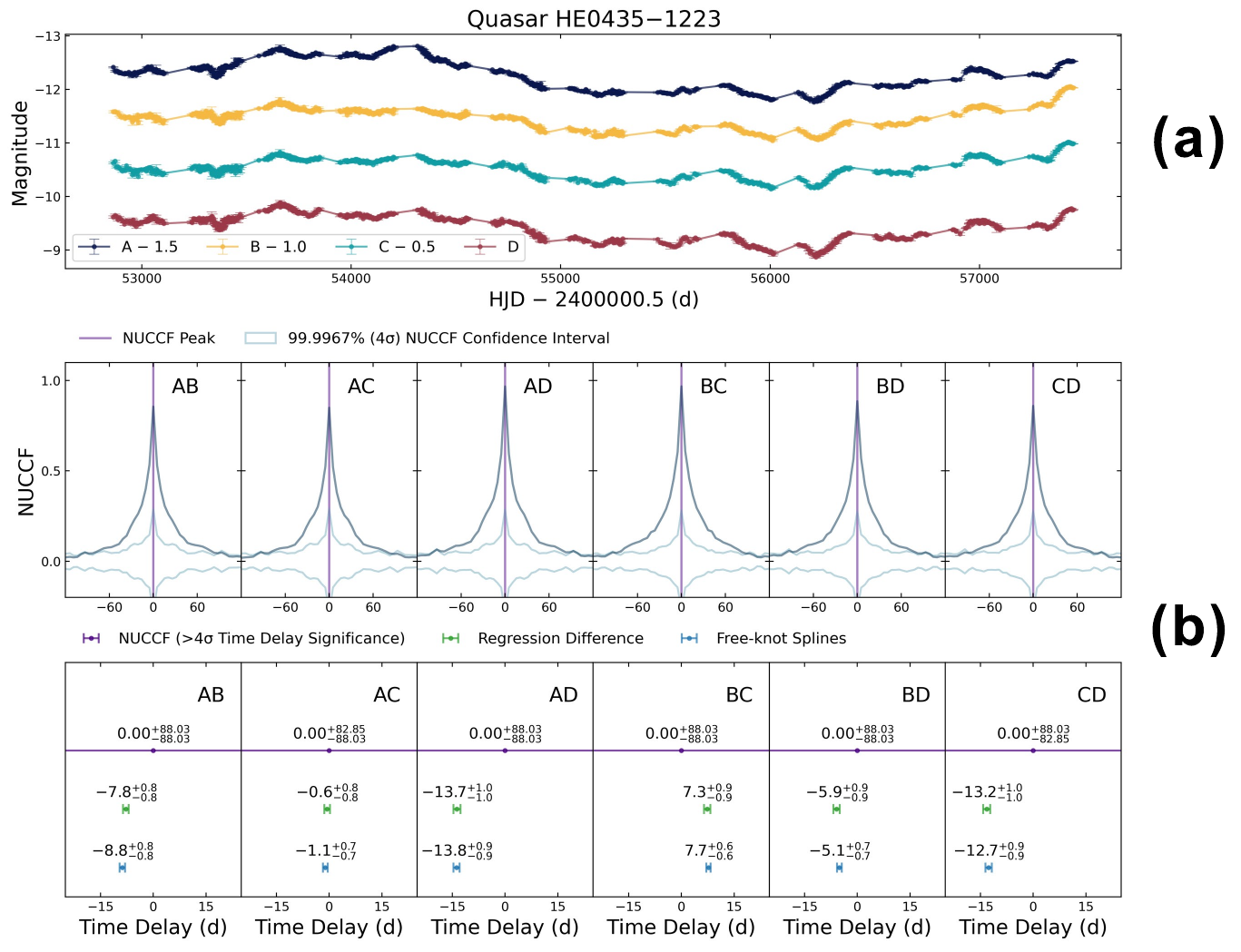}
\caption{Our NUCCF analysis of the lensed quasar HE
0435$-$1223. (a) Light curves of the four lensed
images (A, B, C, D) of the quasar, spanning over a decade of
monitoring with a mean sampling interval of $\sim$5 days (data
from \citet{2017MNRAS.465.4914B}). (b) Results of
different methods for the time delays between image pairs.
Following the original study, the light curves were preprocessed
with \texttt{PyCS3} to mitigate microlensing effects. First row:
Our NUCCF result (dark blue line) with $4\sigma$ confidence
interval (blue region) and MC-derived most probable peak (purple
line). Second row: our NUCCF delay (purple) vs. regression
difference model-fitting (green, original-study method), and
free-knot splines model-fitting (blue, original-study method). In
this long-baseline, sparsely sampled regime, our NUCCF produces
broad peaks centered near zero delay. The width of these
significant peaks defines a reliable, significance-tested range
(indicated by purple horizontal bars) for the true time delay,
providing an independent model-agnostic constraint.}
\label{Fig19}
\end{figure}

\begin{figure}[!htbp]
\centering
\includegraphics[width=0.8\textwidth]{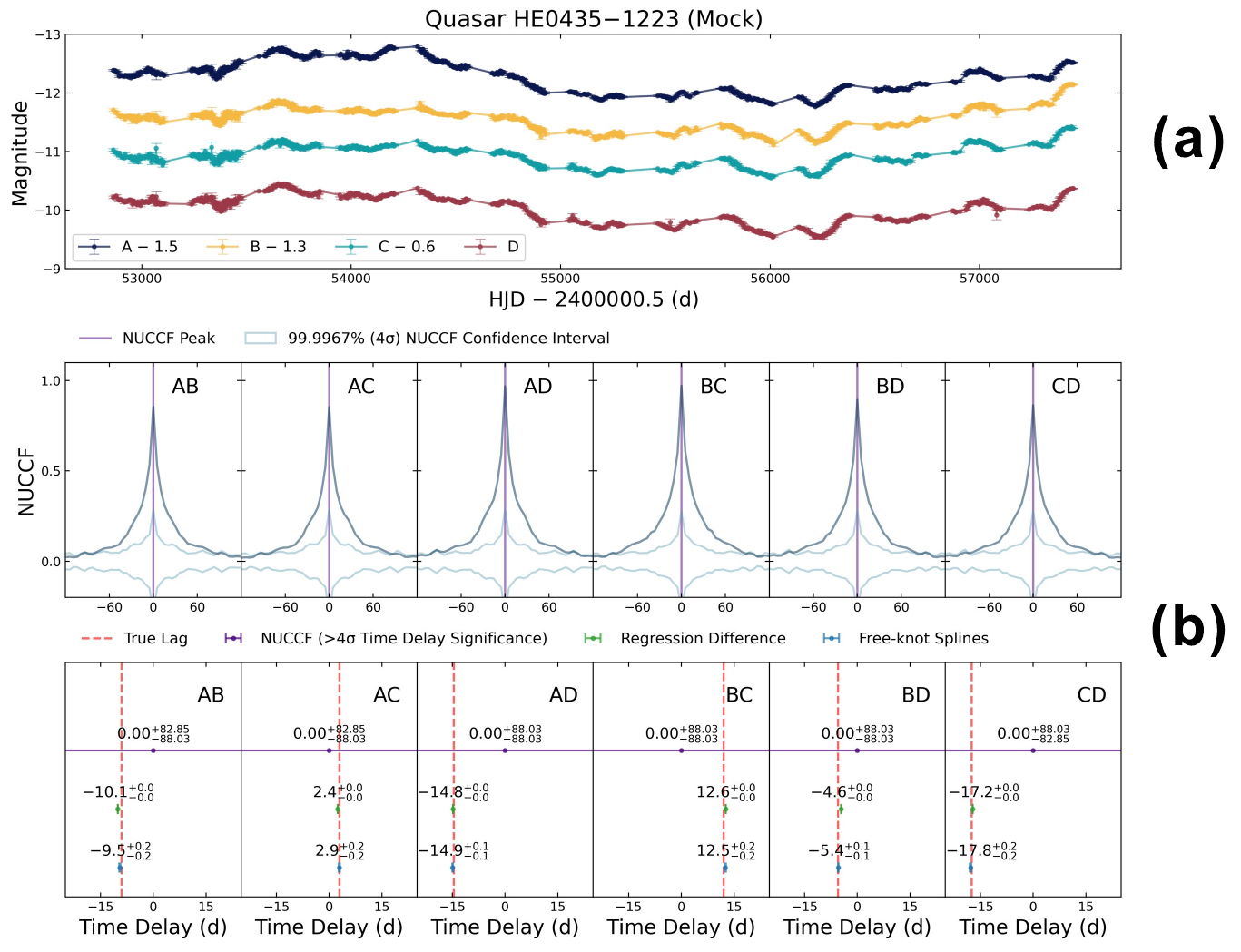}
\caption{Comparing our NUCCF method with other dedicated
model-fitting methods using simulated light curves based on
observational data of HE 0435$-$1223. This figure is similar to
Figure \ref{Fig19}, but using simulated
light curves replicating the sampling and variability
characteristics of HE 0435-1223 [simulated data are taken from
\citet{2017MNRAS.465.4914B}]. The two specialized,
model-fitting techniques (free-knot splines and regression
differences), employed in the original study, provides precise
estimates for the time delays. Both model-fitting estimates fall
within our NUCCF-defined range, offering an independent,
data-driven validation (i.e. $>\!4\sigma$ significance) of their
results. This illustrates the effectiveness of our NUCCF as a
model-independent tool.}
\label{Fig20}
\end{figure}

We also explore the application of NUCCF in gravitational lensing,
in which the time delay between multiple images of a lensed quasar
(created when its light is bent by a foreground massive galaxy)
provides geometric constraints on cosmological parameters
\citep{2013A&A...553A.120T, 2017MNRAS.465.4914B,
2020MNRAS.498.1420W}. Let us take the renowned quadruply lensed
quasar HE 0435$-$1223 \citep{2017MNRAS.465.4914B} as an example,
which is a system with over a decade of monitoring. The results
are shown in Figure \ref{Fig19}, while
Figure \ref{Fig20} presents a comparison
of the performance for different CCF methods. In this
long-baseline but sparsely sampled case (with a mean sampling
interval of $\sim$5 days), our NUCCF yields correlation peaks that
are broad in width and centered near zero delay. We therefore use
the width of these significant peaks to define a reliable,
significance-tested range for the true time delay. This highlights
a key, context-dependent trade-off: while specialized,
model-fitting techniques like free-knot splines and regression
differences \citep{2013A&A...553A.120T}, tailored specifically to
the intrinsic variability of lensed quasars, can leverage the rich
dataset to provide more precise time-delay estimates with smaller
formal errors, the NUCCF provides a robust, model-agnostic bound.
For HE 0435$-$1223, our simulations define a delay range from the
NUCCF that is significant at $>\!4\sigma$. The time-delay
estimates from the two dedicated model-fitting methods both lie
within this range, providing an independent, data-driven
validation of their results at this confidence level. This
illustrates that even in cases where the precision is limited by
sparse sampling, our NUCCF still offers a complementary,
model-independent constraint for validating results from more
specialized techniques.

\section{Summary}
\label{sec5:Summary}

In summary, our NUACF/NUCCF method can effectively solve the
problem induced by nonuniform sampling in astronomical
observations. It performs the correlation analysis directly by
treating the nonuniform sampling as a kind of temporal noise,
incorporating it directly into the confidence-band
construction (which thus naturally renders the band irregular). We
then use such a band to assess whether the repeating variability
patterns exist. In this way, we do not need to eliminate the
sampling irregularity via resampling or interpolation. Our
analytic formulation integrates robust significance assessment and
complete error estimation, both of which are generally
lacking in traditional approaches. This enables the method to
function effectively across diverse astrophysical contexts,
serving as a model-agnostic tool and providing a robust, scalable
foundation for analyzing large and complex datasets.

It is worth noting that our method also has its own
limitations. As demonstrated in our simulation tests, strongly
windowed observations and sparse sampling can degrade NUACF
performance. For the NUCCF, a large disparity in sampling counts
between the two light curves primarily reduces the significance of
the detected peak, while a low total sampling count increases the
uncertainty of the measured time delay. In practice, these
limitations could be mitigated by improving the quality of the
observational data.

In the context of gravitational lens time-delay measurements,
for instance, dedicated model-fitting methods can often provide more
precise time-delay estimates with smaller formal errors, whereas our
method offers a model-independent, analytically grounded cross-check
that can assess the significance of such results and help validate them.

As a natural extension, our framework can also be readily extended to
investigate other nascent time-domain astrophysical sources, including
FRBs, by probing the relationship between persistent radio source
luminosity and the activity of repeating FRBs \citep{2026arXiv260307123L},
or examining the interconnections among the various physical variability
curves intrinsic to FRB radiation.

\clearpage


\section*{Data Availability}

All simulations evaluating the performance of the NUACF and NUCCF
were conducted using synthetic data generated by code, without
employing observational data. The stellar light curves used to
validate the NUACF performance were sourced from the LEAVES
dataset (\citealt{Yu_LEAVES_2024}; available at 
\href{https://nadc.china-vo.org/res/r100962/}
{https://nadc.china-vo.org/res/r100962/}; also see the survey paper of 
\citealt{2024ApJS..275...10F}). Hosted by China's
National Astronomical Data Center, the LEAVES dataset integrates
stellar light curves from several major surveys: ASAS-SN Catalog
of Variable Stars X (\citealt{2023MNRAS.519.5271C};
\href{https://asas-sn.osu.edu/variables}
{https://asas-sn.osu.edu/variables}), Gaia Data Release 3
(\citealt{2023A&A...674A...1G};
\href{https://www.cosmos.esa.int/web/gaia/dr3}
{https://www.cosmos.esa.int/web/gaia/dr3}), and ZTF Data Release 2
(\citealt{2019PASP..131a8003M};
\href{https://irsa.ipac.caltech.edu/Missions/ztf.html}
{https://irsa.ipac.caltech.edu/Missions/ztf.html}). The AGN
Fairall 9 data used to validate the performance of NUCCF were
obtained from \citet{2020MNRAS.498.5399H} (also see 
\citealt{hernandez_santisteban_2020_3956577}; the multi-band light 
curves are available at \href{https://zenodo.org/records/3956577}
{https://zenodo.org/records/3956577}),
while the light curves for the four images of the lensed quasar HE
0435$-$1223 were sourced from \citet{2017MNRAS.465.4914B} and are
publicly available at
\href{https://shsuyu.github.io/H0LiCOW/site/h0licow_data.html}
{\text{https://shsuyu.github.io/H0LiCOW/site/h0licow\_data.html}}.


\section*{Code Availability}

Results can be fully reproduced using the methodology described in
Methods. All analysis scripts and source code, along with all
observational data used in this study, have been deposited in
Zenodo (\doi{10.5281/zenodo.20809667}) for
reproducibility. The package also includes standalone
implementations of the NUACF and NUCCF tools.

Furthermore,
the PyCS3 software toolbox used to preprocess the light curves of
the lensed quasar images by mitigating microlensing effects is
available at
\href{https://gitlab.com/cosmograil/PyCS3/-/blob/master/README.md}
{https://gitlab.com/cosmograil/PyCS3/}. PyCS3 also provides the
two specialized model-fitting approaches (regression difference
and free-knot splines) used in the original study to extract time
delays.


\section*{Acknowledgments}

We are grateful to the anonymous referee for valuable comments and suggestions.
This study is supported by the National Natural Science Foundation
of China (Grant Nos. 12233002, 12622309, 12273113) and by the National Key R\&D Program
of China (2021YFA0718500). Y.-F.H. also acknowledges the support from
the Xinjiang Tianchi Program. J.-J.G. acknowledges support from the Youth
Innovation Promotion Association (2023331). O.A. was also supported by the
Project funded by China Postdoctoral Science Foundation (Grant No. 2025M783225).

\clearpage


\appendix
\vspace{-6mm}

\section{Derivation of the Nonuniform Autocorrelation Function}
\label{appendix A:Derivation of the nonuniform autocorrelation function}

\numberwithin{equation}{section}
\renewcommand{\theequation}{\Alph{section}\arabic{equation}}

Traditional normalized ACF can be conveniently calculated for
uniformly sampled discrete timing sequences. Correspondingly, our
NUACF is a further extension of ACF from the uniform discrete
domain to the nonuniform discrete domain, bridging the gap between
idealized and real observational conditions.

To clarify the logic of this extension, we begin with a familiar concept,
variance, and examine its formulation in both continuous and discrete
domains. For a continuous time series $x\left(t\right)$, the variance is
defined as
\begin{eqnarray}
\label{eq10}
\sigma_{\rm{C}}^2=\lim_{T\to\infty} \frac{1}{2T}\int_{-T}^{T}
{\left[x\left(t\right)-\overline{x\left(t\right)}\right]^2{\rm{d}}t},
\end{eqnarray}
where $\overline{x\left(t\right)}$ denotes the mean of $x\left(t\right)$.
For a finite observation duration, Equation (\ref{eq10}) becomes
\begin{eqnarray}
\label{eq11}
\sigma_{\rm{C}}^2=\frac{1}{T}\int_{0}^{T}{\left[x\left(t\right)-
\overline{x\left(t\right)}\right]^2{\rm{d}}t}.
\end{eqnarray}
Discretizing Equation (\ref{eq11}) yields
\begin{eqnarray}
\label{eq12}
\sigma_{\rm{D}}^2=\frac{1}{t_N-t_1}\sum_{i=1}^{N}{\left(x_i-
\bar{x}\right)^2\Delta t_i},
\end{eqnarray}
where $\Delta t_i$ represents the time interval of the $i$-th
observation, and $N$ is the total number of observations.

For uniformly sampled data where $\Delta t_i=\Delta t$ is
constant, we have $t_N-t_1=\left(N-1\right)\Delta t$, and Equation
(\ref{eq12}) reduces to the common unbiased sample variance:
\begin{eqnarray}
\label{eq13}
\sigma_{\rm{D,U}}^2=\frac{1}{N-1}\sum_{i=1}^{N}\left(x_i-\bar{x}\right)^2.
\end{eqnarray}
For nonuniformly sampled data, we apply the trapezoidal rule to
Equation (\ref{eq12}), obtaining the nonuniform variance
estimator:
\begin{eqnarray}
\label{eq14}
\sigma_{\rm{D,NU}}^2=\frac{1}{t_N-t_1}\left[\sum_{i=2}^{N-1}{\left(x_i-
\bar{x}\right)^2\frac{t_{i+1}-t_{i-1}}{2}}+\left(x_1-
\bar{x}\right)^2\frac{t_2-t_1}{2}+\left(x_N-\bar{x}\right)^2\frac{t_N-t_{N-
1}}{2}\right].
\end{eqnarray}
The variance refers to the spread of a continuous variable
in the time domain (i.e., the $L^2$ norm of a continuous
function), rather than the one in static statistics (e.g., the
spread of student heights in a classroom). This example
illustrates how the transition from the continuous to the discrete
domain differs between uniform and nonuniform sampling. In the
continuous domain, time matters; under uniform sampling, the
time-related terms cancel, yielding the explicitly
time-independent mathematical expression; under nonuniform
sampling, they do not cancel, and the time dependence explicitly
remains.

We now consider the ACF. For a stationary discrete series (i.e., one whose
statistical moments, such as the mean and variance, are time-invariant),
the standard unbiased sample ACF under uniform sampling is
\begin{eqnarray}
\label{eq15}
{\rm{acf_{D,U}}}\left(k\right)=\frac{N}{N-k}\frac{\sum_{i=1}^{N-
k}\left(x_i-\bar{x}\right)\left(x_{i+k}-\bar{x}\right)}
{\sum_{i=1}^{N}\left(x_i-\bar{x}\right)^2},\quad k\in\mathbb{N},k\le N-10,
\end{eqnarray}
where $k$ is the lag in observation number. For uniform sampling, shifting
the series by $k$ points aligns the delayed series perfectly with the
original series, resulting in $N-k$ matched pairs. The corresponding time
delay is $\tau_{\rm{D,U}}\left(k\right)=k\frac{t_n-t_1}{N-1}$.

For nonuniform sampling, however, a simple shift by $k$ positions (i.e.,
$k$ data points) in the sequence does not produce temporally aligned pairs
(see Figure \ref{Fig1}). This presents two challenges: (i) determining an optimal trial
time delay $\tau_{\rm{D,NU}}\left(k\right)$ that minimizes the overall
temporal misalignment between the delayed and original series
for a given trial index lag $k$, and
(ii) robustly comparing the resulting misaligned pairs.

Assuming for the moment that the optimal trial time delay
$\tau_{\rm{D,NU}}\left(k\right)$ is known, we address the second
challenge by introducing a pair-centered time axis. The
representative time for the $i$-th pair is defined as
$\left\{\left[t_i+\tau_{\rm{D,NU}}\left(k\right)\right]+t_{i+k}\right\}/2$
(gray dashed line in Figure \ref{Fig1}). A weight $w_i$ is also
introduced to penalize the residual temporal offset within each
pair,
$\left|\left[t_i+\tau_{\rm{D,NU}}\left(k\right)\right]-t_{i+k}\right|$.
Following the same logic that extends Equation (\ref{eq13}) to
Equation (\ref{eq14}), we generalize the standard sample ACF in
Equation (\ref{eq15}) to the nonuniform domain:
\begin{eqnarray}
\label{eq16}
\begin{aligned}
\hspace{-1cm}
{\rm{acf_{D,NU}}}\left(k\right) = &\ \left[\left\{1/\left[\frac{t_{N-k}+
  \tau_{\rm{D,NU}}\left(k\right)+t_N}{2}-
  \frac{t_1+\tau_{\rm{D,NU}}\left(k\right)+t_{k+1}}{2}\right]\right\}\right. \\
&\ \left(\sum_{i=2}^{N-k-1}{\left(x_i-\bar{x}\right)\left(x_{i+k}-
  \bar{x}\right)\left\{
  {\frac{1}{2}\left[\frac{t_{i+1}+\tau_{\rm{D,NU}}\left(k\right)+t_{i+k+1}}{2}-\frac{t_
  {i-1}+\tau_{\rm{D,NU}}\left(k\right)+t_{i+k-1}}
  {2}\right]}\right\}w_i}\right. \\
&\ +\left(x_1-\bar{x}\right)\left(x_{k+1}-\bar{x}\right)\left\{
  {\frac{1}{2}\left[\frac{t_2+\tau_{\rm{D,NU}}\left(k\right)+t_{k+2}}{2}-
  \frac{t_1+\tau_{\rm{D,NU}}\left(k\right)+t_{k+1}}{2}\right]}\right\}w_1 \\
&\ \left.\left.+\left(x_{N-k}-\bar{x}\right)\left(x_N-\bar{x}\right)\left
  \{{\frac{1}{2}\left[\frac{t_{N-k}+\tau_{\rm{D,NU}}\left(k\right)+t_N}{2}-\frac{t_
  {N-k-1}+\tau_{\rm{D,NU}}\left(k\right)+t_{N-1}}{2}\right]}\right\}w_{N-
  k}\right)\right] \\
&\ /\left(\left[1/\left(t_N-t_1\right)\right]\left\{\sum_{i=2}^{N-1}
  {\left(x_i-\bar{x}\right)^2\left[\frac{1}{2}\left(t_{i+1}-t_{i-
  1}\right)\right]}+\left(x_1-\bar{x}\right)^2\left[\frac{1}{2}\left(t_2-
  t_1\right)\right]+\left(x_N-\bar{x}\right)^2\left[\frac{1}{2}\left(t_N-t_{N-
  1}\right)\right]\right\}\right) \\
  = &\ \left(\left(t_N-t_1\right)\left\{\sum_{i=2}^{N-k-1}{\left(x_i-
  \bar{x}\right)\left(x_{i+k}-\bar{x}\right)\left[\left(t_{i+1}-t_{i-
  1}\right)+\left(t_{i+k+1}-t_{i+k-1}\right)\right]w_i}\right.\right. \\
&\ +\left(x_1-\bar{x}\right)\left(x_{k+1}-
  \bar{x}\right)\left[\left(t_2-t_1\right)+\left(t_{k+2}-
  t_{k+1}\right)\right]w_1 \\
&\ \left.\left.+\left(x_{N-k}-\bar{x}\right)\left(x_N-
  \bar{x}\right)\left[\left(t_{N-k}-t_{N-k-1}\right)+\left(t_N-t_{N-
  1}\right)\right]w_{N-k}\right\}\right) \\
&\ /\left\{\left[\left(t_{N-k}-t_1\right)+\left(t_N-
  t_{k+1}\right)\right]\left[\sum_{i=2}^{N-1}{\left(x_i-
  \bar{x}\right)^2\left(t_{i+1}-t_{i-1}\right)}+\left(x_1-
  \bar{x}\right)^2\left(t_2-t_1\right)+\left(x_N-\bar{x}\right)^2\left(
  t_N-t_{N-1}\right)\right]\right\}, \\
&\ k\in\mathbb{N},k\le N-10,
\end{aligned}
\end{eqnarray}
where the misalignment weight $w_i=\exp{\left\{-\frac{\left(N-
1\right)^2\left[t_i-t_{i+k}+\tau_{\rm{D,NU}}\left(k\right)\right]^2}
{\left(t_N-t_1\right)^2}\right\}}$ is chosen as a Gaussian kernel.
This weight quantifies the specific temporal offset within each
pair, normalized by the mean sampling interval.

Note that the pair-centered time axis we just mentioned is
introduced purely as a virtual computational device to assign a
representative time to each misaligned pair. Its explicit dependence is
algebraically removed in the final expression of Equation (\ref{eq16})
[see the cancellation of $\tau_{\rm{D,NU}}\left(k\right)$ from the
second to the sixth line of Equation (\ref{eq16})].

We then determine the optimal trial time delay
$\tau_{\rm{D,NU}}\left(k\right)$. Under uniform sampling, Equation
(\ref{eq16}) reduces to
\begin{eqnarray}
\label{eq17}
\begin{aligned}
\hspace{-1cm}
{\rm{acf_{D,NU\to U}}}\left(k\right) = &\ \left(\left(N-1\right)\left[\sum_{i=2}^
  {N-k-1}{\left(x_i-\bar{x}\right)\left(x_{i+k}-
  \bar{x}\right)\cdot4}+\left(x_1-\bar{x}\right)\left(x_{k+1}-
  \bar{x}\right)\cdot2+\left(x_{N-k}-\bar{x}\right)\left(x_N-
  \bar{x}\right)\cdot2\right]\right. \\
&\ \left.\exp{\left\{-\frac{\left[k\Delta t-
  \tau_{\rm{D,NU}}\left(k\right)\right]^2}{\left(\Delta
  t\right)^2}\right\}}\right)/\left\{2\left(N-k-
  1\right)\left[\sum_{i=2}^{N-1}{\left(x_i-
  \bar{x}\right)^2\cdot2}+\left(x_1-\bar{x}\right)^2\cdot1+\left(x_N-
  \bar{x}\right)^2\cdot1\right]\right\} \\
  \approx &\ \frac{N-1}{N-k-1}\frac{\sum_{i=1}^{N-k}{\left(x_i-
  \bar{x}\right)\left(x_{i+k}-\bar{x}\right)\exp{\left\{-
  \frac{\left[k\Delta t-\tau_{\rm{D,NU}}\left(k\right)\right]^2}{\left(\Delta
  t\right)^2}\right\}}}}{\sum_{i=1}^{N}\left(x_i-\bar{x}\right)^2},
  \quad k\in\mathbb{N},k\le N-10.
\end{aligned}
\end{eqnarray}
A natural choice that ensures consistency [i.e., that Equation
(\ref{eq17}) reduces to Equation (\ref{eq15}) under uniform
sampling] is $\tau_{\rm{D,NU}}\left(k\right)$ should reduce to
$k\Delta t$ under uniform sampling. Therefore,
$\tau_{\rm{D,NU}}\left(k\right)$ [i.e. Equation (\ref{eq1}) in the
main text] is given by
\begin{eqnarray}
\label{eq18}
\tau_{\rm{D,NU}}\left(k\right)=\overline{t_{i+k}-t_i}=\frac{1}{N-
k}\sum_{i=1}^{N-k}\left(t_{i+k}-t_i\right).
\end{eqnarray}
This definition minimizes the expected value of the residual misalignment
$t_i-t_{i+k}+\tau_{\rm{D,NU}}\left(k\right)$ to zero, thereby providing
the best overall temporal alignment between the series
$\left\{\left.t_i+\tau_{\rm{D,NU}}\left(k\right)\right|i=1,2,\cdots,\right.$
$\left.N-k\right\}$ and $\left\{\left.t_i\right|i=k+1,k+2,\cdots,N\right\}$:
\begin{eqnarray}
\label{eq19}
\begin{aligned}
  E\left[t_i-t_{i+k}+\tau_{\rm{D,NU}}\left(k\right)\right] = &\ E\left[t_i-
  t_{i+k}+\frac{1}{N-k}\sum_{i=1}^{N-k}\left(t_{i+k}-t_i\right)\right] \\
  = &\ E\left(t_i-\frac{1}{N-k}\sum_{i=1}^{N-k}t_i\right)-E\left(t_{i+k}-\frac{1}{N-k}\sum_{i=1}^{N-k}t_{i+k}\right) \\
  = &\ E\left(t_i-\bar{t_i}\right)-E\left(t_{i+k}-\bar{t_{i+k}}\right)=0.
\end{aligned}
\end{eqnarray}
Here we exploit the fact that the mean of a stationary series is
time-invariant.

Substituting Equation (\ref{eq18}) into Equation (\ref{eq16})
leads to the final compact form of the NUACF presented in the main
text [i.e. Equation (\ref{eq2})]:
\begin{eqnarray}
\label{eq20}
{\rm{acf_{D,NU}}}\left(k\right)=\frac{h_{N,1}}
{h_{N-k,1}+h_{N,k+1}}\frac{\sum_{i=1}^{N-k}{\left(x_i-
\bar{x}\right)\left(x_{i+k}-\bar{x}\right)H_i^{\left(2\right)}w_i}}
{\sum_{i=1}^{N}{\left(x_i-\bar{x}\right)^2H_i^{\left(1\right)}}},\quad
k\in\mathbb{N},k\le N-10,
\end{eqnarray}
where we define $h_{m,n}=t_m-t_n$ for conciseness. The discrete weight
factors $H_i^{\left(1\right)}$ and $H_i^{\left(2\right)}$ are derived via
the trapezoidal rule. Along with the misalignment weight $w_i$ in this
final form, they are defined as follows:
\begin{eqnarray}
\label{eq21}
 H_i^{\left(1\right)} = \left\{
    \begin{array}{lc}
         h_{i+1,i},
            \quad i=1, \\
         h_{i+1,i-1},
            \quad 1<i<N, \\
         h_{i,i-1},
            \quad i=N,
    \end{array}
\right.
\end{eqnarray}
\begin{eqnarray}
\label{eq22}
\begin{aligned}
 H_i^{\left(2\right)} = &\ \left\{
    \begin{array}{lc}
         h_{i+1,i}+h_{i+k+1,i+k},
            \quad i=1, \\
         h_{i+1,i-1}+h_{i+k+1,i+k-1},
            \quad 1<i<N-k, \\
         h_{i,i-1}+h_{i+k,i+k-1},
            \quad i=N-k,
    \end{array}
\right. \\
= &\ \left\{
   \begin{array}{lc}
         H_i^{\left(1\right)}+H_{i+k}^{\left(1\right)}-h_{i+k,i+k-1},
            \quad i=1, \\
         H_i^{\left(1\right)}+H_{i+k}^{\left(1\right)},
            \quad 1<i<N-k, \\
         H_i^{\left(1\right)}+H_{i+k}^{\left(1\right)}-h_{i+1,i},
            \quad i=N-k,
    \end{array}
\right.
\end{aligned}
\end{eqnarray}

\begin{eqnarray}
\label{eq23}
w_i=\exp{\left[-\frac{\left(N-1\right)^2\left(h_{i+k,i}-
\overline{h_{i+k,i}}\right)^2}{h_{N,1}^2}\right]}.
\end{eqnarray}
Given an observation-number lag $k$, Equations (\ref{eq18}) and
(\ref{eq20}) directly provide the corresponding physical time
delay $\tau_{\rm{D,NU}}\left(k\right)$ and the NUACF value
${\rm{acf_{D,NU}}}\left(k\right)$. This formulation is
self-consistent in that it degenerates exactly to the standard
sample ACF [Equation (\ref{eq15})] and its associated time delay
when applied to uniformly sampled data.

\section{Confidence Intervals for the Nonuniform Autocorrelation
Function}
\label{appendix B:Confidence intervals for the nonuniform autocorrelation function}

\numberwithin{equation}{section}
\renewcommand{\theequation}{\Alph{section}\arabic{equation}}

We now consider the confidence intervals of our NUACF.
Specifically, to determine whether a peak in the NUACF is
statistically meaningful, we need to assess if such a peak arises from
random fluctuations of the data. The core
idea is to set up a white noise sequence
(i.e. a random signal with a flat power spectrum across all
frequencies) and derive the distribution of the NUACF values at
different lags $k$. As noted in Section \ref{sec1:Introduction},
real observational sampling cannot be adequately
described by an ideal Poisson process. Consequently, the NUACF is
model-independent by construction. This property should naturally
extend to its confidence intervals, a point that will be
demonstrated by the following theoretical derivation.

We begin by reviewing the standard ACF for a uniformly sampled
series. For a white noise sequence, we expect
${\rm{acf_{D,U}}}\left(k\right)=0$ in Equation (\ref{eq15}), since
$x_i$ and $x_{i+k}$ are independent. In practice, due to finite
sample size, the calculated ${\rm{acf_{D,U}}}\left(k\right)$
exhibits fluctuations around zero. From the Central Limit Theorem
(CLT) for independent and identically distributed (i.i.d.)
variables, we have
\begin{eqnarray}
\label{eq24}
\frac{1}{N-k}\sum_{i=1}^{N-k}\left(x_i-\bar{x}\right)\left(x_{i+k}-
\bar{x}\right)\sim\mathcal{N}\left\{E\left[\left(x_i-
\bar{x}\right)\left(x_{i+k}-\bar{x}\right)\right],\frac{D\left[\left(x_i-
\bar{x}\right)\left(x_{i+k}-\bar{x}\right)\right]}{N-k}\right\},
\end{eqnarray}
where
$\mathcal{N}\left(\mu_{\mathcal{N}},\sigma_{\mathcal{N}}^2\right)$
denotes a normal distribution. Leveraging the independence and
stationarity of $\left\{\left.x_i\right|i=1,2,\cdots,N\right\}$,
we have
\begin{eqnarray}
\label{eq25}
E\left[\left(x_i-\bar{x}\right)\left(x_{i+k}-
\bar{x}\right)\right]=E\left(x_i-\bar{x}\right)E\left(x_{i+k}-
\bar{x}\right)=0,
\end{eqnarray}
\begin{eqnarray}
\label{eq26}
\begin{aligned}
  D\left[\left(x_i-\bar{x}\right)\left(x_{i+k}-\bar{x}\right)\right] = &\
  E\left[\left(x_i-\bar{x}\right)^2\left(x_{i+k}-\bar{x}\right)^2\right]-
  \left\{E\left[\left(x_i-\bar{x}\right)\left(x_{i+k}-
  \bar{x}\right)\right]\right\}^2 \\
  = &\ E\left[\left(x_i-\bar{x}\right)^2\right]E\left[\left(x_{i+k}-
  \bar{x}\right)^2\right] \\
  = &\ \left\{D\left(x_i-\bar{x}\right)+\left[E\left(x_i-
  \bar{x}\right)\right]^2\right\}\left\{D\left(x_{i+k}-
  \bar{x}\right)+\left[E\left(x_{i+k}-\bar{x}\right)\right]^2\right\} \\
  = &\ D\left(x_i-\bar{x}\right)D\left(x_{i+k}-
  \bar{x}\right)=\sigma_{\rm{D,U}}^4.
\end{aligned}
\end{eqnarray}
Substituting Equations (\ref{eq25}) and (\ref{eq26}) into
(\ref{eq24}) yields
\begin{eqnarray}
\label{eq27}
\frac{1}{N-k}\sum_{i=1}^{N-k}\left(x_i-\bar{x}\right)\left(x_{i+k}-
\bar{x}\right)\sim\mathcal{N}\left(0,\frac{\sigma_{\rm{D,U}}^4}
{N-k}\right).
\end{eqnarray}

Combining Equation (\ref{eq27}) with Equation (\ref{eq15}) and
using the approximation
$\frac{1}{N}\sum_{i=1}^{N}\left(x_i-\bar{x}\right)^2\approx
\sigma_{\rm{D,U}}^2$, we obtain the distribution for the ACF value
of a white-noise, uniformly sampled series:
\begin{eqnarray}
\label{eq28}
{\rm{acf_{D,U}^{noise}}}\left(k\right)\sim\mathcal{N}\left(0,\frac{1}{N-k}\right).
\end{eqnarray}
For a normal distribution
$\mathcal{N}\left(\mu_{\mathcal{N}},\sigma_{\mathcal{N}}^2\right)$, the
$100\left(1-\alpha\right)\%$ confidence interval is given by
$\left[\mu_{\mathcal{N}}-z_{\alpha/2}\sigma_\mathcal{N},\right.$
$\left.\mu_\mathcal{N}+z_{\alpha/2}\sigma_{\mathcal{N}}\right]$, where
$z_{\alpha/2}$ is the critical value from the standard normal
distribution, satisfying $P\left(Z>\ z_{\alpha/2}\right)=\frac{\alpha}{2}$
for $Z\sim\mathcal{N}\left(0,1\right)$. Thus, the confidence interval for
the standard sample ACF is
\begin{eqnarray}
\label{eq29}
{\rm{acf_{D,U}^{noise}}}\left(k\right)\in\left[-\frac{z_{\alpha/2}}
{\sqrt{N-k}},\frac{z_{\alpha/2}}{\sqrt{N-k}}\right].
\end{eqnarray}
This interval describes the range within which the ACF values of a
purely white-noise, uniformly sampled series are expected to lie.
Therefore, ACF values falling outside this interval for a real
uniform time series can be considered statistically significant.
For consistency with common terminology, we will still refer to
Equation (\ref{eq29}) as a ``confidence interval,'' while noting
that it actually defines a rejection region against ACF
fluctuations arising from noise.

We extend this reasoning to derive the confidence intervals of our
NUACF. A key quantity is $h_{i+k,i}=t_{i+k}-t_i$. If the sampling
times are generated by a Poisson process, the intervals
$h_{m,m-1}$ follow an exponential distribution. Consequently,
$h_{i+k,i}$, being the sum of $k$ such intervals, follows an
Erlang distribution (a distribution widely used in
queueing theory) with the probability density expressed as
\begin{eqnarray}
\label{eq30}
f_{\rm{Erlang}}\left(h_{i+k,i}\right)=\frac{\lambda^k}{\left(k-
1\right)!}h_{i+k,i}^{k-1}\exp\left(-\lambda h_{i+k,i}\right),\quad h_{i+k,i}>0.
\end{eqnarray}
The mean and variance of the Erlang distribution are given by:
\begin{eqnarray}
\label{eq31}
E\left(h_{i+k,i}\right)=\frac{k}{\lambda}=\overline{h_{i+k,i}}=\frac{1}{N-
k}\sum_{i=1}^{N-k}h_{i+k,i},
\end{eqnarray}
\begin{eqnarray}
\label{eq32}
D\left(h_{i+k,i}\right)=\frac{k}{\lambda^2}=\frac{1}{N-k}\sum_{i=1}^{N-k}\left(h_{i+k,i}-\overline{h_{i+k,i}}\right)^2.
\end{eqnarray}
The Erlang distribution reduces to the exponential distribution
when $k=1$. For the case $k=1$, Equation (\ref{eq31}) and the
stationarity of the series yield
\begin{eqnarray}
\label{eq33}
\frac{1}{\lambda}=\frac{1}{N-1}\sum_{i=1}^{N-1}h_{i+1,i}=\frac{h_{N,1}}{N-1},
\end{eqnarray}
where $\lambda$ is the event rate of the Poisson process, which is
assumed to be constant.

If we extend the observed time series by a preceding instant $t_0$
and a following instant $t_{N+1}$, Equation (\ref{eq20}) can be
rewritten as
\begin{eqnarray}
\label{eq34}
{\rm{acf_{D,NU}}}\left(k\right)=\frac{h_{N,1}}{h_{N-
k,1}+h_{N,k+1}}\frac{\sum_{i=1}^{N-k}{\left(x_i-
\bar{x}\right)\left(x_{i+k}-\bar{x}\right)\left(h_{i+1,i-1}+h_{i+k+1,i+k-
1}\right)\exp{\left[-\frac{\left(N-1\right)^2\left(h_{i+k,i}-
\overline{h_{i+k,i}}\right)^2}{h_{N,1}^2}\right]}}}{\sum_{i=1}^{N}
{\left(x_i-\bar{x}\right)^2h_{i+1,i-1}}}.
\end{eqnarray}
Substituting Equations (\ref{eq31}) and (\ref{eq33}) into
(\ref{eq34}), we get
\begin{eqnarray}
\label{eq35}
{\rm{acf_{D,NU}}}\left(k\right)=\frac{h_{N,1}}{h_{N-
k,1}+h_{N,k+1}}\frac{\sum_{i=1}^{N-k}{\left(x_i-
\bar{x}\right)\left(x_{i+k}-\bar{x}\right)\left(h_{i+1,i-1}+h_{i+k+1,i+k-
1}\right)\exp{\left[-\left(\lambda h_{i+k,i}-k\right)^2\right]}}}
{\sum_{i=1}^{N}{\left(x_i-\bar{x}\right)^2h_{i+1,i-1}}}.
\end{eqnarray}
Let us define $A_i=\left(x_i-\bar{x}\right)\left(x_{i+k}-
\bar{x}\right)\left(h_{i+1,i-1}+h_{i+k+1,i+k-1}\right)\exp{\left[-
\left(\lambda h_{i+k,i}-k\right)^2\right]}$. From the CLT for
i.i.d. variables, we have
\begin{eqnarray}
\label{eq36}
\sum_{i=1}^{N-k}A_i\sim\mathcal{N}\left[\left(N-
k\right)E\left(A_i\right),\left(N-k\right)D\left(A_i\right)\right].
\end{eqnarray}
Given the independence and stationarity of $\left\{
\left.x_i\right|i=1,2,\cdots,N\right\}$, we obtain
\begin{eqnarray}
\label{eq37}
\begin{aligned}
  E\left(A_i\right)= &\ E\left\{\left(x_i-\bar{x}\right)\left(x_{i+k}-
  \bar{x}\right)\left(h_{i+1,i-1}+h_{i+k+1,i+k-1}\right)\exp{\left[-
  \left(\lambda h_{i+k,i}-k\right)^2\right]}\right\} \\
  = &\ E\left(x_i-\bar{x}\right)E\left(x_{i+k}-\bar{x}\right)E\left\{\left(h_{i+1,i-1}+h_{i+k+1,i+k-1}\right)\exp{\left[-\left(\lambda h_{i+k,i}-k\right)^2\right]}\right\} \\
  = &\ 0\cdot0\cdot E\left\{\left(h_{i+1,i-1}+h_{i+k+1,i+k-1}\right)\exp{\left[-\left(\lambda h_{i+k,i}-k\right)^2\right]}\right\}=0,
\end{aligned}
\end{eqnarray}
\begin{eqnarray}
\label{eq38}
\begin{aligned}
  D\left(A_i\right) = &\ E\left(A_i^2\right)-
  \left[E\left(A_i\right)\right]^2 \\
  = &\ E\left[\left(x_i-\bar{x}\right)^2\right]\ E\left[\left(x_{i+k}-
  \bar{x}\right)^2\right]E\left[\left(h_{i+1,i-1}+h_{i+k+1,i+k-
  1}\right)^2\right]E\left\{\exp{\left[-2\left(\lambda h_{i+k,i}-
  k\right)^2\right]}\right\} \\
  = &\ D\left(x_i-\bar{x}\right)D\left(x_{i+k}-
  \bar{x}\right)E\left[\left(h_{i+1,i-1}+h_{i+k+1,i+k-
  1}\right)^2\right]E\left\{\exp{\left[-2\left(\lambda h_{i+k,i}-
  k\right)^2\right]}\right\} \\
  = &\ \sigma_{\rm{D,U}}^4 E\left[\left(h_{i+1,i-1}+h_{i+k+1,i+k-
  1}\right)^2\right]E\left\{\exp{\left[-2\left(\lambda h_{i+k,i}-
  k\right)^2\right]}\right\}.
\end{aligned}
\end{eqnarray}

Employing Equations (\ref{eq31}) and (\ref{eq32}), the expectation
$E\left[\left(h_{i+1,i- 1}+h_{i+k+1,i+k-1}\right)^2\right]$ in
Equation (\ref{eq38}) is evaluated as:
\begin{eqnarray}
\label{eq39}
\begin{aligned}
  E\left[\left(h_{i+1,i-1}+h_{i+k+1,i+k-1}\right)^2\right] = &\
  D\left(h_{i+1,i-1}+h_{i+k+1,i+k-1}\right)+\left[E\left(h_{i+1,i-
  1}+h_{i+k+1,i+k-1}\right)\right]^2 \\
  = &\ D\left(h_{i+1,i-1}\right)+D\left(h_{i+k+1,i+k-
  1}\right)+\left[E\left(h_{i+1,i-1}\right)+E\left(h_{i+k+1,i+k-
  1}\right)\right]^2 = \frac{20}{\lambda^2}.
\end{aligned}
\end{eqnarray}
The expectation $E\left\{\exp{\left[-2\left(\lambda h_{i+k,i}-
k\right)^2\right]}\right\}$ in Equation (\ref{eq38}) is computed
by substituting Equation (\ref{eq30}) and letting $g=\lambda
h_{i+k,i}-\left(k-\frac{1}{4}\right)$:
\begin{eqnarray}
\label{eq40}
\begin{aligned}
  E\left\{\exp{\left[-2\left(\lambda h_{i+k,i}-k\right)^2\right]}\right\}
  = &\ \int_{0}^{\infty}{\exp{\left[-2\left(\lambda h_{i+k,i}-
  k\right)^2\right]}\frac{\lambda^k}{\left(k-1\right)!}h_{i+k,i}^{k-
  1}\exp\left(-\lambda h_{i+k,i}\right){\rm{d}}h_{i+k,i}} \\
  = &\ \frac{\exp{\left(-k+\frac{1}{8}\right)}}{\left(k-1\right)!}\int_{-
  k+\frac{1}{4}}^{\infty}{\left(g+k-\frac{1}{4}\right)^{k-
  1}\exp\left(-2g^2\right){\rm{d}}g} \\
  = &\ \frac{\exp{\left(-k+\frac{1}{8}\right)}}{\left(k-1\right)!}\int_{-
  k+\frac{1}{4}}^{\infty}\left[\sum_{r=0}^{k-1}{\frac{\left(k-1\right)!}
  {r!\left(k-r-1\right)!}g^{k-r-1}\left(k-\frac{1}
  {4}\right)^r}\right]\exp\left(-2g^2\right){\rm{d}}g \\
  = &\ \exp{\left(-k+\frac{1}{8}\right)}\sum_{r=0}^{k-
  1}\left[\frac{\left(k-\frac{1}{4}\right)^r}{r!\left(k-r-1\right)!}\int_{-
  k+\frac{1}{4}}^{\infty}{g^{k-r-1}\exp\left(-2g^2\right){\rm{d}}g}\right].
\end{aligned}
\end{eqnarray}
The integral can be expressed in terms of the Gamma function
$\Gamma\left(m\right)=\int_{0}^{\infty}{a^{m-1}{\rm{e}}^{-a}{\rm{d}}a}$
and the lower incomplete Gamma function
$\gamma\left(m,n\right)=\int_{0}^{n}{a^{m-1}{\rm{e}}^{-a}{\rm{d}}a}$:
\begin{eqnarray}
\label{eq41}
\begin{aligned}
  \int_{-k+\frac{1}{4}}^{\infty}{g^{k-r-1}\exp\left(-2g^2\right){\rm{d}}g}
  = &\ \int_{0}^{\infty}{g^{k-r-1}\exp\left(-2g^2\right){\rm{d}}g}+\int_{-
  k+\frac{1}{4}}^{0}{g^{k-r-1}\exp\left(-2g^2\right){\rm{d}}g} \\
  = &\ 2^{-\frac{k-r+2}{2}}\left[\int_{0}^{\infty}
  {\left(2g^2\right)^\frac{k-r-2}{2}\exp\left(-2g^2\right)
  {\rm{d}}\left(2g^2\right)}\right. \\
&\ \left.-\left(-1\right)^{k-r}\int_{0}^{2\left(k-
  \frac{1}{4}\right)^2}{\left(2g^2\right)^\frac{k-r-2}
  {2}\exp\left(-2g^2\right){\rm{d}}\left(2g^2\right)}\right] \\
  = &\ 2^{-\frac{k-r+2}{2}}\left\{\mathrm{\Gamma}\left(\frac{k-r}
  {2}\right)-\left(-1\right)^{k-r}\gamma\left[\frac{k-r}{2},2\left(k-
  \frac{1}{4}\right)^2\right]\right\}.
\end{aligned}
\end{eqnarray}
Substituting Equations (\ref{eq39})--(\ref{eq41}) into
(\ref{eq38}), we have
\begin{eqnarray}
\label{eq42}
\hspace{-0.5cm}
D\left(A_i\right)=\frac{20}{\lambda^2}\sigma_{\rm{D,U}}^4\exp{\left(-
k+\frac{1}{8}\right)}\sum_{r=0}^{k-1}\left(2^{-\frac{k-r+2}
{2}}\frac{\left(k-\frac{1}{4}\right)^r}{r!\left(k-r-1\right)!}\left
\{\mathrm{\Gamma}\left(\frac{k-r}{2}\right)-\left(-1\right)^{k-
r}\gamma\left[\frac{k-r}{2},2\left(k-\frac{1}
{4}\right)^2\right]\right\}\right).
\end{eqnarray}

Inserting Equations (\ref{eq37}) and (\ref{eq42}) into
(\ref{eq36}), and combining the results of Equations (\ref{eq11}),
(\ref{eq33}) and (\ref{eq35}), we obtain the distribution of the
NUACF value under the white noise hypothesis:
\begin{eqnarray}
\label{eq43}
\begin{aligned}
  {\rm{acf_{D,NU}^{noise}}}\left(k\right)\sim &\
  \mathcal{N}\left[0,5\frac{N-k}{\left(N-1\right)^2}\frac{h_{N,1}^2}
  {\left(h_{N-k,1}+h_{N,k+1}\right)^2}\frac{\sigma_{\rm{D,U}}^4}
  {\sigma_{\rm{D,NU}}^4}\exp{\left(-k+\frac{1}{8}\right)}\right. \\
&\ \left.\sum_{r=0}^{k-1}\left(2^{-\frac{k-r+2}{2}}\frac{\left(k-\frac{1}
  {4}\right)^r}{r!\left(k-r-1\right)!}\left\{\mathrm{\Gamma}\left(\frac{k-
  r}{2}\right)-\left(-1\right)^{k-r}\gamma\left[\frac{k-r}{2},2\left(k-
  \frac{1}{4}\right)^2\right]\right\}\right)\right].
\end{aligned}
\end{eqnarray}
Hence, the $100\left(1-\alpha\right)\%$ confidence interval for
the NUACF [i.e. Equation (\ref{eq3}) in the main text] is
\begin{eqnarray}
\label{eq44}
{\rm{acf_{D,NU}^{noise}}}\left(k\right)\in\left[-z_{\alpha/2}V\left(k\right),z_{\alpha/2}V\left(k\right)\right],
\end{eqnarray}
where $V\left(k\right)$ is the square root of the variance
expression in Equation (\ref{eq43}), i.e., the right-hand side of
Equation (\ref{eq43}) is treated as
$\mathcal{N}\left\{0,\left[V\left(k\right)\right]^2\right\}$.
Unlike the uniform-sampling case, the NUACF confidence interval
depends explicitly on the observed time stamps
$\left\{\left.t_i\right|i=1,2,\cdots,N\right\}$. Furthermore,
while the NUACF itself degenerates to the standard sample ACF
under uniform sampling, its confidence interval [Equation
(\ref{eq44})] does not degenerate to Equation (\ref{eq29}). This
is because its derivation employs the Erlang distribution for
temporal intervals, which does not collapse to a distribution
described by the Dirac function under uniform sampling.

To avoid numerical overflow and ensure stability, $V\left(k\right)$ can be
calculated using
\begin{eqnarray}
\label{eq45}
\begin{aligned}
  V\left(k\right) = &\ \left[5\frac{N-k}{\left(N-1\right)^2}\frac{h_{N,1}^2}{\left(h_{N-k,1}+h_{N,k+1}\right)^2}\frac{\sigma_{\rm{D,U}}^4}{\sigma_{\rm{D,NU}}^4}\right. \\
&\ \sum_{r=0}^{k-1}\left(\exp{\left\{\frac{1}{8}-k-\frac{k-r+2}
  {2}\ln{2}+r\ln{\left(k-\frac{1}{4}\right)}-
  \ln{\left[\mathrm{\Gamma}\left(r+1\right)\right]}-
  \ln{\left[\mathrm{\Gamma}\left(k+r\right)\right]}\right\}}\right. \\
&\ \left.\left.\left\{\mathrm{\Gamma}\left(\frac{k-r}{2}\right)-
  \left(-1\right)^{k-r}\gamma\left[\frac{k-r}{2},2\left(k-\frac{1}
  {4}\right)^2\right]\right\}\right)\right]^{0.5}.
\end{aligned}
\end{eqnarray}
The Gamma functions $\Gamma\left(m\right)$ and
$\ln{\left[\Gamma\left(m\right)\right]}$ can be computed via
\texttt{scipy.special.gamma} and \texttt{scipy.special.gammaln},
respectively. The lower incomplete gamma function $\gamma\left(m,n\right)$
is obtained using \texttt{scipy.special.gammainc}.

However, since real observation times are fixed and not strictly
Poissonian, a more practical approach is to derive the confidence
interval via MC simulations of white-noise fluxes at the fixed
observation times. As detailed in the main text (see the
discussion accompanying Figures \ref{Fig2} \& \ref{Fig3}), while the
theoretical derivation provides a useful guide, the fixed-time MC
calculation proves more reliable for practical analysis.
Consequently, we employ MC-based NUACF confidence intervals for
all analyses in this work.

\section{Uncertainty of Time Delays}
\label{appendix C:Uncertainty of time delays}

\numberwithin{equation}{section}
\renewcommand{\theequation}{\Alph{section}\arabic{equation}}

When a significant peak or trough is identified through our NUACF
method, determining the corresponding time delay and its
uncertainty becomes crucial. The uncertainty of time delay stems
from two independent sources: flux measurement errors and temporal
irregularity in the sampling. Traditional methods, such as the
interpolated/resampled ACF or the time-delay-binned DCF typically
assess the contribution from flux errors through MC simulations.
However, they inherently fail to account for the uncertainty
introduced by the temporal irregularity. These conventional
approaches suppress the impact of irregular sampling by
introducing artificial regularization, yet they cannot adequately
incorporate the associated biases into an effective error budget.

Our NUACF framework, in contrast, naturally accommodates a full
uncertainty estimation. We first address the error originating
from temporal irregularity. Based on Equation (\ref{eq18}) and the
CLT, we have:
\begin{eqnarray}
\label{eq46}
\tau_{\rm{D,NU}}\left(k\right)=\overline{h_{i+k,i}}=\frac{1}
{N-k}\sum_{i=1}^{N-
k}h_{i+k,i}\sim\mathcal{N}\left[E\left(h_{i+k,i}\right),\frac{D\left(h_{i+k
,i}\right)}{N-k}\right].
\end{eqnarray}
Substituting the expectation and variance of $h_{i+k,i}$ into
Equation (\ref{eq46}) yields
\begin{eqnarray}
\label{eq47}
\tau_{\rm{D,NU}}\left(k\right)
\sim\mathcal{N}\left[\overline{h_{i+k,i}},\frac{1}{\left(N-k\right)^2}\sum_{i=1}^{N-k}\left(h_{i+k,i}-\overline{h_{i+k,i}}\right)^2\right].
\end{eqnarray}
Consequently, the uncertainty in $\tau_{\rm{D,NU}}\left(k\right)$ due to
temporal irregularity [i.e. Equation (\ref{eq4}) in the main text] is
\begin{eqnarray}
\label{eq48}
\varepsilon_t\left[\tau_{\rm{D,NU}}\left(k\right)\right]=\frac{1}{N-
k}\sqrt{\sum_{i=1}^{N-k}\left(h_{i+k,i}-\overline{h_{i+k,i}}\right)^2}.
\end{eqnarray}
Next, we consider the uncertainty propagated from flux measurement
errors. If, within the error bounds of each flux measurement, a
new flux value is randomly generated to create a new light curve,
the estimated lag $k$ corresponding to the same underlying
physical time delay may fluctuate. This leads to a distribution of
estimated time delays, representing the error contribution from
flux uncertainties. To combine this with the temporal irregularity
error, we employ MC simulations. The procedure is as follows: in
each simulation run, a new signal sequence is generated by
randomizing the fluxes based on their central values and errors.
The NUACF and its confidence band are then used to locate
significant peaks, and Equation (\ref{eq48}) is applied to compute
the temporal irregularity error $\varepsilon_t$ for the time delay
associated with each peak. After completing all runs, for the
$\varphi$-th underlying time delay (corresponding to the
$\varphi$-th significant NUACF peak, ${\rm{P}}_\varphi)$, we
obtain ensembles of time delay estimates
$\left\{\left.\tau_{\rm{D,NU}}^{(\xi)}\left(k_{{\rm{P}}_
\varphi}\right)\right|\xi=1,2,\cdots,S\right\}$ and their
associated temporal irregularity errors
$\left\{\left.\varepsilon_t\left[\tau_
{\rm{D,NU}}^{(\xi)}\left(k_{{\rm{P}}_\varphi}\right)\right]\right|\xi=1,2,\cdots,S\right\}$.

The uncertainty for the sample estimate of the $\varphi$-th
underlying time delay [i.e. Equation (\ref{eq5}) in the main text]
is then given by:
\begin{eqnarray}
\label{eq49}
\varepsilon_{\rm{total}}\left[\tau_{\rm{D,NU}}\left(k_{{\rm{P}}_\varphi}
\right)\right]=\sqrt{\frac{\sum_{\xi}\left[\tau_{\rm{D,NU}}^{(\xi)}\left(k_
{{\rm{P}}_\varphi}\right)-\overline{\tau_{\rm{D,NU}}^{(\xi)}
\left(k_{{\rm{P}}_\varphi}\right)}\right]^2}{S\left(S-1\right)}
+\overline{\left\{\varepsilon_t\left[\tau_{\rm{D,NU}}^
{(\xi)}\left(k_{{\rm{P}}_\varphi}\right)\right]\right\}^2}}.
\end{eqnarray}
Here, $\overline{\tau_{\rm{D,NU}}^{(\xi)}\left(k_
{{\rm{P}}_\varphi}\right)}$ serves as the final sample estimate for the
$\varphi$-th time delay, and Equation (\ref{eq49}) represents its complete
uncertainty. Three important points should be noted in this process: \\
(1) Since the observation times are fixed, the MC-based NUACF
confidence intervals need to be prepared only once for the entire
analysis and can be reused in each simulation run for the complete
error calculation,
eliminating the need for nested MC simulations. \\
(2) As repetitive variability patterns may occur, multiple significant
NUACF peaks can be present. After all simulation runs, the time delays
corresponding to all identified significant peaks can be plotted in a
histogram. The clustering in this histogram (e.g., a Gaussian distribution
around a specific underlying time delay) can then be used to group delays
belonging to the same physical time delay. \\
(3) Increasing the confidence level of the NUACF confidence
interval can help filter out less significant peaks that exhibit
poor clustering in the simulations, thereby focusing the analysis
on robustly detected time delays.

 \section{Extending to the Nonuniform Cross-correlation Function}
\label{appendix D:Extending to the nonuniform cross-correlation function}

\numberwithin{equation}{section}
\renewcommand{\theequation}{\Alph{section}\arabic{equation}}

Building upon the NUACF framework, we now introduce the NUCCF to quantify
the similarity between two irregularly sampled time series,
$\left\{\left.\left(x_i,t_i^x\right)\right|i=1,2,\cdots,N\right\}$ and
$\left\{\left.\left(y_i,t_i^y\right)\right|i=1,2,\cdots,M\right\}$.
As a foundation, for uniformly sampled data, the standard sample CCF is
given by
\begin{eqnarray}
\label{eq50}
{\rm{ccf_{D,U}}}\left(k\right)=\frac{\sqrt{NM}}{i_{\rm{max}}-
i_{\rm{min}}+1}\frac{\sum_{i=i_{\rm{min}}}^{i_{\rm{max}}}\left(x_i-
\bar{x}\right)\left(y_{i+k}-\bar{y}\right)}{\sqrt{\sum_{i=1}^{N}\left(x_i-
\bar{x}\right)^2\sum_{i=1}^{M}\left(y_i-\bar{y}\right)^2}},\quad
k\in\mathbb{Z},-\left(N-10\right)\le k\le M-10,
\end{eqnarray}
where $i_{\rm{min}}={\rm{max}}\left(1,1-k\right)$ and $i_{\rm{max}}=
{\rm{min}}\left(N,M-k\right)$.

Following the same logic used to extend the sample ACF from uniform to
nonuniform sampling [Equations (\ref{eq15})--(\ref{eq23})], we generalize
Equation (\ref{eq50}) to obtain the NUCCF:
\begin{eqnarray}
\label{eq51}
\begin{aligned}
\hspace{-1.65cm}
  {\rm{ccf_{D,NU}}}\left(k\right) = &\
  \left[\left\{1/\left[\frac{t_{i_{\rm{max}}}^x+
  \tau_{\rm{D,NU}}^{xy}\left(k\right)+t_{i_{\rm{max}}+k}^y}{2}-
  \frac{t_{i_{\rm{min}}}^x+\tau_{\rm{D,NU}}^{xy}\left(k\right)+
  t_{i_{\rm{min}}+k}^y}{2}\right]\right\}\right. \\
&\ \left(\sum_{i=i_{\rm{min}}+1}^{i_{\rm{max}}-1}{\left(x_i-
  \bar{x}\right)\left(y_{i+k}-\bar{y}\right)\left\{{\frac{1}{2}\left[
  \frac{t_{i+1}^x+\tau_{\rm{D,NU}}^{xy}\left(k\right)+t_{i+k+1}^y}{2}-
  \frac{t_{i-1}^x+\tau_{\rm{D,NU}}^{xy}\left(k\right)+t_{i+k-1}^y}
  {2}\right]}\right\}w_i}\right. \\
&\ +\left(x_{i_{\rm{min}}}-\bar{x}\right)\left(y_{i_{\rm{min}}+k}-
  \bar{y}\right)\left\{{\frac{1}{2}\left[\frac{t_{i_{\rm{min}}+1}^x+
  \tau_{\rm{D,NU}}^{xy}\left(k\right)+t_{i_{\rm{min}}+k+1}^y}{2}-
  \frac{t_{i_{\rm{min}}}^x+\tau_{\rm{D,NU}}^{xy}\left(k\right)+
  t_{i_{\rm{min}}+k}^y}{2}\right]}\right\}w_{i_{\rm{min}}} \\
&\ \left.\left.+\left(x_{i_{\rm{max}}}-\bar{x}\right)\left(
  y_{i_{\rm{max}}+k}-\bar{y}\right)\left\{{\frac{1}{2}\left[\frac{t_{i_{\rm{max}}}^x+
  \tau_{\rm{D,NU}}^{xy}\left(k\right)+t_{i_{\rm{max}}+k}^y}{2}-
  \frac{t_{i_{\rm{max}}-1}^x+\tau_{\rm{D,NU}}^{xy}\left(k\right)+
  t_{i_{\rm{max}}+k-1}^y}{2}\right]}\right\}
  w_{i_{\rm{max}}}\right)\right], \\
&\ /\left[\left(\left[1/\left(t_N^x-t_1^x\right)\right]\left\{
  \sum_{i=2}^{N-1}{\left(x_i-\bar{x}\right)^2\left[\frac{1}{2}\left(t_{i+1}^x-t_
  {i-1}^x\right)\right]}+\left(x_1-\bar{x}\right)^2\left[\frac{1}{2}\left(t_2^x-
  t_1^x\right)\right]+\left(x_N-\bar{x}\right)^2\left[\frac{1}{2}\left(t_N^x-t_{N-
  1}^x\right)\right]\right\}\right)^{0.5}\right. \\
&\ \left.\left(\left[1/\left(t_M^y-t_1^y\right)\right]\left\{\sum_{i=2}^
  {M-1}{\left(y_i-\bar{y}\right)^2\left[\frac{1}{2}\left(t_{i+1}^y-t_{i-
  1}^y\right)\right]}+\left(y_1-\bar{y}\right)^2\left[\frac{1}{2}\left(t_2^y-
  t_1^y\right)\right]+\left(y_M-\bar{y}\right)^2\left[\frac{1}{2}\left(t_M^y-t_{M-
  1}^y\right)\right]\right\}\right)^{0.5}\right] \\
  = &\ \left(\left[\left(t_N^x-t_1^x\right)\left(t_M^y-
  t_1^y\right)\right]^{0.5}\left\{\sum_{i=i_{\rm{min}}+1}^{i_{\rm{max}}-1}{\left(x_i-
  \bar{x}\right)\left(y_{i+k}-\bar{y}\right)\left[\left(t_{i+1}^x-t_{i-
  1}^x\right)+\left(t_{i+k+1}^y-t_{i+k-
  1}^y\right)\right]w_i}\right.\right. \\
&\ +\left(x_{i_{\rm{min}}}-\bar{x}\right)\left(y_{i_{\rm{min}}+k}-
  \bar{y}\right)\left[\left(t_{i_{\rm{min}}+1}^x-
  t_{i_{\rm{min}}}^x\right)+\left(t_{i_{\rm{min}}+k+1}^y-
  t_{i_{\rm{min}}+k}^y\right)\right]w_{i_{\rm{min}}} \\
&\ \left.\left.+\left(x_{i_{\rm{max}}}-\bar{x}\right)\left(y_{i_{\rm{max}}+k}-
  \bar{y}\right)\left[\left(t_{i_{\rm{max}}}^x-
  t_{i_{\rm{max}}-1}^x\right)+\left(t_{i_{\rm{max}}+k}^y-t_{i_{\rm{max}}+k-
  1}^y\right)\right]w_{i_{\rm{max}}}\right\}\right) \\
&\ /\left(\left[\left(t_{i_{\rm{max}}}^x-t_{i_{\rm{min}}}^x\right)+
  \left(t_{i_{\rm{max}}+k}^y-t_{i_{\rm{min}}+k}^y\right)\right]\left\{\left[
  \sum_{i=2}^{N-1}{\left(x_i-\bar{x}\right)^2\left(t_{i+1}^x-t_{i-
  1}^x\right)}+\left(x_1-\bar{x}\right)^2\left(t_2^x-
  t_1^x\right)+\left(x_N-\bar{x}\right)^2\left(t_N^x-t_{N-
  1}^x\right)\right]\right.\right. \\
&\ \left.\left.\left[\sum_{i=2}^{M-1}{\left(y_i-
  \bar{y}\right)^2\left(t_{i+1}^y-t_{i-1}^y\right)}+\left(y_1-
  \bar{y}\right)^2\left(t_2^y-t_1^y\right)+\left(y_M-
  \bar{y}\right)^2\left(t_M^y-t_{M-1}^y\right)\right]\right\}
  ^{0.5}\right), \\
&\  k\in\mathbb{Z},-\left(N-10\right)\le k\le M-10,
\end{aligned}
\end{eqnarray}
where $w_i=\exp{\left\{-\frac{\left(N-1\right)\left(M-1\right)\left[t_i^x-
t_{i+k}^y+\tau_{\rm{D,NU}}^{xy}\left(k\right)\right]^2}{\left(t_N^x-
t_1^x\right)\left(t_M^y-t_1^y\right)}\right\}}$ and
$\tau_{\rm{D,NU}}^{xy}\left(k\right)$ denotes the time delay at lag $k$.
When the sampling is uniform, with constant intervals $\Delta t^x$ for
series $\left\{x_i\right\}$ and $\Delta t^y$ for series $\left\{
y_i\right\}$, Equation (\ref{eq51}) simplifies to
\begin{eqnarray}
\label{eq52}
\begin{aligned}
  {\rm{ccf_{D,NU\to U}}}\left(k\right) = &\ \left(\left[\left(N-
  1\right)\Delta t^x\left(M-1\right)\Delta t^y\right]^{0.5}\left\{\sum_
  {i=i_{\rm{min}}+1}^{i_{\rm{max}}-1}{\left(x_i-
  \bar{x}\right)\left(y_{i+k}-\bar{y}\right)\left[2\left(\Delta t^x+
  \Delta t^y\right)\right]w_i}\right.\right. \\
&\ \left.\left.+\left(x_{i_{\rm{min}}}-\bar{x}\right)\left(y_{i_{\rm{min}}+
  k}-\bar{y}\right)\left(\Delta t^x+\Delta t^y\right)w_
  {i_{\rm{min}}}+\left(x_{i_{\rm{max}}}-
  \bar{x}\right)\left(y_{i_{\rm{max}}+k}-\bar{y}\right)\left(\Delta
  t^x+\Delta t^y\right)w_{i_{\rm{max}}}\right\}\right) \\
&\ /\left(\left[\left(i_{\rm{max}}-i_{\rm{min}}\right)\left(\Delta t^x+
  \Delta t^y\right)\right]\left\{\left[\sum_{i=2}^{N-1}{\left(x_i-
  \bar{x}\right)^2\left(2\Delta t^x\right)}+\left(x_1-
  \bar{x}\right)^2\Delta t^x+\left(x_N-\bar{x}\right)^2\Delta
  t^x\right]\right.\right. \\
&\ \left.\left.\left[\sum_{i=2}^{M-1}{\left(y_i-\bar{y}\right)^2
  \left(2\Delta t^y\right)}+\left(y_1-\bar{y}\right)^2\Delta t^y+\left(y_M
  -\bar{y}\right)^2\Delta t^y\right]\right\}^{0.5}\right) \\
  \approx &\ \frac{\sqrt{\left(N-1\right)\left(M-1\right)}}{i_{\rm{max}}-
  i_{\rm{min}}}\frac{\sum_{i=i_{\rm{min}}}^{i_{\rm{max}}}{\left(x_i-
  \bar{x}\right)\left(y_{i+k}-\bar{y}\right)\exp{\left\{-\frac{\left[t_i^x-
  t_{i+k}^y+\tau_{\rm{D,NU}}^{xy}\left(k\right)\right]^2}{\Delta t^x\Delta
  t^y}\right\}}}}{\sqrt{\sum_{i=1}^{N}\left(x_i-
  \bar{x}\right)^2\sum_{i=1}^{M}\left(y_i-\bar{y}\right)^2}}, \\
&\ k\in\mathbb{Z},-\left(N-10\right)\le k\le M-10.
\end{aligned}
\end{eqnarray}

To ensure that Equation (\ref{eq52}) degenerates exactly to the uniform sampling
form Equation (\ref{eq50}), it is natural to define the time delay
$\tau_{\rm{D,NU}}^{xy}\left(k\right)$ as the average temporal offset of
the matched pairs [i.e. Equation (\ref{eq7}) in the main text]:
\begin{eqnarray}
\label{eq53}
\tau_{\rm{D,NU}}^{xy}\left(k\right)=\overline{t_{i+k}^y-t_i^x}=\frac{1}
{i_{\rm{max}}-i_{\rm{min}}+1}\sum_{i=i_{\rm{min}}}^
{i_{\rm{max}}}\left(t_{i+k}^y-t_i^x\right).
\end{eqnarray}
This guarantees that the argument of the exponential in $w_i$ has zero expectation,
\begin{eqnarray}
\label{eq54}
\begin{aligned}
  E\left[t_i^x-t_{i+k}^y+\tau_{\rm{D,NU}}^{xy}\left(k\right)\right] = &\
  E\left[t_i^x-t_{i+k}^y+\frac{1}{i_{\rm{max}}-
  i_{\rm{min}}+1}\sum_{i=i_{\rm{min}}}^{i_{\rm{max}}}\left(t_{i+k}^y-
  t_i^x\right)\right] \\
  = &\ E\left(t_i^x-\frac{1}{i_{\rm{max}}-
  i_{\rm{min}}+1}\sum_{i=i_{\rm{min}}}^{i_{\rm{max}}}t_i^x\right)-
  E\left(t_{i+k}^y-\frac{1}{i_{\rm{max}}-
  i_{\rm{min}}+1}\sum_{i=i_{\rm{min}}}^{i_{\rm{max}}}t_{i+k}^y\right) \\
  = &\ E\left(t_i^x-\overline{t_i^x}\right)-E\left(t_{i+k}^y-
  \overline{t_{i+k}^y}\right)=0,
\end{aligned}
\end{eqnarray}
thereby preserving the unbiased character of the estimator in the
uniform sampling limit.

Substituting Equation (\ref{eq53}) into Equation (\ref{eq51}), we arrive at the final,
self consistent form of the NUCCF that appears in the main text
[i.e. Equation (\ref{eq6})]:
\begin{eqnarray}
\label{eq55}
\begin{gathered}
{\rm{ccf_{D,NU}}}\left(k\right)=\frac{\sqrt{h_{N,1}^xh_{M,1}^y}}
{h_{i_{\rm{max}},i_{\rm{min}}}^x+h_{i_{\rm{max}}+k,i_{\rm{min}}+k}^y}\frac{
\sum_{i=i_{\rm{min}}}^{i_{\rm{max}}}{\left(x_i-\bar{x}\right)\left(y_{i+k}-
\bar{y}\right)H_i^{xy,\left(2\right)}w_i^{xy}}}{\sqrt{\sum_{i=1}^{N}
{\left(x_i-\bar{x}\right)^2H_i^{x,\left(1\right)}}\sum_{i=1}^{M}{\left(y_i-
\bar{y}\right)^2H_i^{y,\left(1\right)}}}}, \\
k\in\mathbb{Z},-\left(N-10\right)\le k\le M-10,
\end{gathered}
\end{eqnarray}
where $h_{m,n}^x=t_m^x-t_n^x$ and $h_{m,n}^y=t_m^y-t_n^y$. The discrete
weight factors $H_i^{x,\left(1\right)}$, $H_i^{y,\left(1\right)}$ and
$H_i^{xy,\left(2\right)}$, derived via the trapezoidal rule, along with
the misalignment weight $w_i^{xy}$, are defined as:
\begin{eqnarray}
\label{eq56}
 H_i^{x,\left(1\right)} = \left\{
    \begin{array}{lc}
         h_{i+1,i}^x,
            \quad i=1, \\
         h_{i+1,i-1}^x,
            \quad 1<i<N, \\
         h_{i,i-1}^x,
            \quad i=N,
    \end{array}
\right.
\end{eqnarray}
\begin{eqnarray}
\label{eq57}
 H_i^{y,\left(1\right)} = \left\{
    \begin{array}{lc}
         h_{i+1,i}^y,
            \quad i=1, \\
         h_{i+1,i-1}^y,
            \quad 1<i<M, \\
         h_{i,i-1}^y,
            \quad i=M,
    \end{array}
\right.
\end{eqnarray}
\begin{eqnarray}
\label{eq58}
 H_i^{xy,\left(2\right)} = \left\{
    \begin{array}{lc}
         h_{i+1,i}^x+h_{i+k+1,i+k}^y,
            \quad i=i_{\rm{min}}, \\
         h_{i+1,i-1}^x+h_{i+k+1,i+k-1}^y,
            \quad i_{\rm{min}}<i<i_{\rm{max}}, \\
         h_{i,i-1}^x+h_{i+k,i+k-1}^y,
            \quad i=i_{\rm{max}},
    \end{array}
\right.
\end{eqnarray}
\begin{eqnarray}
\label{eq59}
w_i^{xy}=\exp{\left\{-\frac{\left(N-1\right)\left(M-
1\right)\left[t_{i+k}^y-t_i^x-\overline{t_{i+k}^y-t_i^x}\right]^2}
{\left(t_N^x-t_1^x\right)\left(t_M^y-t_1^y\right)}\right\}}.
\end{eqnarray}

The uncertainty in $\tau_{\rm{D,NU}}^{xy}\left(k\right)$ again
comprises contributions from temporal irregularity and flux
measurement errors. To estimate the uncertainty arising from
temporal irregularity alone, we appeal to the Lindeberg-Feller
CLT \citep{Billingsley1995} for independent, non-identically
distributed variables, which yields the asymptotic distribution:
\begin{eqnarray}
\label{eq60}
\tau_{\rm{D,NU}}^{xy}\left(k\right)\sim\mathcal{N}\left[\overline{t_{i+k}^y
-t_i^x},\frac{1}{\left(i_{\rm{max}}-i_{\rm{min}}+1\right)\left(i_
{\rm{max}}-i_{\rm{min}}\right)}\sum_{i=i_{\rm{min}}}^
{i_{\rm{max}}}\left(t_{i+k}^y-t_i^x-\overline{t_{i+k}^y-
t_i^x}\right)^2\right].
\end{eqnarray}
Consequently, the temporal irregularity error [i.e. Equation (\ref{eq9}) in the
main text] is
\begin{eqnarray}
\label{eq61}
\varepsilon_t\left[\tau_{\rm{D,NU}}^{xy}\left(k\right)\right]=\sqrt{\frac{
\sum_{i=i_{\rm{min}}}^{i_{\rm{max}}}\left(t_{i+k}^y-t_i^x-
\overline{t_{i+k}^y-t_i^x}\right)^2}{\left(i_{\rm{max}}-
i_{\rm{min}}+1\right)\left(i_{\rm{max}}-i_{\rm{min}}\right)}}.
\end{eqnarray}

The error component from flux measurement uncertainties is assessed via MC
simulations, following a procedure analogous to that described for the
NUACF in Appendix \ref{appendix C:Uncertainty of time delays}. For each underlying
time delay corresponding to a significant NUCCF peak ${\rm{P}}_\varphi$,
the ensemble of estimates $\left\{\left.\tau_{\rm{D,NU}}^{xy,
(\xi)}\left(k_{{\rm{P}}_\varphi}\right)\right|\xi=1,2,\cdots,S\right\}$
and their associated temporal errors $\left\{\left.\varepsilon_t\left[\tau_
{\rm{D,NU}}^{xy,(\xi)}\left(k_{{\rm{P}}_\varphi}\right)
\right]\right|\xi=1,2,\cdots,\right.$
$S\left.\right\}$ are combined to give the complete
uncertainty [i.e. Equation (\ref{eq8}) in the main text]:
\begin{eqnarray}
\label{eq62}
\varepsilon_{\rm{total}}\left[\tau_{\rm{D,NU}}^{xy}\left(k\right)\right]=
\sqrt{\frac{\sum_{\xi}\left[\tau_{\rm{D,NU}}^{xy,
(\xi)}\left(k_{{\rm{P}}_\varphi}\right)-\overline{\tau_{\rm{D,NU}}^{xy,
(\xi)}\left(k_{{\rm{P}}_\varphi}\right)}\right]^2}{S\left(S-
1\right)}+\overline{\left\{\varepsilon_t\left[\tau_{\rm{D,NU}}^{xy,
(\xi)}\left(k_{{\rm{P}}_\varphi}\right)\right]\right\}^2}},
\end{eqnarray}
where $\overline{\tau_{\rm{D,NU}}^{xy,(\xi)}\left(k_{{\rm{P}}
_\varphi}\right)}$ is the final sample estimate (i.e., the mean of the
ensemble) for the $\varphi$-th time delay. In contrast to
the NUACF case, where multiple significant peaks
may be analyzed, a single time delay is often sought in NUCCF analysis. In
such cases, within a given physically acceptable range, the most frequent
(i.e., maximum-likelihood) peak identified across the MC simulations can
be selected as the final result.

As argued in Section \ref{sec2:Nonuniform autocorrelation function},
confidence intervals for the NUCCF should also be derived
via MC simulations, because real observation times are fixed and
not fully amenable to analytic modeling. The procedure, however,
differs from that for the NUACF. To construct the NUCCF confidence
interval, we hold one light curve fixed and replace the flux
values of the other with a randomly generated white-noise
sequence. This process is then repeated with the roles of the two
series swapped, yielding two distinct MC-based confidence
intervals. A conservative, envelope-based final interval is
obtained by taking, at each lag $k$, the larger absolute bound
from the two individual intervals.

Importantly, the construction of confidence intervals for the
NUCCF requires consideration of flux measurements, in contrast to
the NUACF. Consequently, when estimating time-delay errors via MC
simulations, where each run regenerates both light curves based on
their flux uncertainties, the NUCCF confidence interval itself
must be re-simulated within every individual MC run.

\clearpage


\bibliography{Ref}{}
\bibliographystyle{aasjournal}

\end{CJK*}

\end{document}